\documentclass[12pt]{article}
\usepackage{graphicx}
\usepackage{epstopdf}
\usepackage{amsmath}
\usepackage{amsfonts}
\usepackage{amssymb}
\usepackage{color}
\usepackage{mathrsfs}

\newcommand{\be}{\begin{equation}}
\newcommand{\ee}{\end{equation}}
\newcommand{\bea}{\begin{eqnarray}}
\newcommand{\eea}{\end{eqnarray}}
\newcommand{\barr}{\begin{array}}
\newcommand{\earr}{\end{array}}

\newcommand{\vk}{\vec k}

\newcommand{\di}{\partial}

\usepackage[colorlinks,bookmarks]{hyperref}
\definecolor{linkblue}{rgb}{0,0,0.8}
\definecolor{linkgreen}{rgb}{0,0.5,0}

\hypersetup{pdfpagemode=UseNone, pdfstartview=FitH, linkcolor=linkblue, %
            citecolor=linkgreen, urlcolor=linkblue}

\def\beq{\begin{equation}}
\def\eeq{\end{equation}}
\def\be{\begin{equation}}
\def\ee{\end{equation}}
\def\bea{\begin{eqnarray}}
\def\eea{\end{eqnarray}}
\def\d{{\partial}}

\def\mpl{M_{\rm Pl}}
\def\nn{\nonumber}

\newcommand{\tin}{t_{\rm in} } 

\newcommand{\uint}{U_{\rm int}}
\newcommand{\freevacright}{| 0 \rangle}
\newcommand{\freevacleft}{ \langle 0 |}
\newcommand{\vacright}{| \Omega \rangle}
\newcommand{\vacleft}{ \langle \Omega |}

\def\ba{\begin{align}}
\def\ea{\end{align}}

\begin{document}

%\begin{titlepage}

\setcounter{page}{1} \baselineskip=15.5pt \thispagestyle{empty}

\begin{flushright}
%hep-th/yymmnnn\\
\end{flushright}
%\vfil

\begin{center}

\def\thefootnote{\fnsymbol{footnote}}

{\Large \bf Lectures on Inflation}
\\[0.5cm]

{\large Leonardo Senatore}
\\[0.5cm]

{\normalsize {\sl Stanford Institute for Theoretical Physics\\Department of Physics, Stanford University, Stanford, CA 94306}}\\
\vspace{.3cm}

{\normalsize { \sl Kavli Institute for Particle Astrophysics and Cosmology, \\ Stanford University and SLAC, Menlo Park, CA 94025}}\\
\vspace{.3cm}

\end{center}

\vspace{.8cm}

\hrule \vspace{0.3cm}
{\small  \noindent \textbf{Abstract} \\[0.3cm]
Planning to explore the beginning of the Universe? A lightweight {\it guide du routard} for you.  \noindent 
}
 \vspace{0.3cm}
\hrule
%\vfil
%\begin{flushleft}
%\today
%March 20, 2008
%\end{flushleft}

%\end{titlepage}

%\newpage
%\tableofcontents
%\newpage

\section*{Introduction}

The purpose of these lectures on Inflation is to introduce you to the currently preferred theory of the beginning of the universe: the theory of Inflation. This is one of the most fascinating theories in Physics. Starting from the shortcomings of the standard big bang theory, we will see how a period of accelerated expansion solves these issues. We will then move on to explain how inflation can give such an accelerated expansion ({\bf lecture 1}). We will then move on to what is the most striking prediction of inflation, which is the possibility that quantum fluctuations during this epoch are the source of the cosmological perturbations that seed galaxies and all structures in the universe ({\bf lecture 2}). We will then try to generalize the concept of inflation to develop a more modern description of this theory. We will introduce the Effective Field Theory of Inflation. We will learn how to compute precisely the various cosmological observables, and how to simply get the physics out of the Lagrangians~({\bf lecture 3}). Finally, in the last lecture ({\bf lecture 4}), we will discuss one of the most important observational signatures of inflation: the possible non-Gaussianity of the primordial density perturbation. We will see how a detection of a deviation from Gaussianity would let us learn about the inflationary Lagrangian and make the sky a huge particle detector. Time permitting  ({\bf lecture 5}), we will introduce one of the conceptually most beautiful regimes of inflation, the regime of eternal inflation, during which quantum effects become so large to change the asymptotics of the whole space-time.

A video of these lecture, apart for lecture 5, is available at~\cite{video}. These notes are written as a natural complement to those lectures. The language is highly informal. \\
{\bf Notation}
\bea
c=\hbar=1\ ,\qquad \mpl^2=\frac{1}{8\pi G}\ .
\eea

\tableofcontents

%\begin{figure}[h!]
%\begin{center}
%\includegraphics[width=16cm]{routard}
%\caption{\label{fig:routard} \small The dream of every cosmology lecturer: being recommended by the `{\it guide du routard}', the famous French tourist guides. Of course I have not been officially recommended $\ldots$ yet.}
%\end{center}
%\end{figure}

\section{Lecture 1}

One-sentence intro on Inflation: it was incredibly brave in the early 1980's, when the initial formulation of Inflation was made, to apply the most advanced theories from particle physics to the early universe. The results, as you will see, are beautiful.

Notice that we will perform calculations more explicitly when they are less simple. So in this first lecture we will skip some passages. General homework of this class: fill in the gaps. 

\subsection{FRW cosmology}
We begin by setting up the stage with some basic concepts in cosmology to highlight the shortcoming of the standard big bang picture.\\

The region of universe that we see today seems to be well described by an homogenous and isotropic metric. The most general metric satisfying these symmetries can be put in the following form
\be\label{eq:metric1}
ds^2=-dt^2+a(t)^2\left(\frac{dr^2}{1-k r^2}+r^2\left(d\theta^2+\sin^2\theta d\phi^2\right)\right)
\ee
We see that this metric represents a slicing of space-time with spatial slices $\Sigma$ that are simply rescaled by the scale factor $a$ as time goes on.  If $k=0$, we have a flat space, if $k=+1$, the space $\Sigma$ describes a sphere, while if $k=-1$ we have an hyperbolic space. A fundamental quantity is of course the Hubble rate
\be
H=\frac{\dot a}{a}
\ee
which has units of inverse time. It is useful for us to put the metric (\ref{eq:metric1}) into the following form
\be\label{eq:metric2}
ds^2=-dt^2+a(t)^2\left(d\chi^2+S_k(\chi^2)\left(d\theta^2+\sin^2\theta d\phi^2\right)\right)
\ee
where
\be
r^2=S_k(\chi^2)=\left\{
\begin{array}{rl}
\sinh^2\chi & \text{if } k=-1,\\
\chi^2 & \text{if } k= 0,\\
\sin^2\chi & \text{if } k=+1.
\end{array} \right.
\ee
$\chi$ plays the role of a radius.
Let us now change coordinates in time (it is General Relativity at the end of the day!) to something called conformal time
\be
\tau=\int^\tau \frac{dt}{a(t)}\ .
\ee
Now the FRW metric becomes
\be\label{eq:metric3}
ds^2=a(\tau)^2\left[-d\tau^2+d\chi^2+S_k(\chi^2)\left(d\theta^2+\sin^2\theta d\phi^2\right)\right]
\ee
In these coordinates it is particularly easy to see the casual structure of space-time. This is determined by how light propagates on null geodesic $ds^2=0$. Since the space is isotropic, geodesic solutions have constant $\theta$ and $\phi$. In this case we have
\be
\chi(\tau)=\pm \tau+ {\rm const.}
\ee
These geodesics move at $45$ degrees in the $\tau-\chi$ plane, as they would in Minkowski space. This is so because apart for the angular part, the metric in (\ref{eq:metric3}) is conformally flat: light propagates as in Minwkoski space in the coordinates $\tau-\chi$. Notice that this is not so if we had used $t$, the proper time for comoving (i.e. fixed FRW-slicing spatial coordinates) observers.

\begin{figure}[h!]
\begin{center}
\includegraphics[width=10cm]{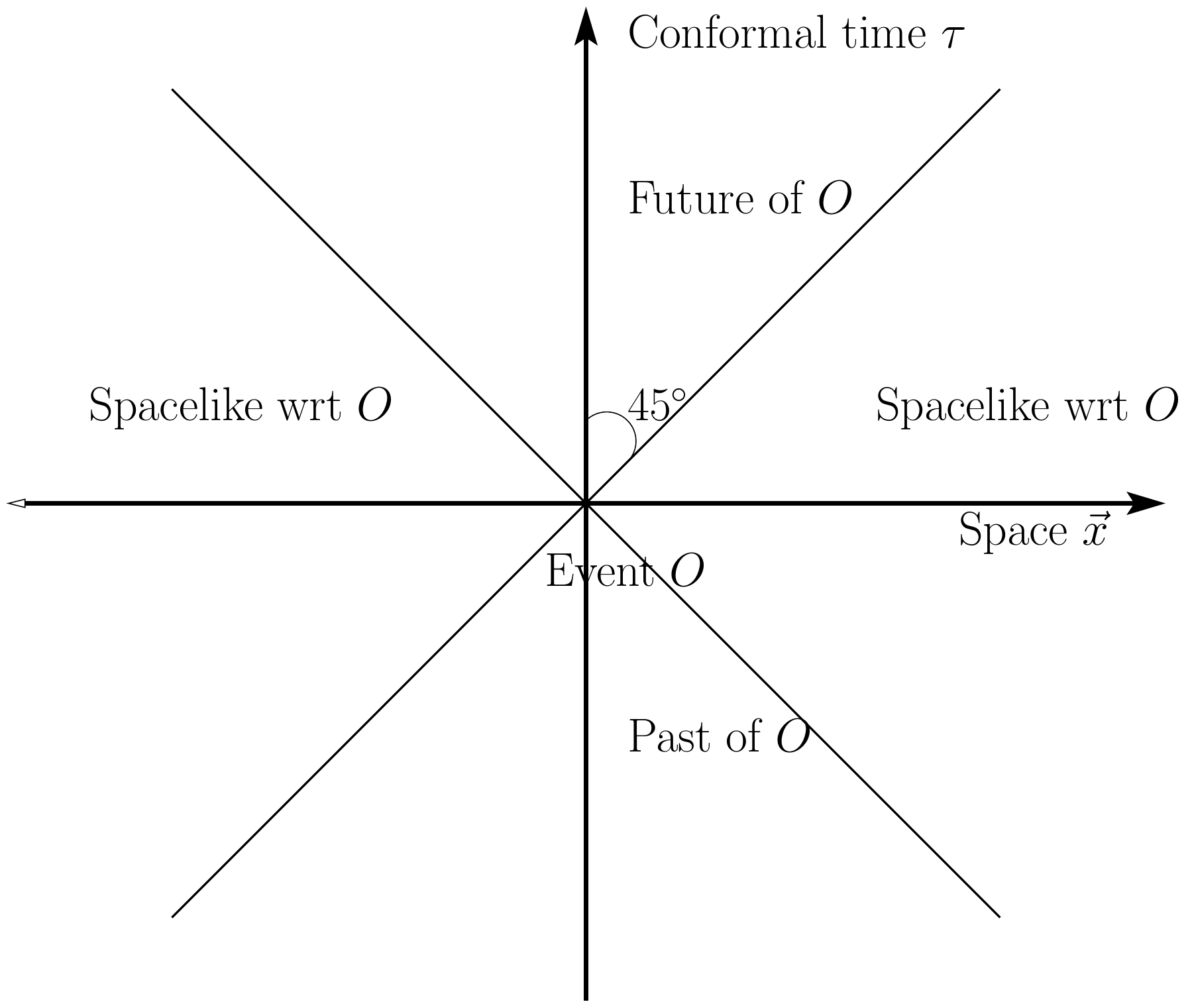}
\caption{\label{fig:horizon} \small Propagation of signals in the $\tau-\chi$ plane.}
\end{center}
\end{figure}

It is interesting to notice that if we declare that the universe started at some time $t_i$, then there is a maximum amount of time for light to have travelled. A point sitting at the origin of space (remember that we are in a space-translation invariant space), by the time $t$ could have sent a signal at most to a point at coordinate $\chi_p$ given by
\be
\chi_p(\tau)=\tau-\tau_i=\int_{t_i}^t \frac{dt}{a(t)}
\ee
The difference in conformal time is equal to the maximum coordinate-separation a particle could have travelled. Notice that the geodesic distance on the spacial slice between two point one particle-horizon apart is obtained by multiplying the coordinate distance with the scale factor:
\be
d_p(t)=a(\tau)\chi_p(\tau)
\ee
The presence of an horizon for cosmologies that begin at some definite time will be crucial for the motivation of inflation.

It will be interesting for us to notice that there is a different kind of horizon, called event horizon. If we suppose that time ends at some point $t_{end}$ (sometimes this $t_{end}$ can be taken to $\infty$), then there is a maximum coordinate separation between two points beyond which no signal can be sent from the first point to reach the second point by the time $t_{end}$. This is called event horizon, and it is the kind of horizon associated to a Schwartshild black hole. From the same geodesic equation, we derive 
\be
\chi_e(\tau)=\tau_{end}-\tau=\int_\tau^{\tau_{\rm end}} \frac{dt}{a(t)}
\ee
Clearly, as $\tau\rightarrow \tau_{\rm end}$, $\chi_e\to 0$.

We have seen that the casual structure of space-time depends on when space-time started and ended, and also on the value of $a(t)$ at the various times, as we have to do an integral. In order to understand how $a(t)$ evolves with time, we need to use the equations that control the {\it dynamics} of the metric. These are the Einstein equations
\be
G_{\mu\nu}=\frac{T_{\mu\nu}}{\mpl^2}\ .
\ee
These in principle 10 equations reduce for an FRW metric to just two. Indeed, by the symmetries of space-time, in FRW slicing, we must have 
\be
T^{\mu}{}_{\nu}=\left(\begin{array}{rrrr}
\rho &0 & 0 & 0\\
0 &-p &0 &0\\
0 & 0 & -p &0\\
0 & 0 &0 & -p\\
\end{array} \right)
\ee
and the Einstein equations reduce to
\bea
&&H^2=\left(\frac{\dot a}{a}\right)^2=\frac{1}{3 \mpl^2}\rho-\frac{k}{a^2}\\
&&\dot H+H^2=\frac{\ddot a}{a}=-\frac{1}{6}\left(\rho+3p\right)\ .
\eea
The first equation is known as Friedamnn equation.
These two equations can be combined to give the energy conservation equation (this follows from the Bianchi identity $0=\nabla_\mu G^{\mu}_\nu=\nabla_\mu T^{\mu}_\nu$):
\be
\frac{d\rho}{d t}+3 H(\rho+p)=0
\ee
This is a general-relativistic generalization of energy conservation. (Homework:  make sense of it by considering dilution of energy and work done by pressure.)
By defining a constant equation of state $w$
\be
p=w\rho\ ,
\ee
energy conservation gives
\be
\rho\propto a^{-3(1+w)}
\ee
and 
\be
a(t)\propto\left\{\begin{array}{rl}
t^{\frac{2}{3(1+w)}} & w\neq 1\\
e^{H t} & w=-1\ .
\end{array}\right.
\ee
Notice that indeed $\rho_{matter}\propto a^{-3}$, $\rho_{radiation}\propto a^{-4}$. Notice also that if $a$ is power low with $t$ to an order one power, than $H\sim 1/t$. That is, the proper time sets the scale of $H$ at each time.

The standard big bang picture is the one in which it is hypothesized that the universe was always dominated by `normal' matter, with $w> 0$. In order to see the shortcomings of this picture, it is useful to define
the present energy fractions of the various constituents of the universe. If we have various components in the universe
\be
\rho=\sum_i\rho_i\ ,\qquad p=\sum_i p_i\  , \quad  w_i=\frac{p_i}{\rho_i}.
\ee
We can define the present energy fraction of the various components by dividing each density by the `critical density' $\rho_{cr}$ (the density that would be required to make the universe expand with rate $H_0$ without the help of anything else)
\be
\Omega_{i,0}=\frac{\rho_0^i}{\rho_{cr,0}}
\ee
We also define
\be
\Omega_{k,0}=-\frac{k}{a(t_0)^2 H_0^2}
\ee
as a measure of the relative curvature contribution.
By setting as it is usually done $a(t_0)=a_0=1$, we can recast the Friedmann equation in the following form
\be\label{eq:omegafrww}
\left(\frac{H^2}{H_0^2}\right)=\sum_i\Omega_{i,0} a^{-3(1+w_i)}+\Omega_{k,0} a^{-2}
\ee
At present time we have $\sum_i \Omega_{i,0}+\Omega_{k,0}=1$.

One can define also time dependent energy fractions
\be
\Omega_{i}(a)=\frac{\rho_i(a)}{\rho_{cr}(a)}\ ,\qquad  \Omega_k(a)=-\frac{k}{a^2 H^2(a)}
\ee
Notice that $\rho_{cr}=3\mpl^2 H^2$ is indeed time dependent. The Friedmann equation becomes 
 \be
 \Omega_{k}(a)=1-\sum_i \Omega_{i}(a)
 \ee

\subsection{Big Bang Shortcomings}

We are now going to highlight some of the shortcoming of the big bang picture that appear if we assume that its history has always been dominated by some form of matter with $w\geq 0$. We will see that upon this assumptions, we are led to very unusual initial conditions. Now, this leads us to a somewhat dangerous slope, which catches current physicists somewhat unprepared. Apart for Cosmology, Physics is usually the science that predicts the evolution of a certain given initial state. No theory is generally given for the initial state. Physicists claim that if you tell them on which state you are, they will tell you what will be your evolution (with some uncertainties). The big bang puzzles we are going to discover are about the very peculiar initial state  the universe should have been at the beginning of the universe if `normal' matter was always to dominate it. Of course, it would be nice to see that the state in which the universe happens to begin in is a natural state, in some not-well defined sense. Inflation was indeed motivated by providing an attractor towards those peculiar looking initial conditions~\footnote{Luckly, we will see that inflation does not do just this, but it is also a predictive theory.}. We should keep in mind that there could be other reasons for selecting a peculiar initial state for the universe.

\subsubsection{Flatness Problem}

Let us look back at 
\be
\Omega_k(a)=-\frac{k}{a^2 H^2(a)}\ ,
\ee
and let us assume for simplicity that the expansion is dominated by some form of matter with equation of state equal to $w$. We have then $a\sim t^{\frac{2}{3(1+w)}}$ and we have
\be
\dot\Omega_k=H \Omega_k (1+3w)\ ,\qquad \frac{\d\Omega_k}{\d \log a}=\Omega_k (1+3w)
\ee
If we assume that $w>-1/3$, then this shows that the solution $\Omega_k=0$ is un unstable point. If $\Omega_k>0$ at some point, $\Omega_k$ keeps growing. Viceversa, if $\Omega_k<0$ at some point, it keeps decreasing. Of corse at most $\Omega_k=\pm1$, in which case $w \to-1/3$ if $k<0$, or otherwise the universe collapses if $k>0$. 

The surprising fact is that $\Omega_k$ is now observed to be smaller than about $10^{-3}$: very close to zero. Given the content of matter of current universe, this mean that in the past it was even closer to zero. For example, at the BBN epoch, it has to be $|\Omega_k|\lesssim 10^{-18}$, at the Planck scale $|\Omega_k|\lesssim 10^{-63}$. In other words, since curvature redshifts as $a^{-2}$, it tends to dominate in the future with respect to other forms of matter (non relativistic matter redshifts as $a^{-3}$, radiation as $a^{-4}$). So, if today curvature is not already dominating, it means that it was very very very negligible in the past. The value of $\Omega_k$ at those early times represents a remarkable small number. Why at that epoch $\Omega_k$ was so small?

Of course one solution could be that $k=0$ in the initial state of the universe. It is unknown why the universe should choose such a precise state initially, but it is nevertheless a possibility. 
A second alternative would be to change at some time the matter content of the universe, so that we are dominated by some matter content with $w<-1/3$. We will see that inflation provides this possibility in a very simple way~\footnote{Another possibility would be to imagine the universe underwent a period of contraction, like in the bouncing cosmologies. Curvature becomes subdominant in a contracting universe.}.

\subsubsection{ Horizon Problem}

An even more dramatic shortcoming of the standard big bang picture is the horizon problem.
Let us assume again that the universe is dominated by some form of matter with equation of state $w$. Let us compute the particle horizon:
\be
\chi_p(\tau)=\tau-\tau_i=\int^{\tau(t)}_{\tau_i(t_i)}\frac{dt'}{a(t')}=\int^a_{a_i}\frac{da}{H a^2}\sim a^{(1+3w)/2}-a_i^{(1+3w)/2}
\ee
We notice that if $w>-1/3$ (É notice, the same $-1/3$ as in the flatness problem), then in an expanding universe the horizon grows with time and is dominated by the latest time contribution. This is very bad. It means that at every instant of time, new regions  that had never been in causal contact before come into contact for the first time. This means that they should look like very different from one another (unless the universe did not decide to start in a homogenous state). But if we look around us, the universe seems to be homogenous on scales that came into causal contact only very recently. Well, maybe they simply equilibrate very fast? Even if this unlikely possibility were to be true, we can make the problem even sharper when we look at the CMB. In this case we can take a snapshot of casually disconnected regions (at the time at which they were still disconnected), and we see that they look like the same. This is the horizon problem.

Notice that if $w>-1/3$ the particle horizon is dominated by late times, and so we can take $a_i\simeq 0$ in its expression. In this way we have that the current physical horizon is
\be
d_p\sim a\tau\sim t\sim \frac{1}{H}\ .
\ee
For this kind of cosmologies where $w>-1/3$ at all times, the Hubble length is of order of the horizon. This is what has led the community to often use the ill-fated name `horizon' for `Hubble'. `Hubble is the horizon' is parametrically true only for standard cosmologies, it is not true in general. We will try to avoid calling Hubble as the horizon in all of these lectures, even though sometimes habit will take a toll.

Notice however that the horizon problem goes away if we assume the universe sit there for a while at the singularity.

Let us look again at the CMB. Naive Horizon scale is one degree ($l\sim  200$), and fluctuations are very small on larger scales. How was that possible?

Apart for postulating an ad hoc initial state, we would need also to include those perturbations in the initial state$\ldots$ This is getting crazy! (though in principle possible) We will see that inflation will provide an attractive solution.

The problem of the CMB large scale fluctuations is a problem as hard as the horizon one.

\begin{figure}[h!]
\begin{center}
\includegraphics[width=10cm]{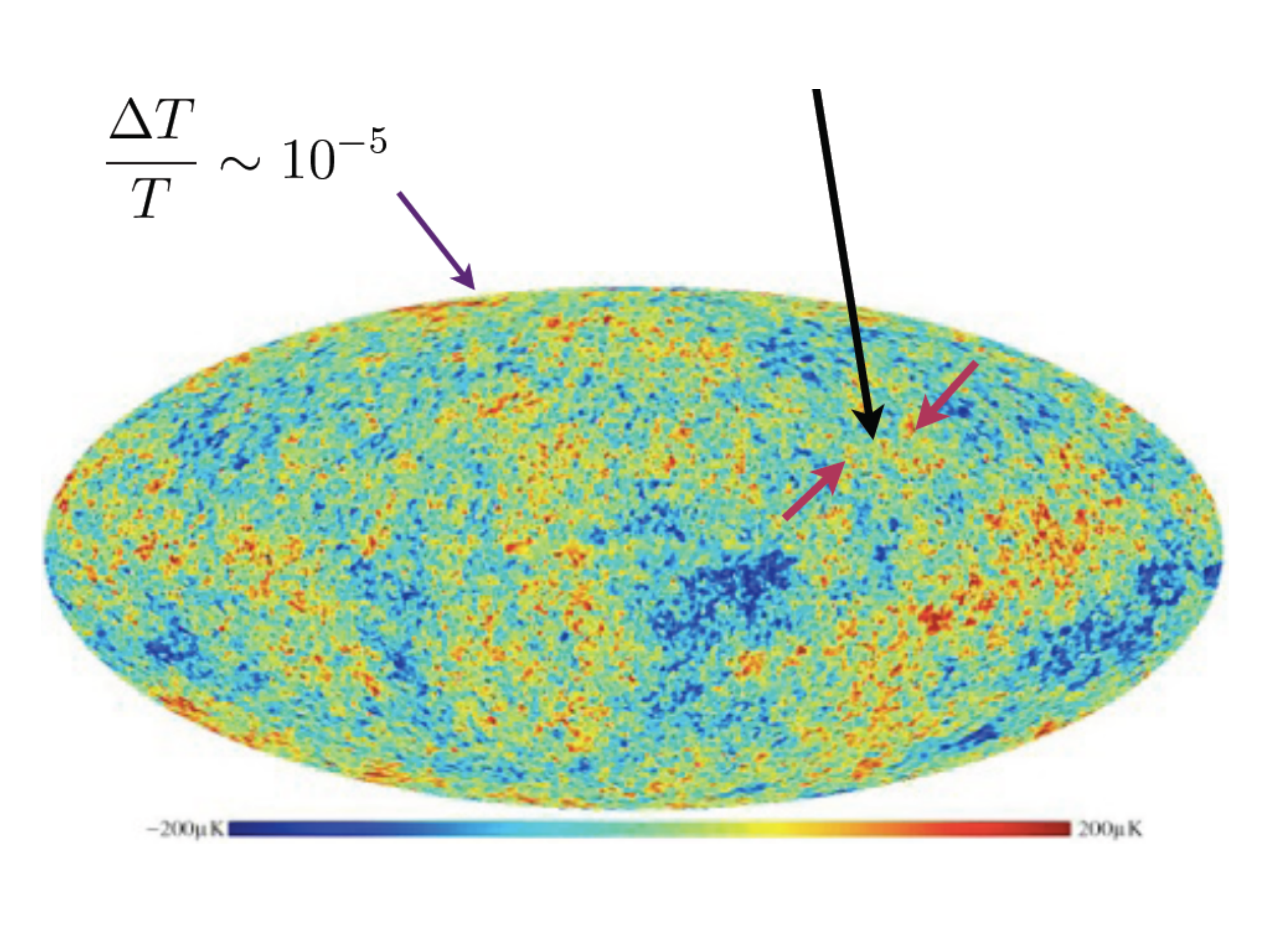}
\caption{\label{fig:horizon} \small The naive horizon $H^{-1}$ at the time of recombination (among the two purple arrows), is much smaller than the scale over which we see statistical homogeneity.}
\end{center}
\end{figure}

\subsubsection{ Solving these problems: conditions}

In order to solve these two problems, we need to have some form of energy with $w<-1/3$. We can say it somewhat differently, by noticing that in order for $\Omega_k$ to decrease with time, since
\be
\Omega_k=-\frac{k}{(a H)^2}
\ee
we want an epoch of the universe in which $a H$ increases with time. Equivalently, $1/(a H)$ decreases with time. $1/(a H)$ is sometimes called `comoving Horizon', $\ldots$ a really bad name in my humble opinion. You can notice that since $1/H$ is the particle horizon in standard cosmologies, $1/(a H)$ identifies the comoving coordinate distance between  two points one naive-Horizon apart. If this decreases with time, then one creates a separation between the true particle horizon, and the naive particle horizon. Two points that naively are separated by a $1/(aH)$ comoving distantce are no more separated by a particle horizon. Even more simply, the formula for the particle horizon reads
\be
\tau=\int_{t_i}^t\frac{da}{(a H)^2}
\ee
If $(a H)^{-1}$ is large in the past, then the integral is dominated by the past, and the actual size of the horizon has nothing to do with present time quantities such as the Hubble scale at present. In standard cosmologies the opposite was happening: the integral was dominated by late times. 

Let us formulate the condition for $(a H)^{-1}$ to decrease with time in equivalent forms. 
\begin{itemize}
\item Accelerate expansion: it looks like that this condition implies that the universe must be accelerating in that epoch:
\be
\frac{\d \frac{1}{(a H)}}{\d t}< 0 \quad\Rightarrow\quad \ddot a>0
\ee
This implies that $k/(aH)$ decreases: physical wavelengths become longer than $H^{-1}$.
\item As we stressed, this should imply $w<-1/3$. Let us verify it. From Friedman equation
\be
0<\ddot a=-\frac{a}{6}(\rho+3p)=-\frac{a\,\rho}{6}(1+3w)\quad\Rightarrow\quad w<-1/3\quad {\rm if\ } \rho>0
\ee

\end{itemize}

Inflation, in its most essential definition, is the postulation of a phase with $w<-1/3$ in the past of our universe~\footnote{If there is only one field involved, than scale invariance of the perturbations and the requirement that the solution is an attractor forces $w\simeq -1$. This is a theorem~\cite{Baumann:2011dt}.}.

Is it possible to see more physically what is going on?
In a standard cosmology, the scale factor goes to zero at finite conformal time. For $w>-1/3$, we have that
\be
a\sim \tau^{2/(1+3w)}
\ee
implying the existence of a singularity $a\rightarrow0, H\rightarrow\infty$ as $\tau\rightarrow 0$. This is why we had to stop there. This is the big bang moment in standard cosmology. This however implies that there is a beginning of time, and that the particle horizon is order $\tau$. This is the source of the problems we discussed about.

However, if we have a phase in which $w<-1/3$, then the singularity in the past is pushed way further back, and the actual universe is much longer than what $\tau$ indicates. For example, for inflation $H\sim$ const. and $a(\tau)=-\frac{1}{H\tau}$, with $\tau\in[-\infty,\tau_{end}]$, $\tau_{end}\leq0$. In general $\tau$ can be extended to negative times, in this way making the horizon much larger than $1/H$.

\begin{figure}[h!]
\begin{center}
\includegraphics[width=17cm]{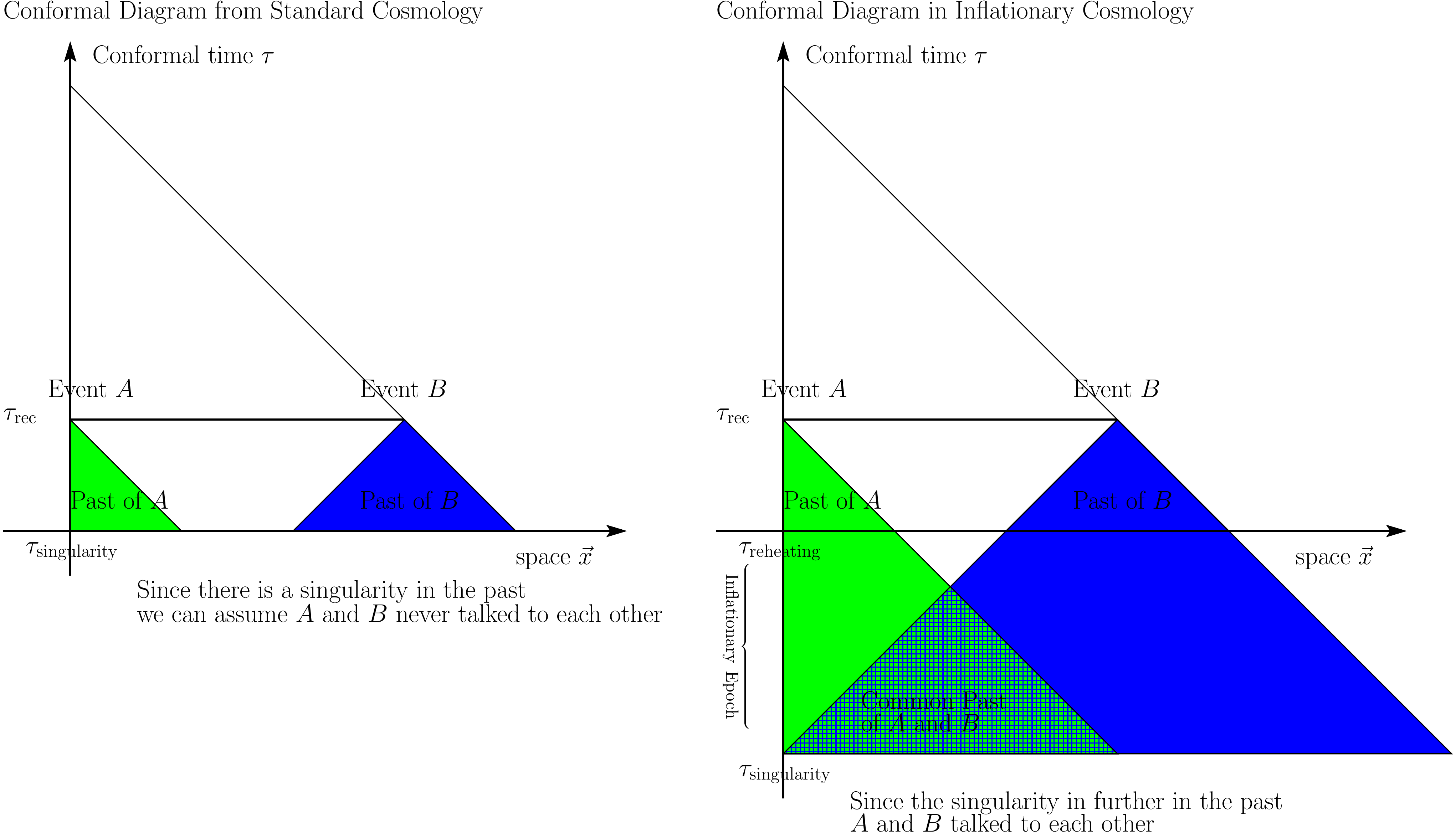}
\caption{\label{fig:inflation} \small How inflation solve the horizon problem: in the past, there is much more time than what there would have naively been without inflation.}
\end{center}
\end{figure}

\subsection{The theory of Inflation}

Inflation is indeed a period of the history of the universe that is postulated to have happened before the standard big bang history. Direct observation of BBN products tell us that the universe was radiation dominated at $t\sim 1-100$ sec,  which strongly suggests that inflation had to happen at least earlier than this.  More specifically, inflation is supposed to be a period dominated by a form of energy with $w\simeq-1$, or equivalently $H\simeq$ const. How can this be achieved by some physical means?

\subsubsection{Simplest example}

The simplest example of a system capable of driving a period of inflation is a scalar field on top a rather flat potential. These kinds of models are called `slow roll inflation' and were the ones initially discovered to drive inflation. Let us look at this 

\begin{figure}[h!]
\begin{center}
\includegraphics[width=10cm]{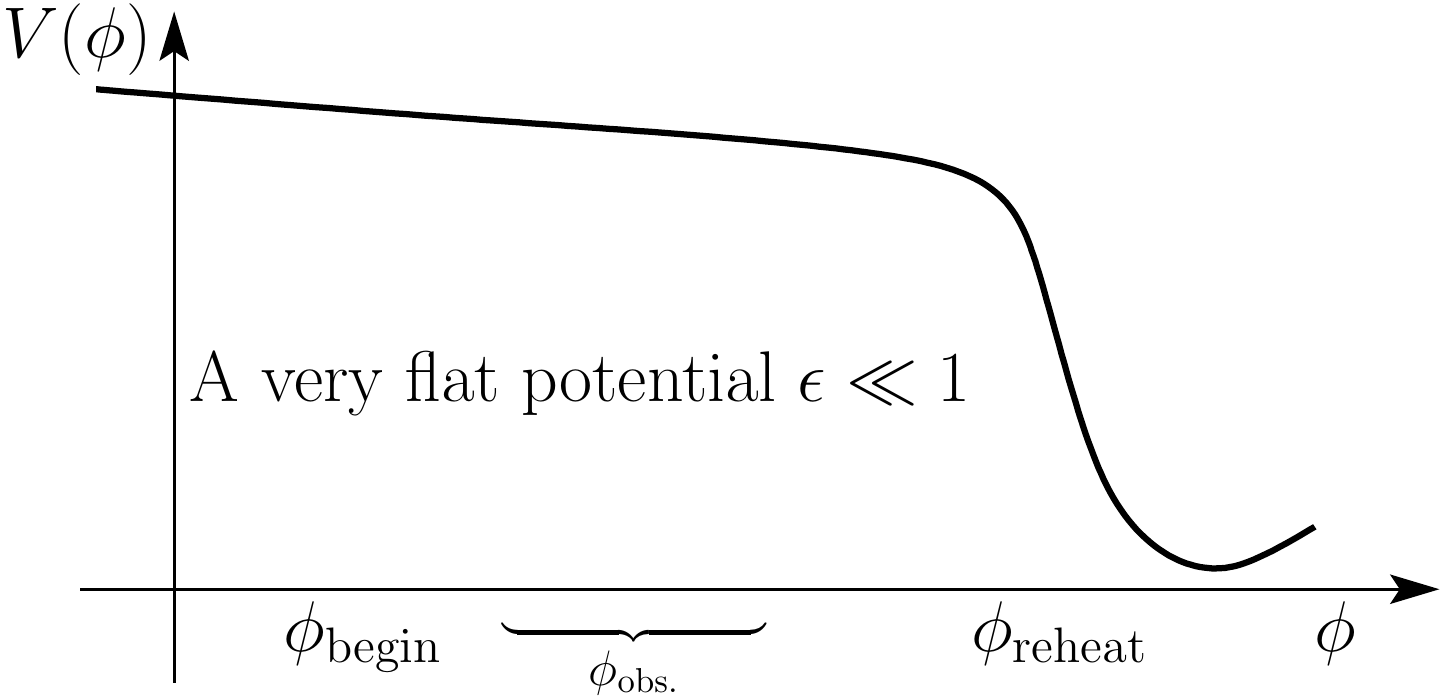}
\caption{\label{fig:inflation} \small A simple inflationary model.}
\end{center}
\end{figure}

The scalar field plus gravity has the following action
\be
S=\int d^4x \sqrt{-g}\left[\frac{\mpl^2}{2}R+\frac{1}{2}g^{\mu\nu}\d_\mu\phi\d_\nu\phi-V(\phi)\right]
\ee
The first term is the Einstein Hilbert term of General Relativity (GR). The second and third terms represent the action of a scalar field $S_\phi$. The idea of inflation is to fill a small region of the initial universe with an homogeneously distributed scalar field sitting on top of its potential $V(\phi)$. Let us see what happens, by looking at the evolution of the space-time. We need the scalar field stress tensor:
\be
T_{\mu\nu}^{(\phi)}=-\frac{2}{\sqrt{-g}}\frac{\delta S_\phi}{\delta g^{\mu\nu}}=\d_\mu\phi\d_\nu\phi-g_{\mu\nu}\left(\frac{1}{2}\d_\rho\phi \d^\rho\phi+V(\phi)\right)
\ee
For an homogenous field configuration, this leads to the following energy density and pressure
\bea
&&\rho_\phi=\frac{1}{2}\dot\phi^2+V(\phi)\quad {\rm obviously}\\
&&p_\phi=\frac{1}{2}\dot\phi^2-V(\phi) \quad {\rm notice\ the\ sign\ of} \ V
\eea
Therefore the equation of state is
\be
w_\phi=\frac{p_\phi}{\rho_\phi}=\frac{\frac{1}{2}\dot\phi^2-V(\phi) }{\frac{1}{2}\dot\phi^2+V(\phi)}.
\ee
We see that if the potential energy dominates over the kinetic energy, we have
\be
\dot\phi^2\ll V(\phi)\quad\Rightarrow\quad w_\phi\simeq-1<-\frac{1}{3}
\ee
as we wished. Notice that this means that
\be
\epsilon=\epsilon_H=-\frac{\dot H}{H^2}\sim \frac{\dot\phi^2}{V}\ll 1\ .
\ee
The equation of motion for the scalar field is
\be
\frac{\delta S}{\delta \phi}=\frac{1}{\sqrt{-g}}\d_\mu(\sqrt{-g} \d^\mu\phi)+V_{,\phi}=0 \quad\Rightarrow\quad \ddot\phi+3H\dot\phi+V_{,\phi}=0
\ee
This equation of motion is the same as the one of a particle rolling down its potential. This particle is subject to friction though the $H\dot\phi$ term. Like for a particle trajectory, this means  that the solution where $\dot\phi\simeq V_{\,\phi}/(3H)$ is an attractor `slow-roll' solution if friction is large enough. Being on this trajectory requires
\be
\eta_H=-\frac{\ddot\phi}{H\dot\phi}\ll 1
\ee

We have therefore found two `slow roll parameters':
\be
\epsilon=-\frac{\dot H}{H^2}\ll1\ ,\qquad \eta_H=-\frac{\ddot\phi}{H\dot\phi}\ll 1
\ee
The first parameters being much smaller than one means that we are on a background solution where the Hubble rate changes very slowly with time. The second parameter means that we are on an attractor solution (so that the actual solution does not depend much from the initial conditions), and also that this phase of accelerated expansion ($w\simeq-1$, $a\sim{\rm \;Exp}(H t)$) will last for a long time. Indeed, one can check that
\be
\frac{\dot\epsilon}{H\epsilon}\sim {\cal O}(\epsilon_H, \eta)\ .
\ee
We will see that the smallness of $\eta_H$ is really forced on us by the scale invariance of the cosmological perturbations.

Once we assume we are on the slow roll solution, then we can express them in terms of the potential terms. We have
\be
\epsilon\simeq \frac{\mpl^2}{2}\left(\frac{V_{,\phi}}{V}\right)^2\ ,\qquad\eta_H\simeq\mpl^2\frac{V_{,\phi\phi}}{V}- \frac{\mpl^2}{2}\left(\frac{V_{,\phi}}{V}\right)^2\ .
\ee
On this solution we also have
\be
\dot\phi\simeq \frac{V_{,\phi}}{3H}\ ,\qquad H^2\simeq \frac{V(\phi)}{3 \mpl^2}\simeq {\rm const}\ ,\qquad a\sim e^{3 H t}\ .
\ee
When does inflation end? By definition, inflation ends when $w$ ceases to be close to $-1$. This means that 
\be
\epsilon\sim \eta_H\sim 1\ . 
\ee
More concretely, we see that the field that starts on top of his potential will slowly roll down until two things will happen: Hubble will decrease, providing less friction, and the potential will become too steep to guaranteed that the kinetic energy is negligible with respect to potential energy. We call the point in field space where this happens $\phi_{end}$. At that point, a period dominated by a form of energy with $w>-1/3$ is expected to begin. We will come back in a second on it. 

{\bf Duration of Inflation:} For the moment, let us see how long inflation needs to last. 
The number of $e$-foldings of inflation is defined as the logarithm of the ratio of the scale factor at the end of inflation and at the beginning of inflation. For a generic initial point $\phi$, we have
\be
N^{to\;end}(\phi)=\log\left(\frac{a_{end}}{a}\right)\simeq\int_t^{t_{end}} H dt =\int_{\phi}^{\phi_{end}} \frac{H}{\dot\phi}d\phi\simeq \int_{\phi_{end}}^{\phi}\frac{V}{V_{,\phi}}d\phi \ ,
\ee
where in the third passage we have used that $a\sim e^{Ht}$, and in the last passage we have used the slow roll solutions.

The horizon and flatness problems are solved in inflation very simply. 
During inflation
\be
\Omega_k=-\frac{k}{a^2 H^2}\propto\frac{1}{a^2}\rightarrow 0.
\ee
So, if we start with $\Omega_k\sim 1$ at the onset of inflation, and we wish to explain why $\Omega_k(a_{BBN})\sim 10^{-18}$, we need about 20 $e$-foldings of inflation. This is so because at the end of inflation we have
\be
\Omega_k(a_{end})\simeq \Omega_k(a_{in}) \frac{a^2_{in}}{a^2_{end}}\sim \frac{a^2_{in}}{a^2_{end}}=e^{-2N}
\ee
and this must be equal to the curvature we expect at the beginning of the FRW phase (that we can assume to be equal to the end of inflation)
\be
\Omega_k(a_{end})=\Omega_k(a_0)\frac{a_0^2 H_0^2}{a_{end}^2 H_I^2} \sim 10^{-2}\frac{a_0^2 H_0^2}{a_{end}^2 H_I^2}\quad\Rightarrow\quad N=\log\left(\frac{a_{end}H_I}{a_0 H_0}\right)\ .
\ee
In this case however we would need the hot-big-bang period to be start after inflation directly with BBN-like temperatures. If the universe started at higher temperatures, say the GUT scale, we would need about 60 $e$-foldings of inflation. So, you see that the required  number of $e$-foldings depends on the starting temperature  of the universe, but we are in the realm of several tens.

The horizon problem is solved by asking that the region we see in the CMB was well inside the horizon. Since the contribution to the particle horizon from the radiation and the matter dominated eras is too small to account for the isotropy of the CMB, we can can assume that the integral that defines the particle horizon is dominated by the period of inflation. If $t_L$ is the time of the last scattering surface, we have
\be
d_p=a(t_L)\int_{\tin}^{t_{end}} \frac{dt}{a(t)}\simeq \frac{a(t_L)}{a_{end} H_I} e^{N}\ ,
\ee
where we have used that $a(t)=a(t_{end}) e^{H_I (t-t_{end})}$. The particle horizon has to be bigger than the region that we can see now of the CMB. This is given by the angular diameter distance of the CMB last scattering surface. It is simply the physical distance between two points that now are one Hubble radius far apart, at the time $t_L$:
\be
d_L=\frac{a(t_L)}{H_0 a_0}
\ee
To solve the horizon problem we need
\be
d_p\gtrsim d_L\quad\Rightarrow\quad N\gtrsim\log\left(\frac{a_{end} H_I}{a_0 H_0}\right)
\ee
This is the same number as we need to solve the flatness problem, so we find the same number of $e$-foldings is needed to solve the horizon problem as are necessary to solve the flatness problems.

\subsection{Reheating}
But we still miss a piece of the story. How inflation ends? So far, we have simply seen that as $\epsilon\sim 1$ the accelerated phase stops.  At this point, typically the inflaton begins to oscillate around the bottom of the potential. In this regime it drives the universe as if it were dominated by non-relativistic matter. The equation for the inflation indeed reads
\be
\frac{\d\rho_\phi}{\d t}+\left(3 H+\Gamma\right)\rho_\phi=0
\ee
(Homework: derive this expression). For $\Gamma=0$, this is the dilution equation for non-relativistic matter. $\Gamma$ represents the inflation decay rate. Indeed, in this period of time the inflation is supposed to decay into other particles. These thermalize and, once the inflation has decayed enough, start dominating the universe. This is the start of the standard big-bang universe.

\subsection{Simplest Models of Inflation}

\subsubsection{Large Field Inflation}
The simplest versions of inflation are based on scalar fields slowly rolling down their potential. These typically fall into two categories: large fields and small fields. Large field models are those characterized by a potential of the form
\be
V(\phi)=\frac{\phi^{\alpha}}{M^{\alpha-4}}. 
\ee

\begin{figure}[h!]
\begin{center}
\includegraphics[width=10cm]{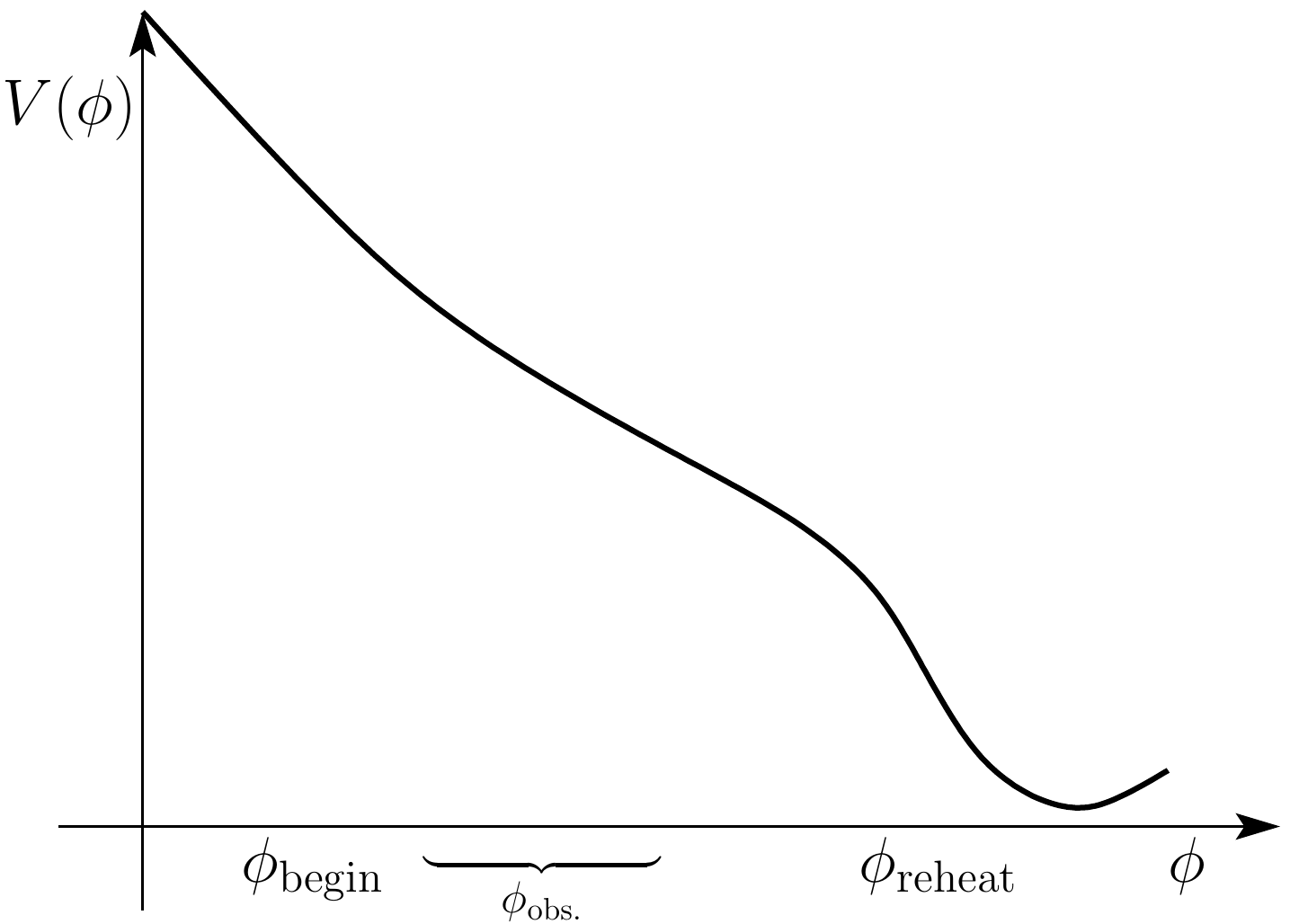}
\caption{\label{fig:inflation} \small A `large-field' inflationary model.}
\end{center}
\end{figure}

For any $M$ and $\alpha$, if we put the scalar field high enough, we can have an inflationary solution. Let us see how this happens by imposing the slow roll conditions.
\be
\epsilon\sim \mpl^2 \left(\frac{V_{,\phi}}{V}\right)^2\sim \alpha^2\frac{\mpl^2}{\phi^2}
\ee
For $\alpha\sim 1$, we have
\be
\epsilon\ll1 \qquad\Rightarrow \qquad \phi\gg \mpl \ .  
\ee
The field vev has to be super planckian. Further, notice that the field travels an amount of order 
\be\label{eq:deltaphi}
\Delta\phi=\int_{\phi_{in}}^{\phi_{end}}d\phi=\int_{\tin}^{t_{end}}\dot\phi dt\simeq\frac{\dot\phi}{H} \int_{H \tin}^{H t_{end}}d(Ht)=\frac{\dot\phi}{H} N_e\sim \epsilon^{1/2}N_e\;\mpl
\ee
For $\epsilon\sim 1/N_e$ and not too small, the field excursion is of order $\mpl$. This is a pretty large field excursion (this explains the name large field models). But notice that in principle there is absolutely nothing bad about this. The energy density of the field is of order $\phi^\alpha/M^{\alpha-4}\sim \left(\frac{\mpl}{M}\right)^\alpha M^4$ and needs to be smaller than $\mpl^4$ for us to be able to trust general relativity and the semiclassical description of space-time. This is realized once $M\gg \mpl$ (for $\alpha=4$ we have $V=\lambda \phi^4$ and we simply require $\lambda\ll 1$). So far so good from the field theory point of view. Now, ideally some of us would like to embed inflationary theories in UV complete theories of gravity such as string theory. In this case the UV complete model need to be able to control all $\mpl$ suppressed operators. This is possible, though sometimes challenging, depending on the scenario considered. This is a lively line of research.

\subsubsection{Small Field Inflation}
From (\ref{eq:deltaphi}) we see that if we wish to have a $\Delta\phi\ll\mpl$, we need to have $\epsilon$ very very small. This is possible to achieve in models of the form
\be
V(\phi)=V_0\left(1-\left(\frac{\phi}{M}\right)^2\right)
\ee

\begin{figure}[h!]
\begin{center}
\includegraphics[width=10cm]{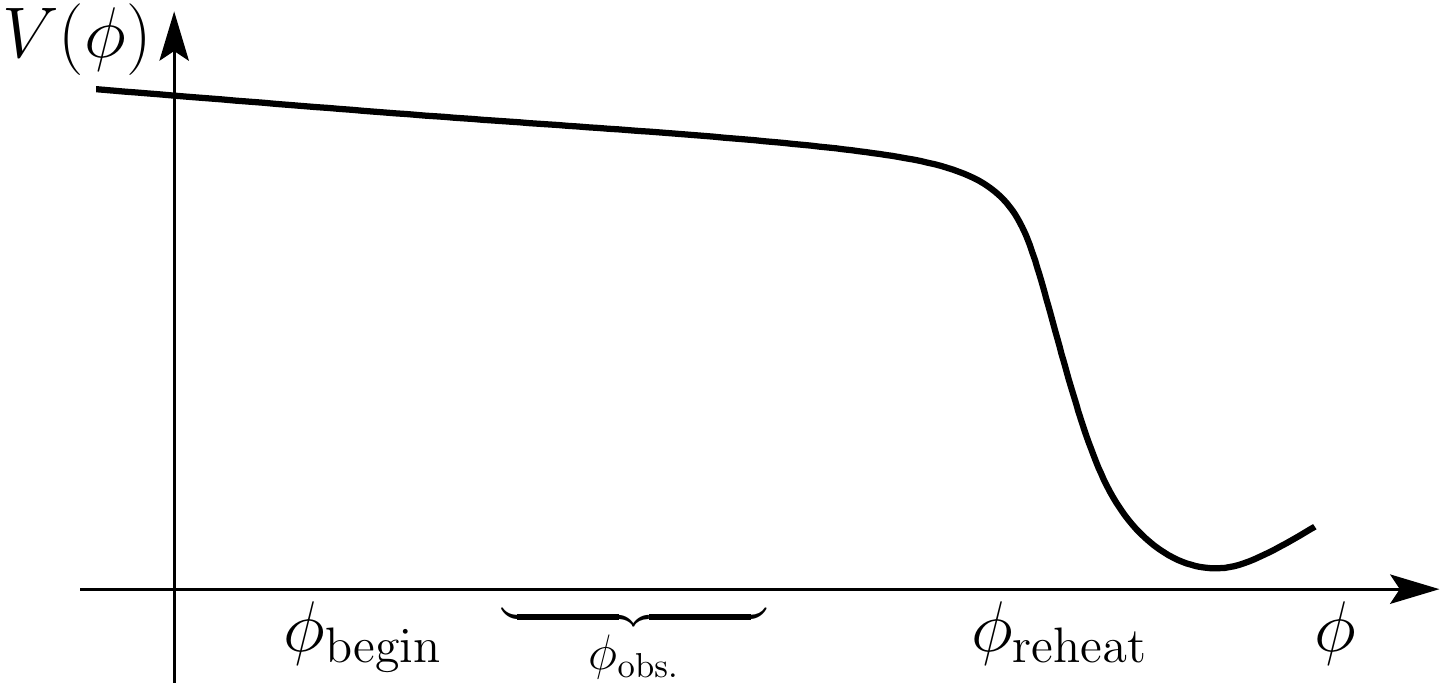}
\caption{\label{fig:inflation} \small A `small-field'inflationary model.}
\end{center}
\end{figure}

In this case, we have
\be
\epsilon\simeq \frac{\mpl^2\phi^2}{M^4}
\ee
that becomes smaller and smaller as we send $\phi\to 0$. Of course, we need to guarantee a long enough duration of inflation, which means that $\phi\sim\Delta\phi\sim \epsilon^{1/2}\mpl N_e$. Both conditions are satisfied by taking $M\gtrsim \mpl N_e$.

\subsubsection{Generalizations}

Over the thirty years since the discovery of the first inflationary models, there have been a very large number of generalizations. From fields with a non-trivial kinetic terms, such as DBI inflation and Ghost Inflation, to theories with multiple fields or with dissipative effects. We will come back to these models later, when we will offer a unified description.

\subsection{Summary of lecture 1}

\begin{itemize}
\item Standard Big Bang Cosmology has an horizon and a flatness problem. Plus, who created the density fluctuations in the CMB?
\item A period of early acceleration solves the horizon and flatness problems.
\item Inflation, here for the moment presented in the simplest form of a scalar field rolling downs a flat potential, solves them.
\end{itemize}

\newpage

\section{Lecture 2: Generation of density perturbations}

This is the most {\bf exciting, fascinating and predicting part}. It is the most predicting part, because we will see that this is what makes inflation predictive. While the former cosmological shortcomings that we saw so far were what motivated scientists such as Guth to look for inflation, cosmological perturbations became part of the story well after inflation was formulated. The fact that inflation could source primordial perturbations was indeed realized only shortly after the formulation of inflation. At that time, CMB perturbations were not yet observed, but the fact that we observed galaxies today, and the fact that matter grows as $\delta\propto a$ in a matter dominated universe predicted that some perturbations had to exist on the CMB. The way inflation produces these perturbations is both exciting and beautiful. It is simply beautiful because it shows that quantum effects, that are usually relegated to the hardly experiencable world of the small distances, can be exponentiated in the peculiar inflationary space-time to become actually the source of all the cosmological perturbations, and ultimately of the galaxies and of all the structures that are present in our universe. With inflation, quantum effects are at the basis of the formation of the largest structures in the universe. This part is also when inflation becomes more intellectually  exciting. We will see that there is a very interesting quantum field theory that happens when we put some field theory in a accelerating space-time. And this is not just for fun, it makes predictions that we are actually testing right now in the universe!

The calculation of the primordial density perturbations can be quite complicated. Historically, it has taken some time to outstrip the description of all the irrelevant parts and make the story simple. This is typical of all parts of science and of all discoveries. Therefore, I will give you what I consider the simplest and most elegant derivation. Even with this, the calculation is quite complicated. Therefore we will first see how we can estimate the most important characteristics of the perturbations without doing any calculations. Only later, we will do the rigorous, and now simple, calculation~\footnote{General lesson I think I have learned from my teachers: always know the answer you have to get before starting a difficult calculation.}.

\subsection{Simple Derivation: real space}
In this simple derivation we will drop all numerical factors. We will concentrate on the physics.

Let us expand the field around the background solution. Since the world is quantum mechanical, if the lowest energy state is not an eigenstate of the field operator $\hat\phi|0\rangle\neq\phi|0\rangle$, then
\be
\phi=\phi_0(t)+\delta\phi(\vec x, t)
\ee
Notice that if we change coordinates 
\be
x^\mu\to x'{}^{\mu}=x^\mu+\xi^\mu
\ee
then
\be
\delta\phi(\vec x,t)\to\widetilde{\delta\phi}(\vec x'{}^\mu)-\dot\phi_0(t)\xi^0
\ee
$\delta\phi$ does not transform as a scalar, it shifts under time diffeomorphisms (diffs.). The actual definition of $\delta\phi$ {\it depends} on the coordinates chosen. This has been the problem that has terrified the community for a long time, and made the treatment of perturbations in inflation very complicated~\footnote{Of course, at the beginning things were new, and it was very justified not to get things immediately in the simplest way.}. Instead, we will simply ignore this subtlety, as it is highly irrelevant. Indeed, we are talking about a scalar field, very much like the Higgs field. When we study the Higgs field we do not bother about specifying the coordinates.

So why we should do it now? For the Higgs we do not even bother of writing down the metric perturbations, so why we should do it now? We will later justify why this is actually possible in more rigorous terms. Let us therefore proceed, and expand the action for the scalar field at quadratic order in an unperturbed FRW metric: 
\be
S=\int d^4x a^3\left[{\cal L}_0+\left.\frac{\delta {\cal L}}{\delta\phi}\right|_0\delta\phi+\frac{1}{2}\left.\frac{\delta^2 {\cal L}}{\delta\phi^2}\right|_0\delta\phi^2\right]=S_0+\int d^4x e^{3Ht} \left[-g^{\mu\nu}\d_\mu\delta\phi\d_\nu\phi\right] \ ,
\ee
Notice that the term linear in $\delta\phi$ is called the tadpole term, and if we expand around the solution of the background equations $\delta S/\delta\phi|_0=0$ it vanishes. We have used that $\sqrt{-g}=a^3=e^{3Ht}$. The action contains simply a kinetic term for the inflation. The potential terms are very small, because the potential is very flat, so that we can neglect it.

\begin{figure}[h!]
\begin{center}
\includegraphics[width=10cm]{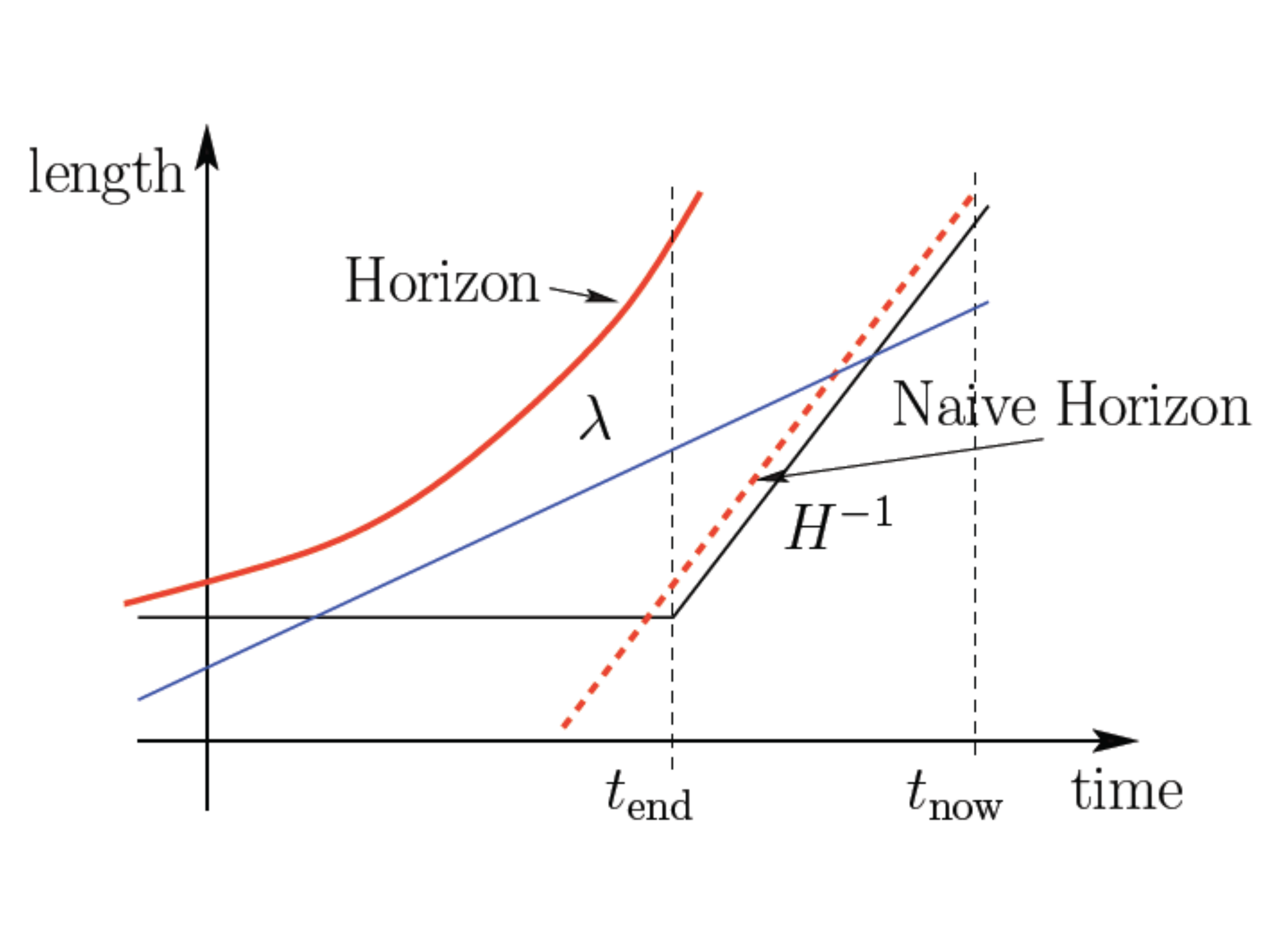}
\caption{\label{fig:scales1} \small Relative ratios of important length scales as a function of time in the inflationary universe.  Modes start shorter than $H^{-1}$ during inflation and become longer than $H^{-1}$ during inflation.}
\end{center}
\end{figure}

\begin{itemize}
\item Let us concentrate on very small wavelengths (high-frequencies). $\omega\gg H$. $\Delta\vec  x\ll H^{-1}$ (see Fig.~\ref{fig:scales1}). In that regime, we can clearly neglect the expansion of the universe, as we do when we do LHC physics (this is nothing but the equivalence principle at work: at distances much shorter than the curvature of the universe we live in flat space). We are like in Minkowski space, and therefore
\be
\langle\delta\phi(\vec x,t)\delta\phi(\vec x',t)\rangle_{vac.}\sim\ {\rm something}\sim[{\rm length}]^{-2}\ ,
\ee
just by dimensional analysis. Since there is no length scale or mass scale in the Lagrangian (remember that $H$ is negligible), then the only length in the system is $\Delta\vec x$. We have
\be
\langle\delta\phi(\vec x,t)\delta\phi(\vec x',t)\rangle_{vac.}\sim\frac{1}{|\Delta\vec x|^2}
\ee
Notice that the two point function decreases as we increase the distance between the two points: this is why usually quantum mechanics is segregated to small distances.

\item But the universe is slowly expanding wrt $1/|\Delta\vec x|$, so the physical distance between to comoving points grows (slowly) with $a$:
\be
|\Delta\vec x|\quad\to\quad |\Delta \vec x(t)|\propto a(t)\qquad\Rightarrow\qquad \langle\delta\phi(\vec x,t)\delta\phi(\vec x',t)\rangle_{vac.}\sim\frac{1}{|\Delta\vec x|^2(t)}
\ee

\item Since $H$ is constant (it would be enough for the universe to be accelerating), at some point we will have
\be
|\Delta\vec x|(t)\sim H^{-1} 
\ee
and keeps growing. At this point, the Hubble expansion is clearly not a slow time scale for the system, it is actually very important. In particular, if two points are one Hubble far apart, then we have~\footnote{Very roughly speaking. In more rigorous terms, one point is beyond the apparent event horizon of the other.}
\be
v_{\rm relative}\gtrsim v_{\rm light}
\ee
Notice that this is not in contradiction with the principle of relativity: the two points simply stop communicating. But then gradients are irrelevant, and the value of $\phi$ and $\vec x$ should be unaffected by the value of $\phi$ at $\vec x'$. Since any value of $\delta\phi$ is as good as the others (if you look at the action, there is no potential term that gives difference in energy to different values of $\delta\phi$). The two point function stops decreasing and becomes constant
\be
\langle\delta\phi(\vec x,t)\delta\phi(\vec x',t)\rangle_{vac.}\sim\frac{1}{|\Delta\vec x|^2=H^{-2}}\sim H^2 \quad {\rm as}\quad \Delta\vec x\to\infty
\ee
So, we see that the two point function stops decreasing and as $\Delta\vec x$ becomes larger than $H^{-1}$, it remains basically constant of order $H^2$. This means that there is no scale in the two point functions, once the distance is larger than $H^{-1}$. An example of a scale dependent two point function that we could have found could have been: $\langle\delta\phi(\vec x,t)\delta\phi(\vec x',t)\rangle\sim H^4|\vec x|^2$. This does not happen here, and we have a scale invariant spectrum.

\end{itemize}

\subsection{Simple Derivation: Fourier space}

Let us look at the same derivation, working this time in Fourier space. The action reads
\be
S=\int d^4x e^{3Ht} \left[-g^{\mu\nu}\d_\mu\delta\phi\d_\nu\delta\phi\right]=\int dtd^3k\; a^3\left(\dot{\delta\phi}_{\vec k}\dot{\delta\phi}_{-\vec k}-\frac{k^2}{a^2}\delta\phi_{\vec k}\delta\phi_{-\vec k}\right) \ ,
\ee
\begin{itemize}
\item Each Fourier mode evolves independently. This is a quadratic Lagrangian!

\item Each Fourier mode represents a quantum mechanical harmonic oscillator (apart for the overall factor of $a^3$), with a time-dependent frequency 
\be
\omega(t)\sim \frac{k}{a(t)}
\ee
The canonically normalized harmonic oscillator is $\delta\phi_{can}\sim a^{-3/2}\delta\phi$
\item Let us focus on one Fourier mode. At sufficiently early times, we have
\be
\omega(t)\simeq \frac{k}{a}\gg H\ .
\ee
In this regime, as before, we can neglect the expansion of the universe and therefore any time dependence. Then we are as if we were in Minkowski space, and therefore we must have, for a canonically normalized scalar field (i.e. harmonic oscillator)
\be
\langle\delta\phi^2_{can,k}\rangle\sim\frac{1}{\omega(t)}\quad\Rightarrow\qquad \langle\delta\phi^2_{k}\rangle\sim\frac{1}{a^3}\cdot\frac{1}{\omega(t)}
\ee

\item While $\omega\gg H$, $\omega$ slowly decreases with time $\dot\omega/\omega\sim H\ll \omega $, so the two point function follows adiabatially the value on the vacuum. This happens until $\omega\sim H$ and ultimately $\omega\ll H$. At this transition, called freeze-out, the adiabatic approximation breaks down. What happens is that no more evolution is possible, because the two points are further away than an Hubble scale, and so they are beyond the event horizon. Equivalently the harmonic oscillator now has an overdamping friction term $\ddot\delta\phi_{\vk}+3H\dot{\delta\phi}_{\vk}=0$ that now is relevant. Since this happen when
\be
\omega\sim\frac{k}{a(t_{f.o.})}\sim H\qquad\Rightarrow\qquad a_{f.o.}\sim \frac{k}{H}
\ee
where $f.o.$ stray for freeze-out.
By substituting in the two point function, we obtain
\be
\langle\delta\phi^2_{k}\rangle\sim\frac{1}{a^3_{f.o.}}\cdot\frac{1}{\omega(t_{f.o.})}\sim \frac{H^2}{k^3}
\ee
This is how a scale invariant two-point function spectrum looks like in Fourier space. It is so because in Fourier space the phase space goes as $d^3k\sim k^3$, so, if the power spectrum goes as $1/k^3$, we have that each logarithmic interval in $k$-space contributes equally to the two-point function in real space. In formulae
\be
\langle\left.\delta\phi(\vec x)^2\right|_{E_1}^{E_2}\rangle\sim\int_{E_1}^{E_2} d^3k \langle\delta\phi^2_{k}\rangle\sim H^2\log\left(\frac{E_1}{E_2}\right)
\ee

\end{itemize}

This is simply beautiful, at least in my opinion. In Minkowski space quantum mechanics is segregated to small distances because
\be
\langle\delta\phi(\vec x,t)\delta\phi(\vec x',t)\rangle_{vac.}\sim\frac{1}{|\Delta\vec x|^2}
\ee
In an inflationary space-time (it locally looks like a de Sitter space, but, contrary to de Sitter space, it ends), we have that on very large distances
\be
\langle\delta\phi(\vec x,t)\delta\phi(\vec x',t)\rangle_{vac.}\sim H^2\gg \frac{1}{|\Delta\vec x|^2} \qquad{\rm for}\qquad \Delta\vec x\gg H^{-1}
\ee
At a given large distance, quantum effects are much larger than what they would have naively been in Minkowski space, and this by a huge amount once we consider that in inflation scales are stretched out of the horizon by a factor of order $e^{60}$. 

Since we are all physicists here, we can say that this is a remarkable story for the universe. 

Further, it tells us that trough this mechanisms, by exploring cosmological perturbations we are studying quantum mechanics, and so fundamental physics.

But still, we need to make contact with observations.

\subsection{Contact with observation: Part 1}

In the former subsection we have seen that the scalar field develops a large scale-invariant two-point function at scales longer than Hubble during inflation. How these become the density perturbations that we see in the CMB and then grow to become the galaxies?

Let us look at what happens during inflation. Let us take a box full of inflation up in the potential, and let inflation happen. In each point in space, the inflaton will roll down the potential and inflation will end when the inflaton at each location will reach a point $\phi(\vec x,t_{end})=\phi_{end}$. We can therefore draw a surface of constant field $\phi=\phi_{end}$. Reheating will start, and in every point in space reheating will happen in the same way: the only thing that changes between the various points is the value of the gradient of the fields, but for the modes we are interested in, these are much much longer than the Hubble scale, and so gradients are negligible; also the velocity of the field matters, but since we are on an attractor solution, we have the same velocity everywhere. At this point there is no difference between the various points, and so reheating will happen in the same way in every location. In the approximation in which re-heating happens instantaneously, the surfaces $\phi=\phi_{end}$ are equal temperature surfaces (if reheating is not instantaneous, then the equal temperature surface will be displaced later, but nothing will change really in the conclusions), and so equal energy density surfaces. Now, is this surface an equal time surface? In the limit in which there no quantum fluctuations for the scalar field, it would be so, but quantum fluctuations make it perturbed. How a quantum fluctuation will affect the duration of inflation at each point? Well, a jump $\delta\phi$ will move the inflaton towards or far away from the end of inflation. This means that the duration of inflation in a given location will be perturbed, and consequently the overall expansion of the universe when $\phi=\phi_{end}$ will be different. We therefore have a  $\phi=\phi_{end}$ surface which locally looks like an unperturbed universe, the only difference is that the have a difference local scale factor~\footnote{Notice that since this surface has the same energy but different overall expansion: by GR, there must be a curvature for space.}. These are the curvature perturbations that we call $\zeta$. In formulas
\be
\delta\phi\quad\Rightarrow\quad\delta t_{inflation}\sim\frac{\delta\phi}{\dot\phi}\quad\Rightarrow \delta{\rm expansion}\sim\zeta\sim\frac{\delta a}{a}\sim H\delta t_{inflation}\sim \frac{H}{\dot\phi}\delta\phi
\ee
Here we defined in an approximate way $\zeta\sim\delta a /a$. We will define it rigorously later on.
So, the power spectrum of the curvature perturbation is given by 
\be
\langle\zeta_{\vec k}\zeta_{\vec k'}\rangle=\frac{H^2}{\dot\phi^2}\langle\phi_{\vec k}\phi_{\vec k'}\rangle=(2\pi)^3\delta^{3}(\vec k+\vec k') \frac{H^4}{\dot\phi^2}\cdot\frac{1}{k^3}\equiv (2\pi)^3\delta^{3}(\vec k+\vec k') P_\zeta
\ee
\be
P_\zeta= \frac{H^4}{\dot\phi^2}\cdot\frac{1}{k^3}\simeq \frac{H^2}{\mpl^2 \epsilon}
\ee
where in the second passage we have used the slow roll expressions.

{ It is the time-delay, stupid!~\footnote{No offense to anybody: this is just a famous quote from Bill Clinton in his campaign to become president in 1992.}}. It is important to realize that the leading mechanism through which inflation generates perturbations is by the time delay induced by the inflation fluctuations, not by the fluctuations in energy during inflation. It took some time for the community to realise this. Let us be sure about this. In slow roll inflation the potential needs to be very flat, we can therefore work by expanding in the smallness of the slow roll parameters. How large are the metric perturbations? Well, the difference in energy associated to a jump of the inflation is about 
\be
\delta\rho\sim V' \delta\phi\sim {\sqrt \epsilon} H^3\mpl \quad\Rightarrow\quad\delta g^{\mu\nu}\sim \frac{\delta\rho}{\rho}\sim {\sqrt \epsilon} \frac{H}{\mpl}
\ee
This means that the curvature perturbation due to this effects has actually an $\epsilon$ {upstairs}, so, in the limit that $\epsilon$ is very small, this is a subleading contribution. Notice indeed that the time-delay effect has an $\epsilon$ {\it downstairs}: the flatter is the potential, the longer it takes to make-up for the lost or gained $\phi$-distance, and so the more $\delta$expansion you get. This is ultimately the justification of why we could do the correct calculation without having to worry {\it at all} about metric perturbations.

\subsubsection{$\zeta$ conservation for modes longer than the horizon} Why we cared to compute the power spectrum of $\zeta\sim \delta a/a$? Why do we care of $\zeta$ and not of something else? The reason is that this is the quantity that it is conserved during all the history of the universe from when a given mode becomes longer than $H^{-1}$, to when it becomes shorter the $H^{-1}$ during the standard cosmology. This is very very important. We know virtually nothing about the history of the universe from when inflation ends to say BBN. In order to trust the predictions of inflation, we need something to be constant during this epoch, so that we can connect to when we know something about the universe. Proving this constancy in a rigorous way requires  some effort, and it is a current topic of research to prove that this conservation holds at quantum level. For the moment, it is easy to give an heuristic argument. The $\zeta$ fluctuation is defined as the component of the metric that represents  the perturbation to the scale factor $a_{\rm eff}=a(1+\zeta)$. Let us consider the regime in which all modes are longer than the Hubble scale. The universe looks locally homogenous, with everywhere the same energy density, exactly the same universe, with the only difference that in each place the scale factor is valued $a(1+\zeta)$ instead of $a$. But remember that the metric, apart for tensor modes, is a constrained variable fully determined by the matter fluctuations. Since matter is locally unperturbed, how can it change in a time dependent way the evolution of the scale factor? Impossible.  The scale factor will evolve as in an unperturbed universe, and therefore $\zeta$ will be constant in time. This will happen until gradients will become shorter than Hubble again, so that local dynamics will be able to feel that the universe is not really unperturbed, and so $\zeta$ will start evolving.

We should think that  it is indeed $\zeta$ that sources directly the temperature perturbations we see in the CMB. We should think that $P_\zeta\sim 10^{-10}$.

The argument above is heuristic. In practice, the proof of the conservation of $\zeta$ is quite complicated. Some proofs of the conservation of $\zeta$ outside of the horizon at tree  level are given in~\cite{Maldacena:2002vr}, while at loop, quantum, level are given in~\cite{Pimentel:2012tw,Senatore:2012ya}.

\subsection{Scale invariance and tilt}
As we saw, the power spectrum of $\zeta$ is given by
\be
P_\zeta(k)= \frac{H^4}{\dot\phi^2}\cdot\frac{1}{k^3}\simeq \frac{H^2}{\mpl^2 \epsilon}\frac{1}{k^3}
\ee
This is a scale invariant power spectrum. The reason why it is scale invariant is because every Foureir mode sees exactly the same history: it starts shorter than $H^{-1}$, becomes longer than $H^{-1}$, and becomes constant. In the limit in which $H$ and $\dot\phi$ are constant (we are in an attractor solution, so $\ddot\phi$ is just a function of $\phi$), then every  Fourier mode sees the same history and so the power in each mode is the same.
In reality, this is only an approximation. Notice that the value of $H$ and of $\dot\phi$ depend slightly on the position of the scalar field. In order to account of this, the best approximation is to evaluate for each mode $H$ and $\dot\phi$ at the time when the mode crossed Hubble and became constant. This happens at the $k$-dependent $t_{f.o.}(k)$ freezing time defined by
\bea
&& \omega(t_{f.o.})\simeq\frac{k}{a(t_{f.o.})}=H(t_{f.o.})\\ \nonumber
&&\quad\Rightarrow\quad t_{f.o.}(k)\simeq \frac{1}{H( t_{f.o.}(k))}\log\left(\frac{H( t_{f.o.}(k))}{k}\right)
\eea
This leads to a deviation from scale invariance of the power spectrum. Our improved version now reads
\be
P_\zeta= \frac{H(t_{f.o.}(k))^4}{\dot\phi(t_{f.o.}(k))^2}\cdot\frac{1}{k^3}
\ee
A measure of the scale dependence of the power spectrum is given by the tilt, defined such that the $k$-dependence of the power spectrum is approximated by the form
\be
P_\zeta\sim \frac{1}{k^3}\left(\frac{k}{k_0}\right)^{n_s-1}
\ee
where $k_0$ is some pivot scale of reference. We therefore have
\be
n_s-1=\frac{d\log(k^3 P_k)}{d\log k}=\left.\frac{d \log\left(\frac{H^4}{\dot\phi^2}\right)}{d \log k}\right|_{k/a\sim H}=\left.\frac{d \log\left(\frac{H^4}{\dot\phi^2}\right) }{H d t} \frac{H d t}{d\log k}\right|_{k/a\sim H}
\ee
where we have used the fact that the solution is a function of $k$ though the ratio $k/a$ as this is the physical wavenumber. At this point we can use that
\be
d\log k=d\log(a H)\simeq H dt
\ee
to obtain
\be
n_s-1\simeq -2\frac{\dot H}{H^2}+2\frac{\ddot \phi}{H\dot\phi}=4\epsilon_H-2\eta_H
\ee
The tilt of the power spectrum is  of order of the slow roll parameters, as expected.
How come we were able to compute the tilt of the power spectrum that is slow roll suppressed, though we neglected metric fluctuations, that are also slow roll suppressed? The reason is that the correction to the power spectrum due to the tilt become larger and larger as $k$ becomes more and more different from $k_0$. Metric fluctuations are expected to give a finite correction of order slow roll to the power spectrum, but not one that is enhanced by the difference of wave numbers considered. This is the same approximation we do in Quantum Field Theory when we use the running of the couplings (which is $\log$ enhanced), without bothering of the finite corrections. The pivot scale $k_0$ is in this context analogous to the renormalization scale.

\subsection{Energy scale of Inflation}

We can at this point begin to learn something about inflation. Remember that the power spectrum and its tilt are of order
\be
P_\zeta\sim \frac{H^2}{\mpl^2\epsilon}\frac{1}{k^3}\ ,\qquad n_s-1=4\epsilon_H-2\eta_H\ ,
\ee
with, for slow roll inflation
\be
H^2\simeq \frac{V(\phi)}{\mpl^2}
\ee
From observations of the CMB, we know that
\be
P_\zeta\sim 10^{-10}\ ,\qquad  n_s-1\sim 10^{-2}\ .
\ee
Knowledge of these two numbers is not enough to reconstruct the energy scale of inflation. However, if we assume for the moment that $\eta\sim \epsilon$, a reasonable assumption that however it is sometimes violated (we could have $\epsilon\ll \eta$), then we get
\be
\frac{H}{\mpl}\sim 10^{-6}\ ,\qquad H\sim 10^{13} {\rm GeV}\ , \qquad V\sim \left(10^{15} {\rm GeV}\right)^4
\ee
These are remarkably large energy scales. This is the energy scale of GUT, not very distant from the Plank scale. Inflation is really beautiful. Not only it has  made quantum fluctuations the origin of all the structures of the universe, but it is likely that these are generated by physics at very high energy scales. These are energy scales that unfortunately we will probably never be able to explore at particle accelerators. But these are energy scales that we really would like to be able to explore. We expect very interesting new physics to lie there: new particles, possibly GUT theories, and even maybe string theory. We now can explore them with cosmological observations!

\subsection{Statistics of the fluctuations: Approximate Gaussianity}

Let us go back to our action of the fluctuations of the scalar field. Let us write again the action in Fourier space, but this time it turns out to be simpler to work in a finite comoving box of volume $V$.
We have
\be
\phi(x)=\frac{1}{V}\sum_{\vec k} \phi_k e^{i \vec k\cdot \vec x}
\ee
Notice that the mass dimensions of $\phi_{\vk}$ is $-2$.
To get the action, we need the following manipulation
\bea
&&\int d^3x\; \phi(x)^2=\frac{1}{V^2}\sum_{k,k'}\int d^3x\; e^{i (\vec k+\vec k')\cdot\vec x}\phi_k\;\phi_{k'}\simeq\frac{1}{V^2}\sum_{k,k'}\delta^{3}(\vec k+\vec k')\phi_k\;\phi_{k'}\\ \nonumber
&&\quad\simeq \frac{1}{V}\sum_{k,k'}\delta_{\vec k,-\vec k'}\phi_k\;\phi_{k'}=\frac{1}{V}\sum_{k,k'}\phi_k\;\phi_{-k'}
\eea
The action therefore reads
\be
S_2=\frac{1}{V}\sum_k a^3 \left(\dot\phi_{\vk}\dot\phi_{-\vk}+\frac{k^2}{a^2}\phi_{\vec k}\phi_{-\vk}\right)
\ee
Let us find the Hamiltonian. We need the momentum conjugate to $\phi_{\vec k}$.
\be
\Pi_{\vk}=\frac{\delta S_2}{\delta\dot\phi_{\vec k}}=\frac{a^3}{V}\dot\phi_{-\vk}
\ee
The Hamiltonian reads
\bea
&&H=\sum_{\vk}\Pi_{\vk}\dot\phi_{\vk}-\frac{1}{V}\sum_k a^3 \left(\dot\phi_{\vk}\dot\phi_{-\vk}+\frac{k^2}{a^2}\phi_{\vec k}\phi_{-\vk}\right) \\ \nonumber
&&\quad=\sum_{\vk} \frac{V}{a^3} \Pi_{\vk} \Pi_{-\vk}+\frac{1}{V}\frac{k^2}{a^2}\phi_{\vec k}\phi_{-\vk}
\eea
If we concentrate on early times where the time dependence induced by Hubble expansion is negligible, we have,
for each $\vk$ mode, the same Hamiltonian as an Harmonic oscillator, which reads (again, remember that I am dropping all numerical factors)
\be
H=\frac{P^2}{m}+m \omega^2 x^2
\ee
We can therefore identify
\be\label{eq:mapping}
m=\frac{a^3}{V}\ , \qquad \phi_{\vk}=x\ ,\qquad \omega=\frac{k}{a} .
\ee
The vacuum wave function for an harmonic oscillator is a Gaussian
\be
|0\rangle=\int dx\; e^{- m \omega x^2 }|x\rangle
\ee
which tells us that the vacuum wave function for each Fourier mode $\vk$ reads 
\be
|0\rangle_{k/a \gg H}=\int d{\phi_{\vk}}\; e^{- \frac{a^3}{V}\frac{k}{a} \phi_{\vk}^2} |\phi_k\rangle
\ee
Since all Fourier mode evolve independently, for the set of Fourier modes that have $k/a\gg H$, we can write
\be
|0\rangle_{k_i/a \gg H}=\prod_{\vk_i\gg H a}\int d{\phi_{\vk_i}}\; e^{- \frac{a^3}{V}\frac{k_i}{a} \phi_{\vk_i}^2} |\phi_{\vk_i}\rangle%=\int {\cal D}_{}{\phi}_{\vk_i\gg H a}\; e^{- \frac{a^3}{V}\frac{k_i}{a} \phi_{\vk_i}^2} |\phi_{\vk_i}\rangle
\ee
For each Fourier mode, at early time we have a Gaussian wave function with width $V^{1/2}/(k^{1/2} a)$.

Let us follow the evolution of the wave function with time. As discussed, at early times when $k/a\gg H$, the wave function follows adiabatically the wave function of the would be harmonic oscillator with those time dependent mass and frequency given by (\ref{eq:mapping}). However, as the frequency drops below the Hubble rate, the natural time scale of the harmonic oscillator becomes too slow to keep up with Hubble expansion. The state gets frozen on the parameters that it had when $\omega(t)\sim H$. Bu substituting $k/a\to H$, $a\to k/H$,the wave function at late times becomes
\be\label{eq:latewavefunction}
|0\rangle_{k_i/a \ll H}=\prod_{\vk_i\ll H a}\int d{\phi_{\vk_i}}\; e^{- \frac{1}{V}\frac{k_i^3}{H^2} \phi_{\vk_i}^2} |\phi_{\vk_i}\rangle%=\int {\cal D}{\phi}_{\vk_i\ll H a}\; e^{- \frac{1}{V}\frac{k_i^3}{H^2} \phi_{\vk_i}^2} |\phi_{\vk_i}\rangle
\ee
This is a Gaussian in field space. Its width is given by
\be
\langle\phi_{\vk}\phi_{\vk'}\rangle=\delta_{\vk,\vk'} V \frac{H^2}{k^3}\simeq (2\pi)^3\delta^3(\vk+\vk') \frac{H^2}{k^3} \quad{\rm as}\quad{V\to\infty}
\ee
We recover the same result of before for the power spectrum. We additionally see that the distribution of values of $\phi_{\vk}$ are Gaussianly distributed. Notice that we are using a quite unusual base of the Hilbert space of a quantum field theory (more used when one talks about the path integral), which is the  $|\phi\rangle$ eigenstates base instead of the usual Fock base with occupation numbers. This base is sometimes more useful, as we see here.

So, we learn that the distribution is Gaussian. This result could have been expected. At the end, (so far!), we started with a quadratic Lagrangian, the field theory is free, and so equivalent to an harmonic oscillator, which, in its vacuum, is Gaussianly distributed. We will see in the last lecture that when we consider interacting field theories the distribution will not be Gaussian anymore! Indeed, the statement that cosmological perturbations are so far Gaussian simply means that the field theory describing inflation is a weakly coupled quantum field theory in its vacuum. We will come back to this.

\subsection{Why does the universe looks classical?}

So far we have seen that the cosmological fluctuations  are produced by the quantum fluctuations of the inflation in its vacuum state. But then, why does the universe looks classical?
The reason is the early vacuum state for  each wave number becomes a very classical looking state at late times. Let us see how this happens.

The situation is very simple. We saw in the former subsection that the vacuum state at early times is the one of an harmonic oscillator with frequency $k/a\gg H$. However the frequency is red shifting, and at some point it becomes too small to keep up with Hubble expansion. At that point, while the frequency goes to zero, the state remains trapped in the vacuum state of the would-be harmonic oscillator with frequency $k/a\sim H$. The situation is very similar to what happens to the vacuum state of an harmonic oscillator when one opens up very abruptly the width of the potential well.

This is an incredibly squeezed state with respect to the ground state of the harmonic oscillator with frequency $\omega\sim e^{-60} H$. This state is no more the vacuum state of the late time harmonic oscillator. It has a huge occupation number, and it looks classical. 

\begin{figure}[h!]
\begin{center}
\includegraphics[width=15cm]{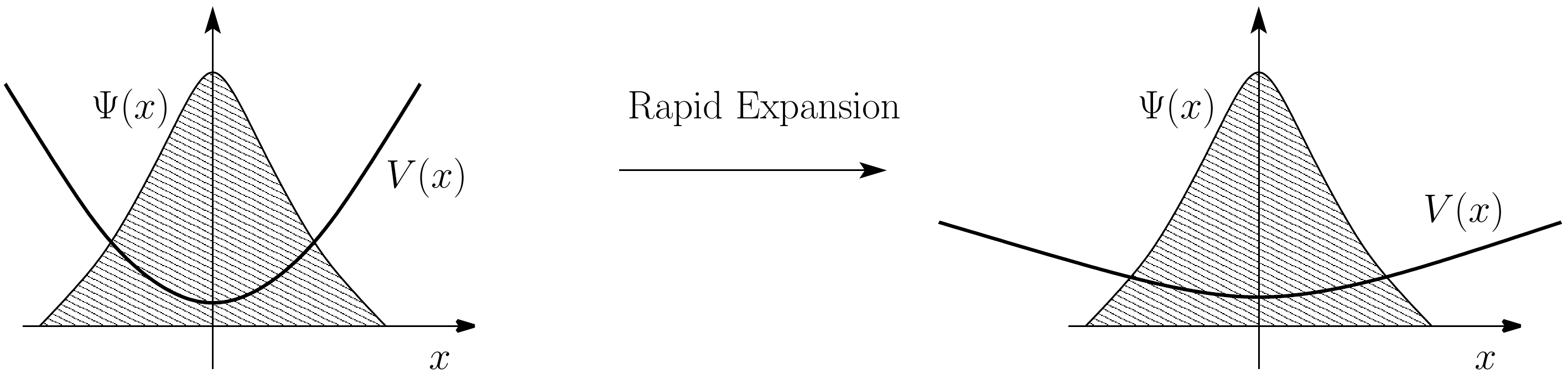}
\caption{\label{fig:scales1} \small Formation of a squeezed state by the rapid expansion of the universe.}
\end{center}
\end{figure}

Let us check that indeed that wave function is semiclassical. The typical condition to check if a wavefunction is well described by a semiclassical approximation is to check if the $\phi$-length scale over which the amplitude of the wavefunction changes is much longer than the $\phi$-length scale over which the phase changes.  To obtain the wavefunction at late times, we performed the sudden approximation of making the frequency instantaneously zero. This corresponds to make an expansion in $k/(a H)$. In our calculation we obtained a real wavefunction (\ref{eq:latewavefunction}). This means that the phase must have been higher order in $k/(aH)\ll 1$, in the sense that it should be much more squeezed than the width of the magnitude, much more certain the outcome: the time-dependent phase has decayed away. We therefore can write approximately
\be\label{eq:latewavefunction}
|0\rangle_{k_i/a \ll H, guess}\sim  \prod_{\vk_i\ll H a}\int d\phi_{\vk_i}\; e^{- \frac{1}{V}\frac{k^3}{H^2} \phi_{\vk_i}^2 \left[1+i \frac{a H}{k}\right]} |\phi_{\vk_i}\rangle %\sim \int {\cal D } \phi_{\vk_i\ll H a}e^{- \frac{1}{V}\frac{k^3}{H^2} \phi_{\vk_i}^2 \left[1+i \frac{a H}{k}\right]} |\phi_{\vk_i}\rangle
\ee
We obtain:
\be
\Delta\phi_{\rm Amplitude\ Variation}\sim \frac{H}{k^{3/2}}\frac{1}{V^{1/2}}\ ,\qquad \Delta\phi_{\rm Phase\ Variation}\sim \frac{H}{k^{3/2}}\frac{1}{V^{1/2}}  \left(\frac{k}{a H}\right)^{1/2}\ ,\qquad
\ee
So
\be
\frac{\Delta\phi_{\rm Phase\ Variation}}{\Delta\phi_{\rm Amplitude\ Variation}}\sim \left(\frac{k}{a H}\right)^{1/2}\to 0
\ee
So we see that the semiclassicality condition is satisfied at late times.

Notice furthermore that the state of the inflation is a very squeezed state. The variance of $\delta\phi$ is huge. Since we have just verified that the system is classical, this means that the system has approached a classical stochastical description. A nice discussion of this, stated not exactly in this language, is given in~\cite{Guth:1985ya}.

Of course, later in the universe, local environmental correlations will develop that will decorrelate the quantum state. But we stress that the system is semiclassical even before decorrelation effects are taken into account.

\subsection{Tensors}

Before moving on, let us discuss briefly the generation of tensor modes. In order to do that, we need to discuss about the metric fluctuations. (Remarkably, this is the first time we have to do that).

\subsubsection{Helicity Decomposition of metric perturbations}

A generically perturbed FRW metric can be put in the following form
\be
ds^2=-(1+2\Phi)dt^2+2a(t) B_i dx^i+a(t)^2\left[(1-2\Psi)\delta_{ij}+E_{ij}\right]
\ee
For background space-times that have simple transformation rules under rotation (FRW for example is invariant), it is useful to decompose these perturbations according to their transformation properties under rotation under one axis. A perturbation of wavenumber $\vk$ has elicity $\lambda$ if under a rotation along the $\hat k$ of angle $\theta$, transforms simply by multiplication by $e^{i\lambda\theta}$:
\be
\delta g\to e^{i\lambda\theta}\delta g
\ee
Scalars have helicity zero, vectors have helicity one, and tensors have helicity two.
It is possible to decompose the various components of $\delta g_{\mu\nu}$ in the following way:
\be
\Phi, \  \Psi 
\ee
have helicity zero. We can then write
\be
B_i=\d_i B_S+\tilde B_{V,i}
\ee
where $\d^i\tilde B_{V,i}=0$. $B_S$ is a scalar, $B_V$ is a vector. Finally
\be
E_{ij}=E_{ij}^S+E_{ij}^V+\gamma_{ij}
\ee
where 
\bea
&&E_{ij}^S=\frac{1}{\d^2}\left(\d_i\d_j-\frac{1}{3}\delta_{ij}\d^2\right)\tilde E^S\\ \nonumber
&&E_{ij}^V=\frac{1}{2\d^2}\left(\d_i \tilde E_j^V+\d_j \tilde E_i^V \right)\ , \quad{\rm with}\quad \d_i\tilde E^{V,i}=0\\ \nonumber
&&\d_i \gamma_{ij}=0, \quad \gamma_i{}^i=0\ .
\eea
with $\d^2=\delta^{ij}\d_i\d_j$. $\tilde E^S$ is a scalar, $\tilde E^V$ is a vector, and $\gamma$ is a tensor.

Now, it is possible to show that at linear level, in a rotation invariant background, scalar, vector and tensor modes do not couple and evolve independently (you can try to contract the vectors togetherÉ it does not work: you cannot make it). 

Under a change of coordinate
\be
x^\mu\to \tilde x^\mu=x^\mu+\xi^\mu
\ee
these perturbations change according to the transformation law of the metric
\be
\tilde g^{\mu\nu}=\frac{\d \tilde x^\mu}{\d x^\rho}\frac{\d \tilde x^\nu}{\d x^\sigma} g^{\rho\sigma}
\ee
The change of coordinates $\xi^\mu$ can also be decomposed into a scalar and a vector component
\bea
&&\xi_S^0\ ,\quad \xi_S^i=\d^i\xi \\
&&\xi_V^0=0\, ,\quad \d_i\xi_V^i=0 
\eea
At linear level, different helicity metric perturbations do not get mixed and they are transformed only by the change coordinates with the same helicity (for the same reasons as before). For this reasons, we see that tensor perturbations are invariant. They are gauge invariant.
This is not so for scalar and vector perturbations. For example, scalar perturbations transform as the following
\bea
&&\Phi\to \Phi-\dot \xi_S^0\\
&&B_S\to B_S+\frac{1}{a}\xi^0_S-a\dot \xi\\ 
&&E\to E-B_S\\
&&\Psi\to\Psi-H\alpha
\eea
The fact that tensor modes are gauge invariant and uncoupled (at linear level!) means that we can write the metric for them as
\be
g_{ij}=a^2\left(\delta_{ij}+\gamma_{ij}\right)\ ,
\ee
and set to zero all other perturbations (including $\delta\phi$). By expanding the action for the scalar field plus GR at quadratic order, one obtains an action of the form (actually only the GR part contributes, and the following action could just be guessed)
\be
S\sim \int d^4x\; a^3\;\mpl^2\left[(\dot \gamma_{ij})^2-\frac{1}{a^2}(\d_l \gamma_{ij})^2\right]\sim\sum_{s=+,\times}\int dt d^3k\;a^3\;\mpl^2 \left[\dot \gamma_{\vec k}^s\dot \gamma_{-\vec k}^s-\frac{k^2}{a^2} \gamma_{\vec k}^s \gamma_{-\vec k}^s\right]\\
\ee
where in the last passage we have decomposed the generic tensor mode in the two possible polarization state 
\be
\gamma_{ij}^{(+,\times)}=\gamma_{(+,\times)}(t)e_{ij}^{(+,\times)}
\ee
In matrix form, for a mode in the $\hat k=\hat z$ direction
\be
\gamma=\left(\begin{array}{rrr}
\gamma^\times & \gamma^+  & 0\\
\gamma^+ & -\gamma^\times  & 0\\
0 & 0 & 0 \\
\end{array}\right)
\ee
\bea
&&{\rm\ }\quad \gamma_{ij}=\int d^3k \sum_{s=+,\times} e^s_{ij}(k) \gamma^s_{\vec k}(t)e^{i \vec k\cdot \vec x}\\
&&{\rm\ }\quad \epsilon_{ii}^s=k^i\epsilon_{ij}^s=0\, \quad \epsilon^s_{il}\epsilon^{s'}_{lj}=\delta_{ij}
\eea
We see that the action for each polarization is the same as for a normal scalar field, just with a different canonical normalization. The two polarization are also independent (of course), and therefore, without having to do any calculation, we obtain the power spectrum for gravity waves to be
\be
\langle\gamma^{s}_{\vk}\gamma^{s'}_{\vk'}\rangle=(2\pi)^3\delta^3(\vk+\vk')\delta_{s,s'}\frac{H^2}{\mpl^2}\frac{1}{k^3}
\ee

Notice that the power spectrum depends only on one unknown quantity $H$. This means that if we detect gravitational waves from inflation, we could measure the energy scale of inflation. $\ldots$ Actually, this was a `theorem' that was believed to hold until last september. At that time new mechanisms further than the vacuum fluctuations have been identified that could dominate the ones produced by vacuum fluctuations and that could be detectable~\cite{Senatore:2011sp}.

By now we are expert: the tilt of gravity waves power spectrum is given by
\be
n_t-1=-2\epsilon_H
\ee
as only the variation of $H$ is involved.

The measurement of this tilt would give us a measurement of $\epsilon$. Again, until recently this was thought to be true, and unfortunately (and luckily) things have changed now, and the above formula for the tilt holds only for the simplest models of inflation.

Notice further that if we were to measure the amplitude of the gravitational waves and their tilt, then, under the hypothesis of standard slow roll inflation, we would know $H$ and $\epsilon$. In this same hypothesis therefore we would therefore predict the size of the $\zeta$ power spectrum. If this would hold, we would discover that inflation happened in the slow roll inflation way. This is called consistency condition for slow roll single field inflation. 

Notice that, in standard slow roll inflation (this is true only for the simple inflationary scenarios), the power in gravity waves is smaller than the one in scalars by a factor of $\epsilon\ll 1$. This means that if gravity waves are detected,  $\epsilon$ cannot be too small, and therefore the field excursion during inflation is over planckian: $\Delta\phi\gtrsim \mpl$. This is known as the Lyth's bound~\cite{Lyth:1996im}.

Finally, notice that this signal is proportional to $\hbar$. Such a measurement would be the first direct evidence that GR is quantized. We have never seen this (frankly there are no doubt that gravity is quantizedÉ but still better to see it in experiments.)

\subsection{Summary of Lecture 2}

\begin{itemize}
\item the quantum fluctuations of the scalar field naturally produce a scale invariant spectrum of perturbations
\item they become curvature perturbations at the end of inflation
\item they look like classical and (quasi) Guassian
\item Quantum mechanical effects are at the source of the largest structures in the universe
\item The Energy scale of inflation could be as high as the GUT scale, opening the possiblity to explore the most fundamental laws of physics from the cosmological observations
\item Tensor modes are also produced. If seen, first evidence of quantization of gravity.
\item Everything is derived without hard calculations
\end{itemize}

Now we are ready to see how we check for this theory in the data.

\newpage
\section{Lecture 3: contact with observations and the Effective~Field~Theory~of~Inflation}

Absolutely, the best way we are testing inflation is by the observation of the cosmological perturbations.

Here I will simply focus on the minimum amount of information that we need to establish what this observations are really telling us about Inflation. I will focus just on CMB, for brevity. The story is very similar also for large scale structures.

\subsection{CMB basics}

For a given perturbation $\delta X(k,\tau)$ at a given time $\tau$ and with Fourier mode $k$, we can define its transfer function for the quantity $X$ at that time $\tau$ and for the Fourier mode $k$ as
\be
\delta X(k,\tau)= T(k,\tau,\tau_{in})\zeta_k(\tau_{in})
\ee
This must be so in the linear approximation. We can take $\tau_{in}$ early enough so that the mode $k$ is smaller than $a H$, in this way $\zeta_k(\tau_{in})$ represents the constant value $\zeta$ took at freeze out during inflation.

For the CMB temperature, we perform a spherical harmonics decomposition
\be
\frac{\delta T}{T}(\tau_0,\hat n)=\sum_{l,m}a_{lm} Y_{lm}(\hat n)
\ee
and the by statistical isotropy the power spectra reads
\be
\langle a_{lm}a_{l'm'}\rangle=C_l^{TT}\delta_{ll'}\delta_{mm'}
\ee
Since the temperature anisotropy are dominated by scalar fluctuations, we have
\be
a_{lm}=\int d^3k \Delta_{l}(k)\zeta_k Y_{lm}(\hat k)\ ,\quad \Rightarrow\quad C_l=\int dk\; k^2\; \Delta_l(k)^2\ \  P_\zeta(k)\ , 
\ee
$\Delta_l(k)$ contains both the effect of the transfer functions and also of the projection on the sky. 

\begin{itemize}
\item {\bf large scales:} If we look at very large scales, we find modes that were still outside $H^{-1}$ at the time of recombination (see Fig.~\ref{fig:scales2}). Nothing could have happened to them.

\begin{figure}[h!]
\begin{center}
\includegraphics[width=10cm]{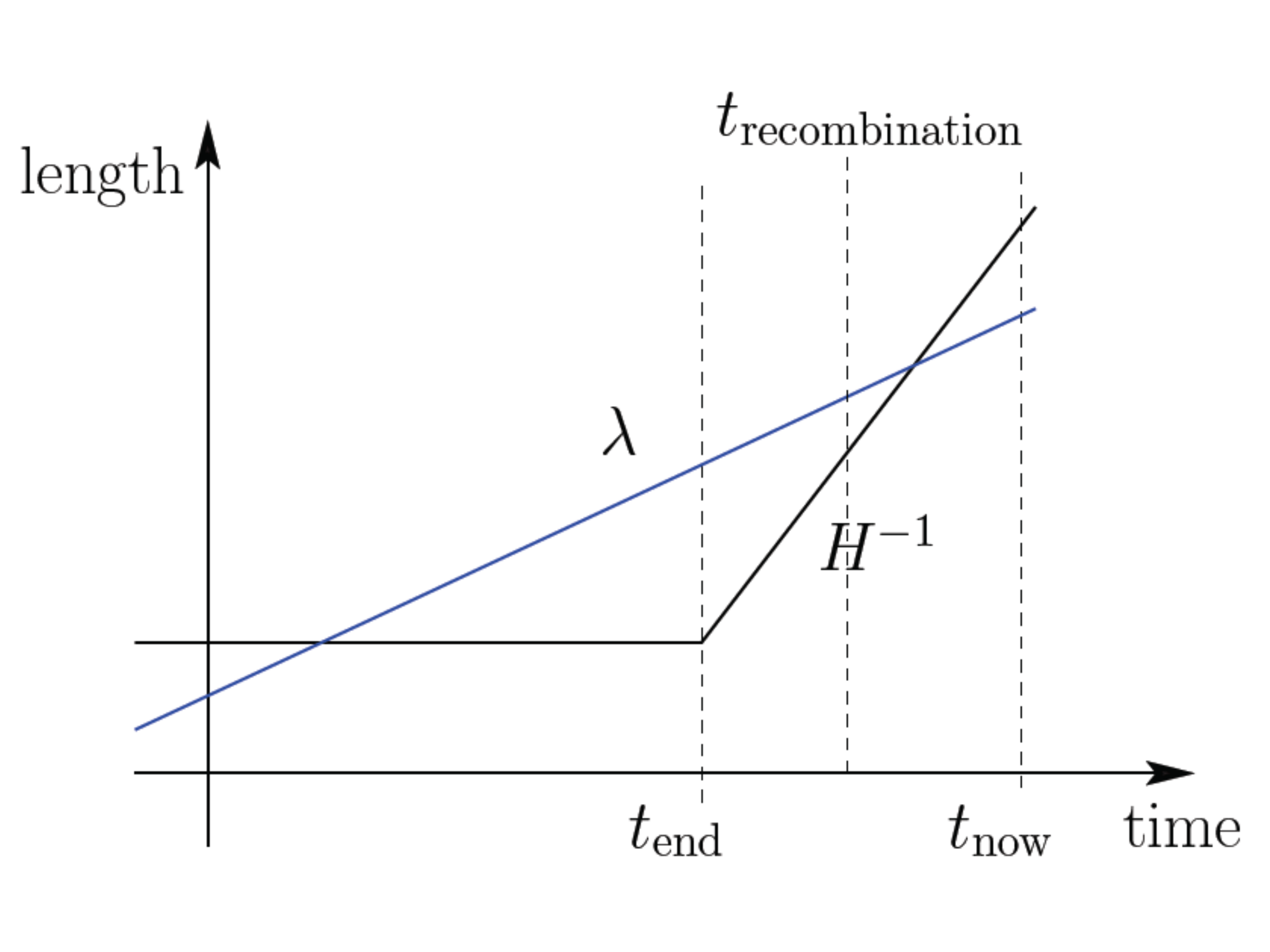}
\caption{\label{fig:scales2} \small Relative ratios of important length scales as a function of time in the inflationary universe. There are length scales that we can see now that were longer than $H^{-1}$ at the time of recombination.}
\end{center}
\end{figure}
There has been no evolution and only projection effects.
\be
 \Delta_{l}(k)\simeq j_l(k(\tau_0-\tau_{\rm rec}))\quad\Rightarrow\quad C_l\simeq\int dk\, k^2 P_\zeta\ j_l^2(k(\tau_0-\tau_{\rm rec}))
\ee
$j_l^2(k(\tau_0-\tau_{rec}))$ is sharply peaked at $k(\tau_0-\tau_{\rm rec})\sim l$, so we can approximately perform the integral, to obtain
\bea
&& C_l\simeq \left.k^3 P_\zeta \right|_{k=l/(\tau_0-\tau_{\rm rec})} \times \int d\log x\; j_l^2(x)\sim \left.k^3 P_\zeta \right|_{k=l/(\tau_0-\tau_{\rm rec})}\times\frac{1}{l(l+1)}\\
&&\Rightarrow \quad l(l+1)C_l\ \  {\rm is\ flat\ , \ equivalently\ }l{\rm-independent}\ .
\eea

\item {\bf small scales}. On short scales, mode entered inside $H^{-1}$ and begun to feel both the gravitational attraction of denser zones, but also their pressure repulsion. This leads to oscillatory solutions.
\bea
&&\ddot\delta T+c_s^2 \nabla^2\delta T\simeq F_{\rm gravity}(\zeta)\\
&&\Rightarrow\qquad \delta T_k\simeq A_{\vec k}\, \cos(k\eta)+B_{\vec k}\,\sin(k\eta)=\tilde A_{\vec k}\,\cos(k\eta+\phi_{\vec k})
\eea
Here $A_{\vec k}$ and $B_{\vec k}$ depend on the initial conditions. In inflation, we have 
\be
\tilde A_{\vec k}\simeq \frac{1}{k^3}\ ,\qquad \phi_{\vec k}=0\ .
\ee
All the modes are in phase coherence. Notice, dynamics and wavenumber force all mode of a fixed wavenumber to have the same frequency. However, they need not have necessarily the same phase. Inflation, or superHubble fluctuations, forces $\zeta\simeq\frac{\delta T}{T}=$const on large scales, which implies $\phi_{\vec k}=0$.
This is what leads to acoustic oscillations in the CMB
\bea
&&\delta T(\vec k,\eta)\sim \delta \tin(\vec k) \times \cos(k\eta)\quad\Rightarrow\quad \delta T(\vec k,\eta_{0})\sim  \delta \tin(\vec k) \times \cos(k\eta_{rec})\\
&&\Rightarrow\quad \langle\delta T(\vec k,\eta_0)\rangle\sim \langle\delta T_{\vk}^2\rangle \cos^2(k\eta_{rec})
\eea
we get the acoustic oscillations.

\begin{figure}[h!]
\begin{center}
\includegraphics[width=15cm]{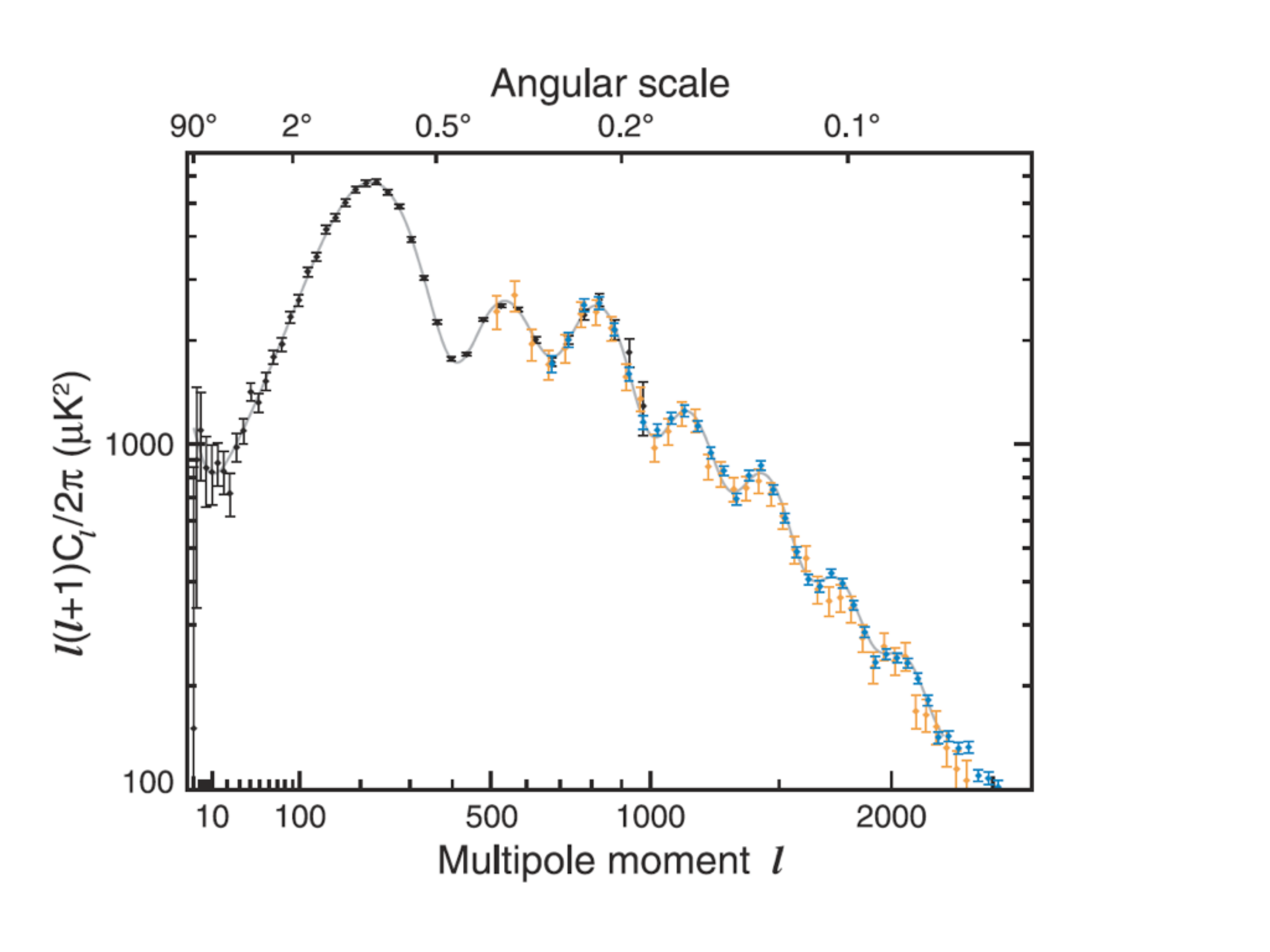}
\caption{\label{fig:phases} \small Power spectrum of the CMB fluctuations. Oscillations are clearly seeable. Picture is taken from~\cite{Hinshaw:2012fq}, which combines the result of several CMB experiments such as WMAP, SPT~\cite{Keisler:2011aw} and ACT~\cite{Das:2011ak}.}
\end{center}
\end{figure}

\begin{figure}[h!]
\begin{center}
\includegraphics[width=15cm]{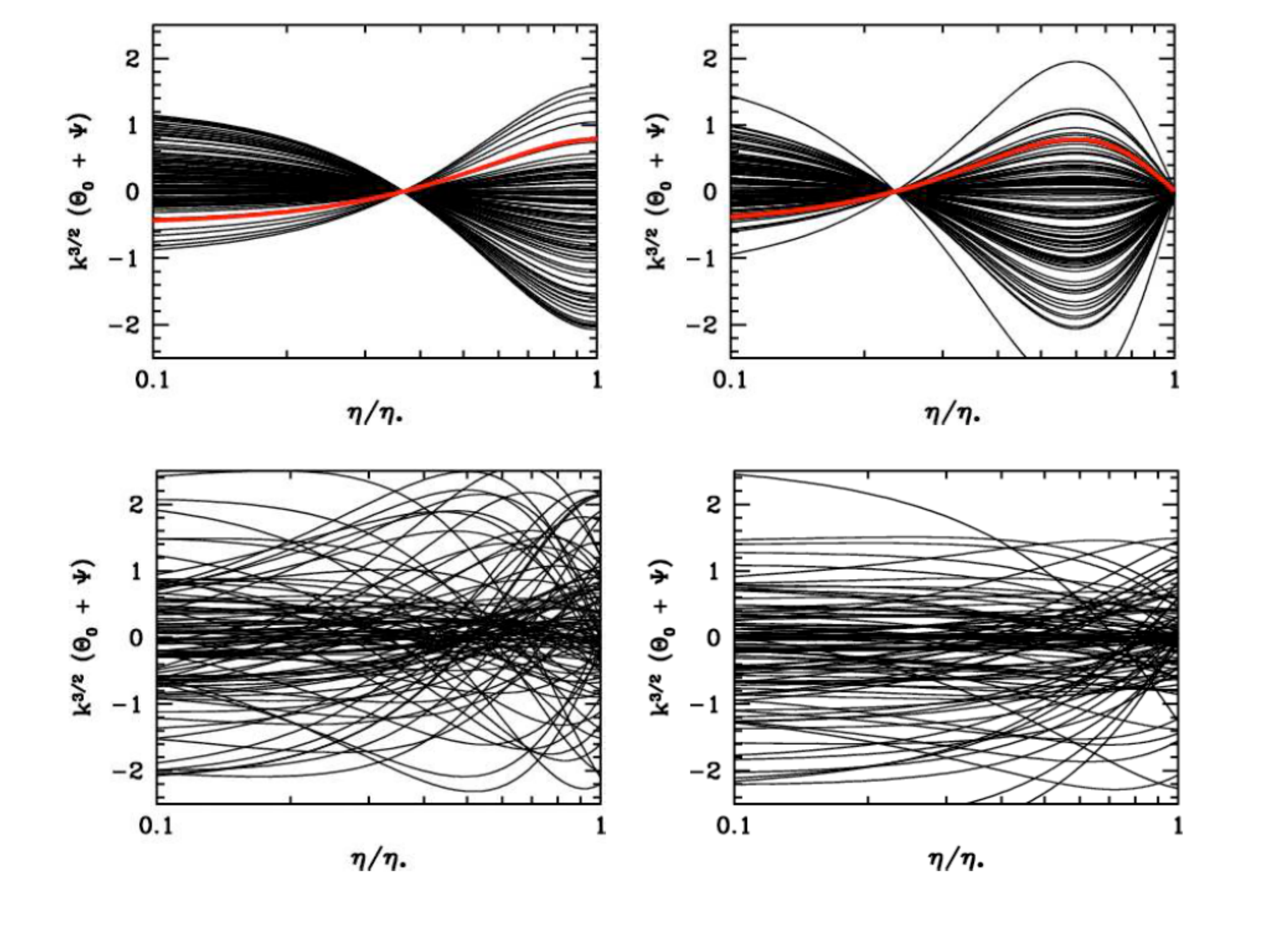}
\caption{\label{fig:phases} \small On top: time evolution of two different modes that have different initial amplitude, but all the same phases. We see that the typical size of the amplitude at the time of recombination is different for different modes. We obtain oscillations in the power spectrum. On bottom: time evolution of two different modes with different amplitudes and phases. We see that the typical size of the fluctuations at the time of recombination is independent of the wavenumber. The power spectrum has not oscillations and is featureless. These pictures are taken from~\cite{Dodelson:2003ip}.}
\end{center}
\end{figure}

This is the greatest qualitative verification of inflation so far. Acoustic oscillations told us that the horizon was much larger than $H^{-1}$ at recombination and that there were constant superHubble perturbation before recombination. This is very very non-trivial prediction of inflation. Notice that scale invariance of the fluctuations was already guessed to be in the sky (Harrison-Zeldovich spectrum) at the time of formulation of inflation, but nobody knew of the acoustic oscillations at that time. CMB experiments found them!

This is a very important  {\it qualitative} verification of inflation that we get from the CMB. But it is not a quantitative confirmation. Information on the quantitative part is very limited.
\end{itemize}

\subsection{What did we verify of Inflation so far?}

\begin{figure}[h!]
\begin{center}
\includegraphics[width=10cm]{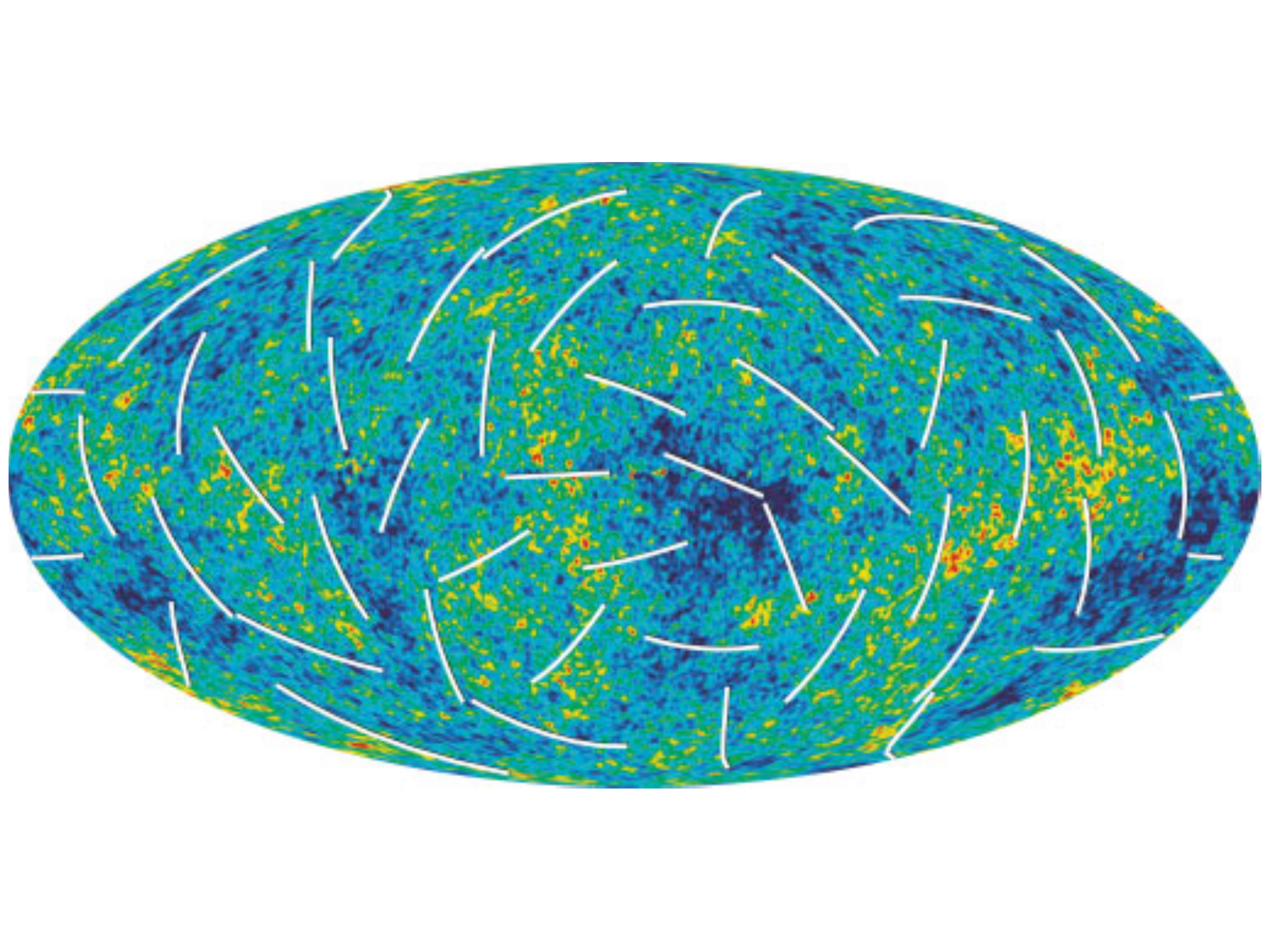}
\caption{\label{fig:CMBpolarization2} \small A nice picture of the CMB as measured by the WMAP experiment~\cite{Bennett:2012fp}. There is a correlation not only in the intensity of the radiation, but also in the its polarization, that can be represented as a bi-dimensional vector living on the 2-sphere.}
\end{center}
\end{figure}

Let us give a critical look at what we learnt about inflation so far form the observational point of view.

There have been three {\it qualitative} theoretical predictions of inflation that have been verified so far. One is the oscillations in the CMB, another is the curvature of the universe, of order $\Omega_k\sim 10^{-3}$. At the time inflation was formulated, $\Omega_k$ could have been of order one. It is a natural prediction of inflation that lasts a little more than the necessary amount to have $\Omega_k\ll1$. The third is that the perturbations are Gaussian to a very good approximation: the signature of a weakly coupled field theory.

But what did we learn at a {\it quantitative} level about inflation so far? Just two numbers, not so much in my opinion unfortunately.  This is so because all the beautiful structures of the peaks in the CMB (and also in Large Scale Structures) is just controlled by well known Standard Model physics at 1 $eV$ of energy. The input from inflation are the qualitative initial conditions for each mode, and quantitatively the power spectrum and its tilt
\be
P_{\zeta}\simeq\frac{H^2}{\mpl^2\epsilon}\sim 10^{-10}\ , \quad n_s-1\simeq 4\epsilon_H-2\eta_H\simeq -4 \times 10^{-2}\ ,
\ee
just two numbers fit it all.

This is a pity, because clearly cosmological data have much more information inside them. Is it there something more to look for?

\subsection{CMB Polarization}

One very interesting observable is the CMB polarization. The CMB has been already observed to be partially polarized. Polarization of the CMB can be represented as the set of vectors  tangent to the sphere, the direction of each vector at each angular point representing the direction of the polarization coming from the point, and its length the fractional amount.

CMB polarization in induced by Thomson scattering in the presence of a quadruple perturbation. Information on cosmological perturbations is carried over by the correlation of polarization (very much the same as the correlation of temperature). It is useful to define two scalar fields that live on the sphere.

Polarization can be decomposed into the sum of the fields, $E$ and $B$, that have very different angular patter.

\begin{figure}[h!]
\begin{center}
\includegraphics[width=10cm]{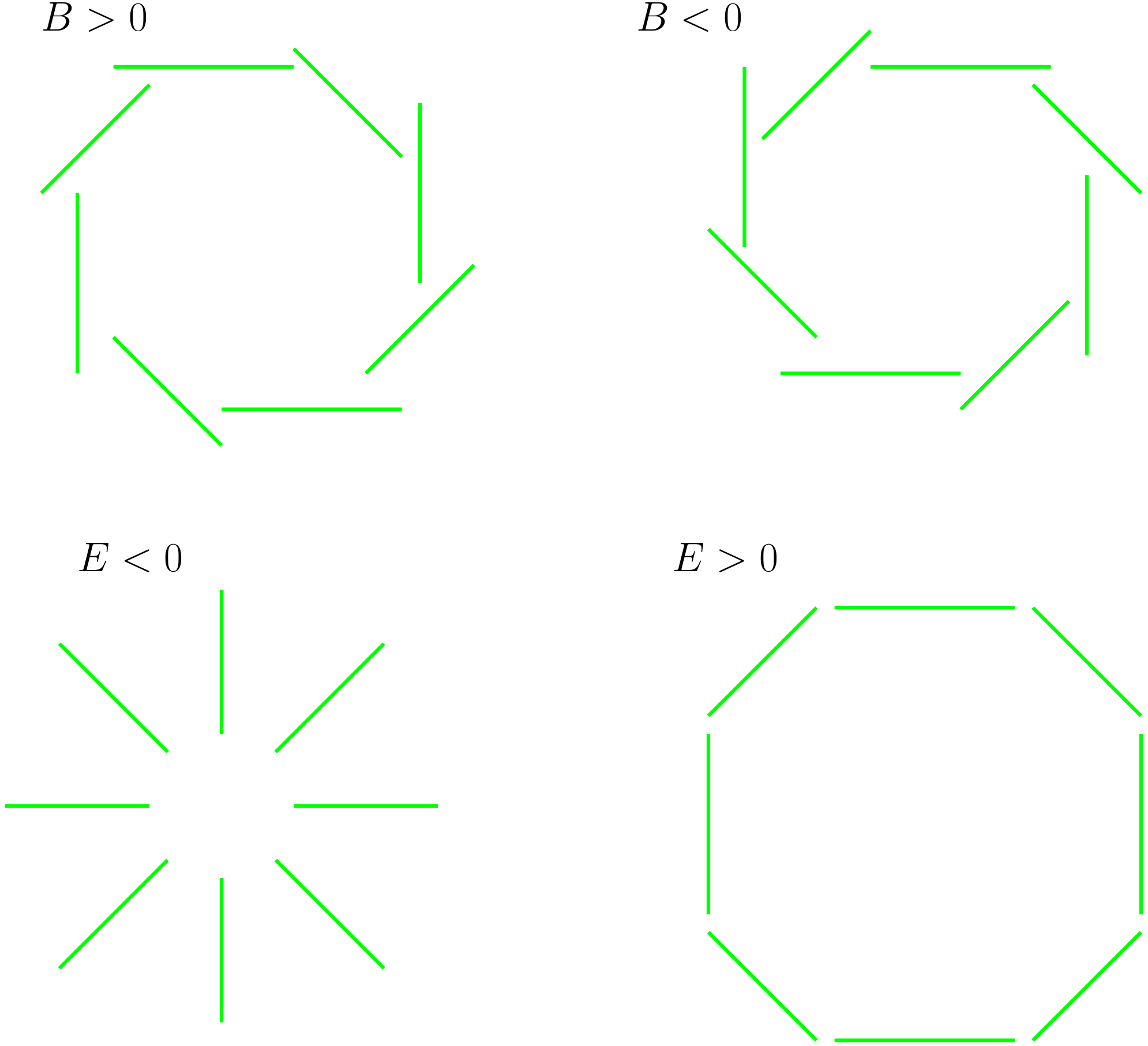}
\caption{\label{fig:EBpattern} \small We normally decompose the vector field on the sphere that represent the polarisation  in terms of $E$ and $B$ vector fields that have the above typical behaviour.}
\end{center}
\end{figure}

Scalar perturbations induce $E$ polarization, and they are being measured with greater and greater accuracy.  However tensor perturbations induce both $E$ and $B$ polarization. See~\cite{Dodelson:2003ft} for more details. This means that a discovery of $B$ modes would be a detection of tensor modes produced during inflation (there are some $B$ modes produced by lensing, but they are only on small angular scales).
So far there is no evidence of them, but even if we saw them, what we would learn about inflation?

We will learn a great qualitative point. Producing scale invariant tensor perturbations is very hard, because tensor perturbation tend to depend only on the nature of the space-time. Measuring scale invariant tensor modes with acoustic oscillations would mean most probably that an early de Sitter phase happened and so that inflation did happen.

At a quantitative level, however, we would just learn two numbers: the amplitude and the tilt of the power spectrum. In the simplest models of inflation, the amplitude of the power spectrum gives us direct information about $H$, and if the signal is detectable, it would teach us about the energy scale of inflation. Its scale invariance would teach us that $H$ is constant with time: this is the definition of inflation.

However recently new mechanism for produce large and detectable tensor modes have been found, which disentangle the measurement of $B$ modes from a measurement of $H$, at least in principle~\cite{Senatore:2011sp}. While the overall size is different, the signal is still scale invariant.

So, the question really remains: is it there something more to look for?

\subsection{Many more models of inflation}

Indeed, there are many more models of inflation than standard slow roll that we discussed. 

{\bf DBI Inflation:} One remarkable example is DBI inflation~\cite{Alishahiha:2004eh}. This described the motion of a brane in ADS space. Since the brane has a speed limit, an inflationary solution happens when the brane is moving at the speed of light. At that point special relativistic effects slow down the brane, and you have inflation, even though the brane is moving at the speed of light. The brane fluctuations in this case play the role of the inflaton.

\begin{figure}[h!]
\begin{center}
\includegraphics[width=7cm]{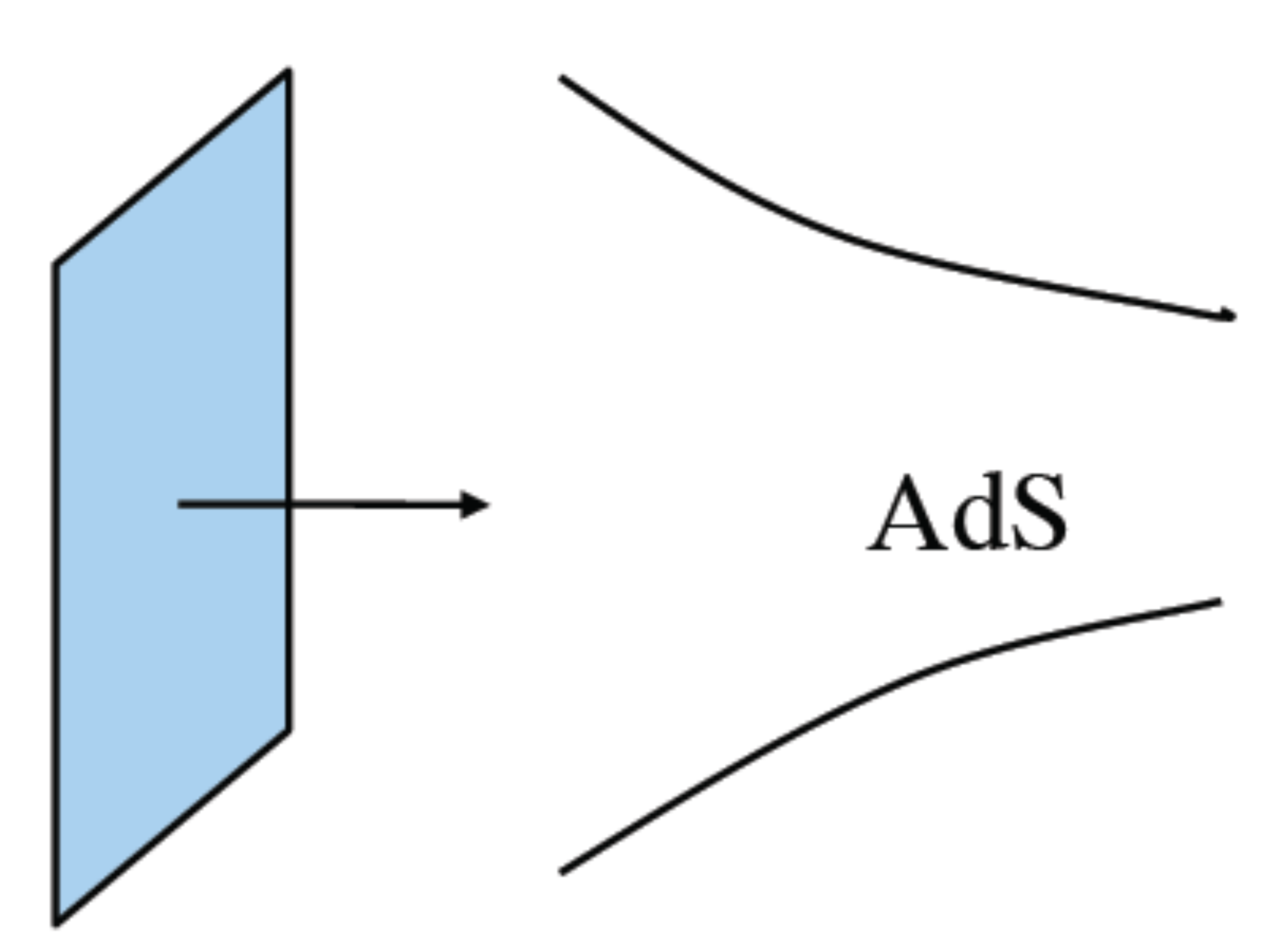}
\caption{\label{fig:DBI} \small Inflation can be realised by a brane moving relativistically in AdS space.}
\end{center}
\end{figure}

This model, though it happens in a totally different regime than slow roll inflation, it is totally fine with the observations we looked at so far. It turns out that the power spectrum scales in a different way that in slow roll models. We have a speed of sound $c_s\ll 1$
\be
\omega^2\sim c_s^2 k^2 \, .
\ee
This affects the power spectrum in the following way
\be
P_\zeta\sim \frac{H^2}{\epsilon \mpl^2 c_s}\sim 10^{-10}\ , \qquad n_s-1\simeq 4\epsilon_H-2\eta_H+\frac{\dot c_s}{H c_s}\sim 10^{-2}
\ee
Given than to match the CMB we need just these two inputs from the inflationary model, it is pretty expectable that they can be fixed. And indeed this happens.  

This inflationary model had the remarkable features that non-gaussianities were detectably large. The skewness of the distribution of the fluctuations was
\be
\frac{\langle\zeta^3\rangle}{\langle\zeta^2\rangle^{3/2}}\sim \frac{1}{c_s^2}\langle\zeta^2\rangle^{1/2}\gg10^{-5}
\ee
where we used that $\langle\zeta^2\rangle^{1/2}\sim 10^{-5}$
For comparison, the same number is standard inflation is of order $\epsilon\langle\zeta^2\rangle^{1/2}\ll10^{-5}$. While for standard slow roll inflation this is undetectably small, it is detectable for DBI inflation. 

This opens up a a totally new possible observational signature, and the possibility to distinguish and to learn about models that would be indistinguishable at the level of the two point function.

Non-Gaussianity!!

{\bf Ghost inflation:} Ghost inflation is another peculiar looking model~\cite{ArkaniHamed:2003uz}. It consists of a scalar field with the wrong sign kinetic term (a ghost).

\begin{figure}[h!]
\begin{center}
\includegraphics[width=10cm]{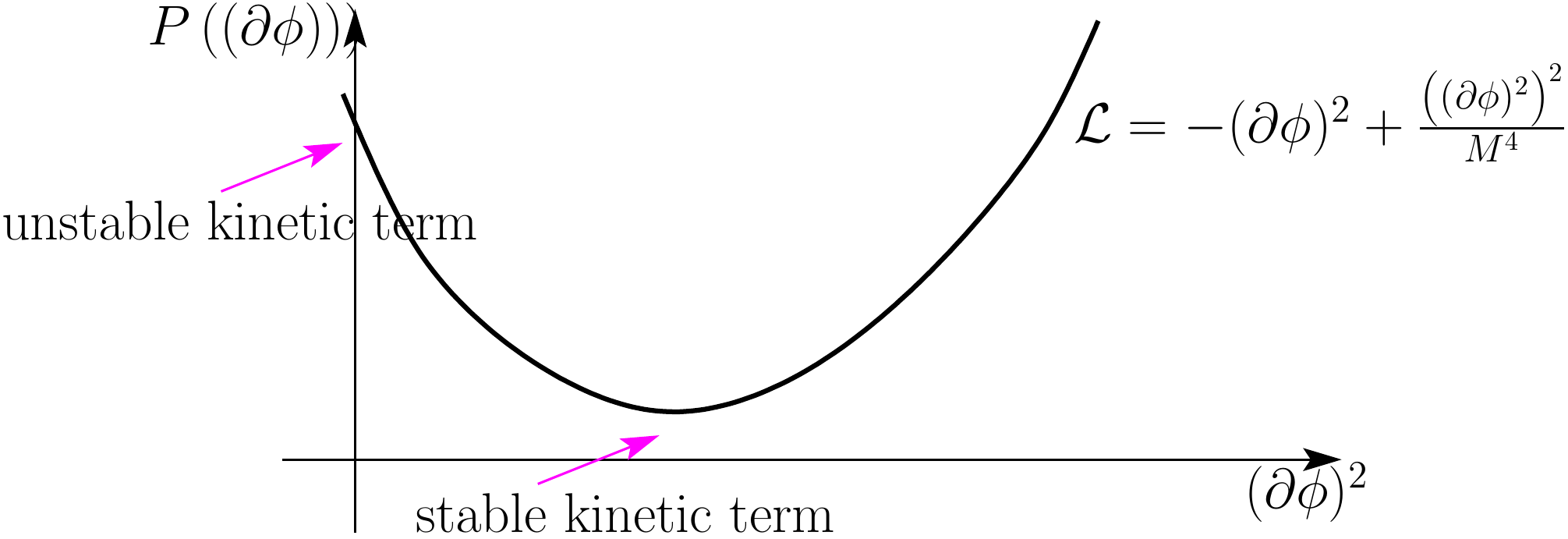}
\caption{\label{fig:scales1} \small The Ghost Inflation  model.}
\end{center}
\end{figure}

This triggers an instability that condensate in a different vacuum, where $\dot\phi=$const even in the absence of potential. This leads to inflation. The fluctuations have a dispersion relation of the form
\be
\omega^2\simeq \frac{k^4}{M^2}
\ee
which is extremely non-relativistic.

Again, this model is totally fine in fitting observations of the power spectrum, but it produces a large and detectable non-Gaussianity.

These are new models, some inspired by string theory. But they have new signatures. So, the question is: how generic are these signatures? What are the generic signatures of inflation? 

In order to do that, we need an approach that is very general, and looks at inflation in its most essential way: we go to the Effective Field Theory approach.

\subsection{The Effective Field Theory of Inflation}

Effective Field Theories (EFTs) have played the role of the guiding principle for particle physics and even condensed matter physics. EFTs have the capacity of synthesizing the relevant physics at the energy scale of interests. Effects of higher energy, largely irrelevant, physics are encoded in the coefficients of the higher dimension operators. It is {\it the way} to explore the phenomenology at a given energy scale. What we are going to do next is to develop the effective field theory of inflation. In doing so, we can look at inflation as the theory of a Goldstone boson: the Goldstone boson of time translations.

{\bf Review of Goldstone bosons:} Goldstone bosons are ubiquitous in particle physics (they got Nambu the well deserved 2008 nobel prize!). Let us consider the simplest theory of a $U(1)$ global symmetry $\phi\to e^{i\alpha}\phi$ that is spontaneously broken because of a mexican hat like potential $\phi\to \langle\phi\rangle$. Then there is Goldstone boson $\pi$ that non-linearly realizes the symmetry $\pi\to\pi+\alpha$.

\begin{figure}[h!]
\begin{center}
\includegraphics[width=7cm]{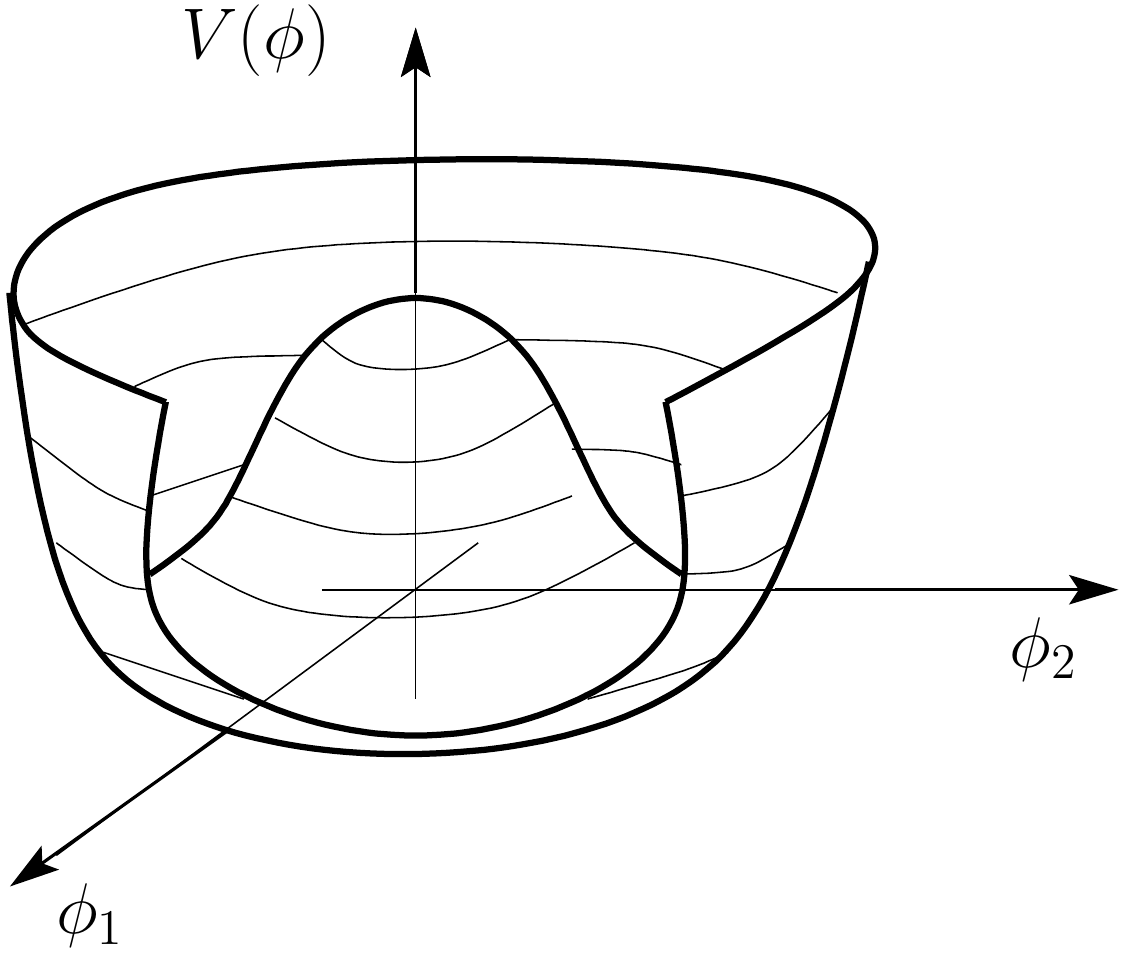}
\caption{\label{fig:mexican_hat} \small Mexican-hat potential for a complex scalar field that leads to spontaneous breaking of a $U(1)$ symmetry.}
\end{center}
\end{figure}

\be
{\cal L}=\d_\mu\phi^\star\d^\mu\phi-m^2 \phi^\star\phi+\lambda\phi^\star{}^2\phi^2\quad\rightarrow \quad\phi=\frac{m}{\lambda^{1/2}} e^{i\;\pi(\vec x,t)}\\
\ee
The action for the field $\pi$ is therefore the one of a massless scalar field endowed with a shift symmetry
\be
{\cal L}_\pi=(\d\pi)^2+\frac{1}{(m/\lambda^{1/2})^4}(\d\pi)^4+\ldots
\ee
the higher derivative operators being suppressed by powers of the high energy scale $m/\lambda^{1/2}$. 

A famous example of Goldstone bosons are the pions of the Chiral Lagrangian, that represent the Goldstone boson that non-linearly realise the $SU(2)$ chiral flavor symmetry, and they represent in the UV theory of QCD bound states of quark and antiquark. Notice that pions represent emergent scalar fields: there is no fundamental scalar field in QCD.

{\bf Inflation as the theory of a Goldstone boson:} How do we build the EFT of Inflation. In order to do that, we need to think of inflation in its most essential way. What we really know about inflation is that it is a period of accelerated expansion, where the universe was quasi de sitter. However, it could not be exactly de Sitter, because it has to end. This means that time-translation is spontaneously broken, and we will therefore consider that there is a physical clock measuring time and forcing inflation to end.

No matter what this clock is, we can use coordinate invariance of GR to go to the frame where these physical clock is set to zero. This can be done by choosing spatial slices where the fluctuations of the clock are zero, by performing a proper time diffs from any coordinate frame. As an example, if the inflaton was a fundamental scalar field (we are not assuming that, but just to make example) and we are in a coordinate frame where $\delta\phi(\vec x,t)\neq 0$, we can perform a time diff. $t\to \tilde t=t+\delta t(\vec x,t) $, such that (at linear order, it can be generalized to arbitrary non-linear order)
\be
0=\tilde{\delta\phi}(\vec x,t)=\delta\phi(\vec x,t)-\dot\phi_0(t) \delta t(\vec x,t)
\ee 
Now, suppose we are in this frame. We follow the rules of EFT. They say we have to write the action with the degrees of freedom that are available to us. This is just the metric fluctuations. We have to expand in fluctuations, and write down all operators compatible with the symmetries of the problem. In our case we can arbitrarily change spatial coordinates within the various spacial slices, on each spatial slice in a different way. This means that the residual gauge symmetry is time-dependent spatial diff.s:
\be\label{eq:spatialdiffs}
x^i\to\tilde x^i=x^i+\xi^i(t,\vec x)\ .
\ee

\begin{figure}[h!]
\begin{center}
\includegraphics[width=8cm]{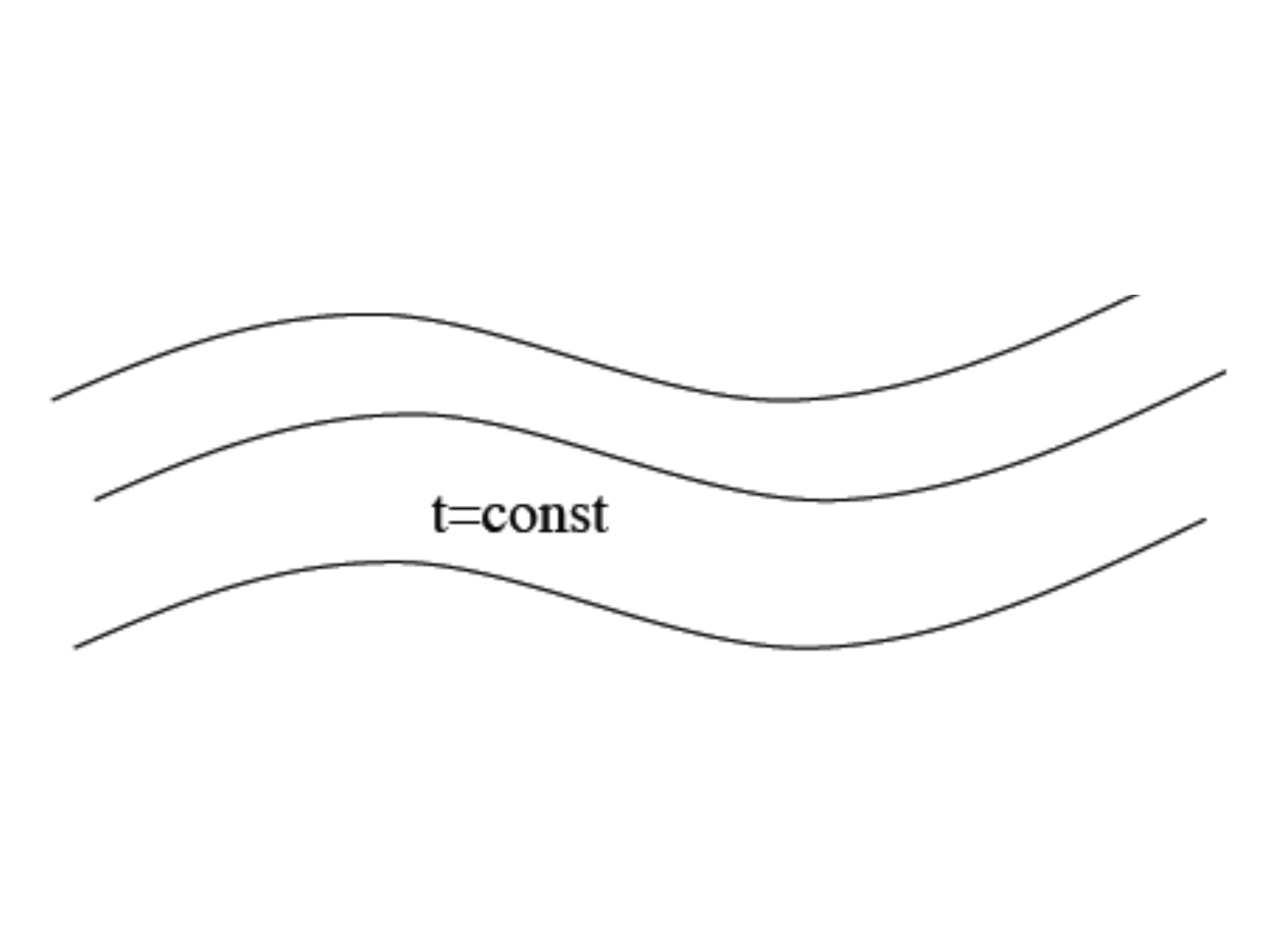}
\caption{\label{fig:time-slicing} \small If there is a clock-field driving inflation, then there is a privileged time-slicing where this clock is taken as uniform.}
\end{center}
\end{figure}

Further, still following the EFT procedure, we expand in perturbations and go to the order up to which we are interested (for example, quadratic order for 2-point functions, cubic order for 3-point functions, quartic order for 4-point functions, and so on), and then expand, at each order in the fluctuations, in derivatives, higher derivative terms being suppressed by the ratio of the energy scale $E$ of the problem versus some high energy scale $\Lambda$.

\subsubsection{Construction of the action in unitary gauge}

What is the most general Lagrangian in this unitary gauge? Here we will follow~\cite{Cheung:2007st} closely. One must write down
operators that are functions of the metric $g_{\mu\nu}$, and that are invariant
under the (linearly realized) time dependent spatial diffeomorphisms $x^i\rightarrow
x^i+\xi^i(t,\vec{x})$. Time-dependent Spatial diffeomorphisms are in fact
unbroken. In words, this amount to saying that Inflation is the theory of spacetime diffs. spontaneously broken to time-dependent spacial diffs.~\footnote{Keep in mind that diff's are a gauge redundancy (sometimes called gauge symmetry, but this is a bit erroneous). This makes the theory of Inflation analogous to the Standard Model of particle physics at low energies, where  $SU(2)\times U(1)$ gauge redundancy is spontaneously broken to $U(1)$ gauge redundancy.}.  Besides the usual terms with the Riemann tensor, which are
invariant under all diffs, many extra terms are now allowed, because
of the reduced symmetry of the system. They describe the additional
degree of freedom eaten by the graviton. For example it is easy to
realize that $g^{00}$ is a scalar under spatial diffs, so that it can
appear freely in the unitary gauge Lagrangian. 
\be
\tilde g^{00}=\frac{\d\tilde t}{\d x^\mu}\frac{\d \tilde t}{\d x^\nu} g^{\mu\nu}=\delta^0_\mu\delta^0_\nu g^{\mu\nu}=g^{00}
\ee
Polynomials of $g^{00}$
are the only terms without derivatives.  Given that there is a preferred slicing of the spacetime, one is also allowed to write geometric objects describing this slicing. For instance the extrinsic curvature $K_{\mu\nu}$ of surfaces at constant time is a tensor under spatial diffs and it can be used in the action. If $n^\mu$ is the vector orthogonal to the equal time slices, we have
\be
K_{\mu\nu}=h_\nu{}^{\sigma}\nabla_\sigma n_\nu\ ,
\ee
with $\nabla$ being the covariant derivative, and $h_{\mu\nu}$ the induced metric on the spatial slices
\be
h_{\mu\nu}=g_{\mu\nu}+n_\mu n_\nu\ .
\ee
  Notice that generic functions of time can multiply any term in the action.
The most generic Lagrangian can be written as (see App.~A and B of ~\cite{Cheung:2007st} for a proof)
\begin{eqnarray}
\label{eq:action}\nonumber
S & = & \int  \! d^4 x \; \sqrt{- g} \Big[ \frac12 M_{\rm Pl}^2 R - c(t)
g^{00} -  \Lambda(t) + \frac{1}{2!}M_2(t)^4(\delta g^{00})^2+\frac{1}{3!}M_3(t)^4 (\delta g^{00})^3+ \\
&& - \frac{\bar M_1(t)^3}{2} (\delta g^{00})\delta K^\mu {}_\mu
-\frac{\bar M_2(t)^2}{2} \delta K^\mu {}_\mu {}^2
-\frac{\bar M_3(t)^2}{2} \delta K^\mu {}_\nu \delta K^\nu {}_\mu + ...
\Big] \; ,
\end{eqnarray}
where the dots stand for terms which are of higher order in the fluctuations or with more derivatives. $\delta g^{00}= g^{00}+1$.
We denote by $\delta K_{\mu\nu}$ the variation of the extrinsic
curvature of constant time surfaces with respect to the unperturbed
FRW: $\delta K_{\mu\nu} = K_{\mu\nu} - a^2 H h_{\mu\nu}$ with
$h_{\mu\nu}$ is the induced spatial metric. Notice that only the first
three terms in the action above contain linear perturbations around
the chosen FRW solution, all the others are explicitly quadratic or
higher. Therefore the coefficients $c(t)$ and $\Lambda(t)$ will be fixed
by the requirement of having a given FRW evolution $H(t)$, {\em i.e.~}requiring that tadpole terms cancel around this solution. Before fixing these coefficients, it is important to realize that this simplification is not trivial. One would expect that there are an infinite number of operators which give a contribution at first order around the background solution. However one can write the action as a polynomial of linear terms like $\delta K_{\mu\nu}$ and $g^{00}+1$, so that it is evident whether an operator starts at linear, quadratic or higher order. All the linear terms besides the ones in eq.~(\ref{eq:action}) will contain derivatives and they can be integrated by parts to give a combination of the three linear terms we considered plus covariant terms of higher order.  We conclude that {\em the unperturbed history fixes $c(t)$ and $\Lambda(t)$, while the difference among different models will be encoded into higher order terms.}

We can now fix the linear terms imposing that a given FRW evolution is
a solution. As we discussed, the terms proportional to $c$ and $\Lambda$ are the only
ones that give a stress energy tensor
\begin{equation}
T_{\mu \nu} = -\frac{2}{\sqrt{-g}}\frac{\delta S_{\rm
matter}}{\delta g^{\mu\nu}}
\end{equation}
which does not vanish at
zeroth order in the perturbations and therefore contributes to the
right hand side of the Einstein equations. During inflation we are mostly interested in a flat FRW Universe 
\be
ds^2 = -dt^2 + a^2(t) d \vec{x}^2 
\ee
so that Friedmann equations are given by
\begin{eqnarray}
H^2 & = & \frac{1}{3 M_{\rm Pl}^2} \big[ c(t)+\Lambda(t)\big]  \\
\frac{\ddot a}{a} = \dot H + H^2 & =  & -\frac{1}{3 M_{\rm Pl}^2} \big[ 2
c(t)-\Lambda(t) \big] \;.
\end{eqnarray}
Solving for $c$ and $\Lambda$ we can rewrite the action (\ref{eq:action}) as
\begin{eqnarray}
\label{eq:actiontad}\nonumber
S & \!\!\!\!\!\!\!\!\!\!\!\!= \!\!\!\!\!\!\!\!\!& \!\!\!\int  \! d^4 x \; \sqrt{- g} \Big[ \frac12 M_{\rm Pl}^2 R + M_{\rm Pl}^2 \dot H
g^{00} - M_{\rm Pl}^2 (3 H^2 + \dot H) + \frac{1}{2!}M_2(t)^4(\delta g^{00})^2+\frac{1}{3!}M_3(t)^4 (\delta g^{00})^3+ \\
&& - \frac{\bar M_1(t)^3}{2} (\delta g^{00})\delta K^\mu {}_\mu
-\frac{\bar M_2(t)^2}{2} \delta K^\mu {}_\mu {}^2
-\frac{\bar M_3(t)^2}{2} \delta K^\mu {}_\nu \delta K^\nu {}_\mu + ...
\Big] \; .
\end{eqnarray}
As we said all the coefficients of the operators in the action above
may have a generic time dependence. However we are interested in
solutions where $H$ and $\dot H$ do not vary significantly in one
Hubble time.  Therefore it is natural to assume that the same holds
for all the other operators. With this assumption the Lagrangian is
approximately time translation invariant \footnote{The limit in which
  the time shift is an exact symmetry must be taken with care because
  $\dot H \to 0$. This implies that the spatial kinetic term for the
  Goldstone vanishes, as we will see in the discussion of Ghost
  Inflation.}. Therefore the time dependence generated by loop effects
will be suppressed by a small breaking parameter \footnote{Notice that
  this symmetry has nothing to do with the breaking of time
  diffeomorphisms. To see how this symmetry appears in the $\phi$
  language notice that, after a proper field redefinition, one can
  always assume that $\dot\phi =$ const. With this choice, invariance
  under time translation in the unitary gauge Lagrangian is implied by
  the shift symmetry $\phi \to \phi $ + const. This symmetry and the
  time translation symmetry of the $\phi$ Lagrangian are broken down
  to the diagonal subgroup by the background. This residual symmetry
  is the time shift in the unitary gauge Lagrangian.}. This assumption
is particularly convenient since the rapid time dependence of the coefficients can win against the friction created by the exponential expansion, so that inflation may cease to be a dynamical attractor, which is necessary to solve the homogeneity problem of standard FRW cosmology.   

It is important to stress that this approach does describe the
most generic Lagrangian not only for the scalar mode, but also for 
gravity. High energy effects will be encoded for example in operators
containing the perturbations in the Riemann tensor $\delta
R_{\mu\nu\rho\sigma}$. As these corrections are of higher order in
derivatives, we will not explicitly talk about them below.

Let us give some examples of how to write simple models of inflation in this language. A model with minimal kinetic term and a slow-roll potential $V(\phi)$ can be written in unitary gauge as 
\be
\int \! d^4x \: \sqrt{-g}
\left[ -\frac 1 2 (\partial \phi)^2 - V(\phi) \right]  \to  \int \!
d^4x \: \sqrt{- g} \left[ -\frac{\dot \phi_0(t)^2}{2} g^{00} - V(\phi_0(t))
\right] \; .
\ee
As the Friedmann equations give 
$\dot\phi_0(t)^2=-2M^2_P \dot{H}$ and $V(\phi(t))=M_{\rm Pl}^2 (3H^2+\dot H$) we see that the action is of the form (\ref{eq:actiontad}) with all but the first three terms set to zero. Clearly this cannot be true exactly as all the other terms will be generated by loop corrections: they encode all the possible effects of high energy physics on this simple slow-roll model of inflation.

A more general case includes all the possible Lagrangians with at most one derivative acting on each $\phi$: $L= P(X,\phi)$, with $X=g^{\mu\nu}\partial_\mu\phi\partial_\nu\phi$. Around an unperturbed solution $\phi_0(t)$ we have
\be
S = \int \! d^4x \: \sqrt{-g} \;  P(\dot\phi_0(t)^2 g^{00}, \phi(t))
\ee
 which is clearly of the form above with $M_n^4(t) =\dot\phi_0(t)^{2n} \partial^n P/\partial X^n$ evaluated at $\phi_0(t)$.
 Terms containing the extrinsic curvature contain more than one
 derivative acting on a single scalar and will be crucial in the limit
 of exact de Sitter, $\dot H \to 0$. They reproduce ghost inflation and new models that are discovered in this set up.

\subsubsection{Action for the Goldstone Boson\label{sec:Goldstone}}

The unitary gauge Lagrangian is very general, but it is clearly not very intuitive. For example, in a particular limit, it  contains standard slow roll inflation. But where is the scalar degree of freedom? This is so complicated because it is the unitary gauge Lagrangian of a spontaneously broken gauge symmetry.

{\bf Goldstone boson equivalence theorem:} The unitary gauge Lagrangian describes three degrees of freedom: the
two graviton helicities and a scalar mode. This mode will become
explicit after one performs a broken time diffeomorphism
(St\"u{}ckelberg trick) as the Goldstone boson which non-linearly realizes this symmetry. In analogy with the equivalence theorem for the longitudinal components of a massive gauge boson \cite{Cornwall:1974km}, we expect that the physics of the Goldstone decouples from the two graviton helicities at short distance, when the mixing can be neglected.
Let us review briefly what happens in a non-Abelian gauge theory before applying the same method in our case.

The unitary gauge action for a non-Abelian gauge group $A_\mu^a$ is
\be
S = \int \! d^4x  \,-\frac{1}{4} {\rm Tr}\, F_{\mu\nu}F^{\mu\nu}-\frac12 m^2 {\rm Tr}\, A_\mu A^\mu \ ,
\ee
where $A_\mu = A_\mu^a T^a$.
Under a gauge transformation we have
\be
\label{eq:AmuU}
A_\mu \to U A_\mu U^\dagger + \frac{i}{g} U \partial_\mu U^\dagger \equiv \frac{i}{g} U D_\mu U^\dagger \;.
\ee
The action therefore becomes
\be
S = \int \! d^4x  \,-\frac{1}{4} {\rm Tr}\, F_{\mu\nu}F^{\mu\nu} - \frac12 \frac{m^2}{g^2} {\rm Tr} D_\mu U^\dagger D^\mu U \;.
\ee
The mass term was not gauge invariant, and so we have factors of $U$ in that term.
The gauge invariance can be ``restored" writing $U=\exp{[i T^a \pi^a(t,\vec x)]}$, where $\pi^a$ are scalars (the Goldstones) which transform non-linearly under a gauge transformation $\Lambda$ as
\be
e^{i T^a \widetilde\pi^a(t, \vec x)} = \Lambda(t, \vec x) \,e^{i T^a \pi^a(t,\vec x)}
\ee
Notice that if for a moment we consider the case in which the gauge theory is a $U(1)$ theory, we would have
\be
\Lambda=e^{i \alpha(\vec x,t)}\ , \qquad \Rightarrow\qquad \pi\to\tilde\pi=\pi+\alpha
\ee
$\pi$ shifts under a gauge transformation. This is a non-linear transformation because 0 is not mapped into 0. Gauge invariance has been restored by reintroducing a dynamical field that however, transforms non-linearly. Gauge invariance is non-linearly realized.

Going to canonical normalization 
\be
\frac{m^2}{g^2}(\d\pi)^2 \quad\Rightarrow\quad \pi_c \equiv m/g \cdot \pi
\ee
we see that the Goldstone boson self-interactions become strongly coupled at the scale $4 \pi m/g$, which is parametrically higher than the mass of the gauge bosons.
The advantage of reintroducing the Goldstones is that for energies $E \gg m$ the mixing between them and the transverse components of the gauge field becomes irrelevant, so that the two sectors decouple.
Mixing terms in eq.~(\ref{eq:AmuU}) are in fact of the form
\be
\frac{m^2}{g} A_{\mu}^a \partial^\mu \pi^a = m A_{\mu}^a \partial^\mu \pi_c^a
\ee
which are irrelevant with respect to the canonical kinetic term $(\partial \pi_c)^2$ for $E \gg m$.  

Notice that from expanding the term $D_\mu U D^\mu U$ we obtain irrelevant (i.e. non-renormalizable) terms of the form
\be
 \frac{m^2}{g^2} \pi^2(\d \pi)^2 \sim \frac{1}{m^2/g^2} \pi_c^2(\d \pi_c)^2
\ee
This is an operator that becomes strongly coupled and leads to unitarity violation at energies $E\sim 4\pi m/g$.

In the window $m \ll E \ll 4 \pi m /g$ the physics of the Goldstone $\pi$ is weakly coupled and it can be studied neglecting the mixing with transverse components.

Let us follow the same steps for our case of broken time diffeomorphisms. Let us concentrate for instance on the two operators:
\begin{equation}
\int d^4x\;  \sqrt{-g} \left[A(t)+B(t)g^{00}(x)\right] \ .
\end{equation}
Under a broken time diff.  $t \to \widetilde t= t + \xi^0(x)$, $\vec{x} \to \vec{\widetilde{x}}=\vec{x}$,  $g^{00}$ transforms as:
\begin{equation}
g^{00}(x)\to \widetilde g^{00}(\widetilde x(x))=\frac{\partial \widetilde x^0(x)}{\partial x^\mu}\frac{\partial \widetilde x^0(x)}{\partial x^\nu} g^{\mu\nu}(x) \,  .
\end{equation}
The action written in terms of the transformed fields is given by:
\begin{eqnarray}
\int d^4x\;  \sqrt{-\widetilde g(\widetilde x(x))} \left|\frac{\partial \widetilde x}{\partial x} \right| \left[A(t)+B(t) \frac{\partial x^0}{\partial \widetilde x^\mu}\frac{\partial x^0}{\partial \widetilde x^\nu} \widetilde g^{\mu\nu}(\widetilde x(x))\right]\ .
\end{eqnarray}
Changing integration variables to $\widetilde x$, we get:
\begin{eqnarray}
\int d^4\widetilde x\;  \sqrt{-\widetilde g(\widetilde x)} \left[A(\widetilde t-\xi^0(x(\widetilde x)))+B(\widetilde t-\xi^0(x(\widetilde x))) \frac{\partial (\widetilde t-\xi^0(x(\widetilde x)))}{\partial \widetilde x^\mu}\frac{\partial (\widetilde t-\xi^0(x(\widetilde x)))}{\partial \widetilde x^\nu} \widetilde g^{\mu\nu}(\widetilde x)\right].
\end{eqnarray}
The procedure to reintroduce the Goldstone is now similar to the gauge theory case. Whenever $\xi^0$ appears in the action above, we make the substitution
\begin{equation} 
\xi^0(x(\widetilde x)) \to - \widetilde \pi(\widetilde x ) \, .
\end{equation}
This gives, dropping the tildes for simplicity:
\begin{eqnarray}
\int d^4x\;  \sqrt{- g(x)} \left[A( t+\pi(x))+B(t+\pi(x)) \frac{\partial (t+\pi(x))}{\partial x^\mu}\frac{\partial (t+\pi(x))}{\partial x^\nu} g^{\mu\nu}(x)\right].
\end{eqnarray}
One can check that the action above is invariant under diffs at all orders (and not only for infinitesimal transformations) upon assigning to $\pi$ the transformation rule
\begin{equation}
\pi(x) \to \widetilde\pi(\widetilde x(x))=\pi(x)-\xi^0(x) \ .
\end{equation}
With this definition $\pi$ transforms as a scalar field plus an additional shift under time diffs. Notice that diff. invariant terms did not get a $\pi$. 

Applying this procedure to the unitary gauge action (\ref{eq:actiontad}) we obtain
\begin{eqnarray}\label{Smixed}
S = \int \! d^4 x  \: \sqrt{- g} &&\left[\frac{1}{2}M_{\rm Pl}^2 R
- M^2_{\rm Pl} \left(3H^2(t+\pi) +\dot{H}(t+\pi)\right)+ \right.\\ \nn
&&+M^2_{\rm
Pl} \dot{H}(t+\pi)\left(
(\d_\mu(t+\pi)\d_\nu(t+\pi)g^{\mu\nu}\right) + \\ \nonumber
&&\frac{M_2(t+\pi)^4}{2!}\left(\d_\mu(t+\pi)\d_\nu(t+\pi)g^{\mu\nu}+1\right)^2 + \nonumber\\
\nonumber && \left. \frac{M_3(t+\pi)^4}{3!}\left(\d_\mu(t+\pi)\d_\nu(t+\pi)g^{\mu\nu}+1\right)^3+ ... \right] \; ,
\end{eqnarray}
where for the moment we have neglected for simplicity terms that involve the extrinsic curvature.

This action is rather complicated, and at this point it is not clear what is the advantage of reintroducing the Goldstone $\pi$ from the unitary
gauge Lagrangian. In  analogy with the gauge theory case, the simplification occurs because, at sufficiently short distances, the physics of the Goldstone can be studied neglecting metric fluctuations (this is nothing but the equivalence principle). As for the gauge theory case, the regime for which this is possible can be estimated just looking at the mixing terms in the Lagrangian above. In eq.(\ref{Smixed}) we see in fact that quadratic terms which mix $\pi$ and $g_{\mu\nu}$ contain fewer derivatives than the kinetic term of $\pi$ so that they can be neglected above some high energy scale. In general the answer will depend on which operators are present.
Let us here just do the simplest case in which only the tadpole terms are relevant ($M_2=M_3=\ldots=0$). This includes the standard slow-roll inflation case. The leading mixing
with gravity will come from a term of the form
\begin{equation}
\sim M_{\rm Pl}^2 \dot H \d_i\pi \delta g^{0i} \ .
\end{equation}

We see that
\bea \nonumber
&&{\rm Kinetic\ term}\sim \mpl^2\dot H \delta g^{00}\quad\to\quad\mpl^2\dot H \left(\d_\mu(t+\pi)\d_\nu(t+\pi)g^{\mu\nu}\right) \quad\supset\quad \mpl^2\dot H\dot\pi^2\\ \nonumber
&&{\rm Mixing\ term}\sim \mpl^2\dot H \delta g^{00}\quad\to\quad\mpl^2\dot H \left(\d_\mu(t+\pi)\d_\nu(t+\pi)g^{\mu\nu} \right)\quad\supset\quad \mpl^2\dot H\delta g^{0i}\d_i\pi\\
\eea
$\delta g^{0i}$ is a constrained variable, it is a sort of gravitational potential, and it is determined by $\pi$. At short distances, the Newtonian approximation holds:
\be
\mpl^2  \d_j^2 \delta g^{0i}\sim \dot H\mpl^2\d_i \pi \quad\Rightarrow\quad \delta g^{0i}\sim \dot H \frac{\d^i}{\d_j^2}\pi\ .
\ee
We have
\be
\frac{{\rm Mixing\ term}}{{\rm Kinetic\ term}}\sim \frac{\delta g^{0i} \d_i\pi}{\dot\pi}\sim \frac{\dot H\frac{\d^i}{\d_j^2}\pi \d_i\pi}{\dot \pi}\sim \frac{\dot H \pi}{\dot\pi}\sim\frac{\dot H}{E H}\ll 1\quad \Rightarrow\quad E\gg \epsilon H\ ,
\ee
where in the next to last step we have integrated the spatial derivative by parts, and in the last step we have used that at energies of order $E$, $\d_t\sim E$. The mixing term is negligible in the UV (GR equivalence principle). The actual scale $E_{\rm mix}$ at which the mixing can be neglected depends on the actual operators turned on, but it is guaranteed that at energies $E\gg E_{{\rm mix}}$ we can neglect the mixing terms. 

In the regime $E\gg E_{\rm mix}$ the action dramatically simplifies to
\begin{eqnarray}\label{Spi} 
\! \! S_{\rm \pi} = \int \!  d^4 x   \sqrt{- g} \left[\frac12 M_{\rm Pl}^2 R -M^2_{\rm Pl}
\dot{H} \left(\dot\pi^2-\frac{ (\partial_i \pi)^2}{a^2}\right)
+2 M^4_2
\left(\dot\pi^2+\dot{\pi}^3-\dot\pi\frac{(\partial_i\pi)^2}{a^2}
\right) -\frac{4}{3} M^4_3 \dot{\pi}^3+ ... \right].
\end{eqnarray}

Notice that the non-linear realization of time-diffs forces $\pi$ to appear in non-linear `blocks':
\bea
&&\left[\dot\pi^2-\frac{ (\partial_i \pi)^2}{a^2}\right]\ , \\ \nonumber
&&\left[\dot\pi^2+\dot{\pi}^3-\dot\pi\frac{(\partial_i\pi)^2}{a^2}+\frac{(\d_i\pi)^2(\d_j\pi)^2}{a^4}\right]\ , \\ \nonumber
&&\ldots\ .
\eea
This offers a precise relationship among different operators.

Given an inflationary model, one is interested in computing
predictions for present cosmological observations. From this point of
view, it seems that the decoupling limit (\ref{Spi}) is completely
irrelevant for these extremely infrared scales. However, as for
standard single field slow-roll inflation, one can prove that there
exists a quantity, the usual $\zeta$ variable, 
which is constant out of the horizon at any order in perturbation
theory

Therefore the problem is reduced to calculating correlation functions just after horizon crossing. We are therefore interested in studying our Lagrangian with an IR energy cutoff of order $H$.  If the decoupling scale $E_{\rm mix}$ is smaller than $H$, the Lagrangian for $\pi$ (\ref{Spi}) will give the correct predictions up to terms suppressed by $E_{\rm mix}/H$. When this is not the case, nothing dramatic happens: we simply have to keep also the metric fluctuations.

This is the justification of the calculations we did in lecture 2.

As we discussed, we are assuming that the time dependence of the coefficients in the unitary gauge Lagrangian is slow compared to the Hubble time, that is, suppressed by some generalized slow roll parameters. This implies that  the additional $\pi$ terms coming from the Taylor expansion of the coefficients are small. In particular, the relevant operators, {\it i.e.} the ones which dominate moving towards the infrared, like the cubic term, are unimportant at the scale $H$ and have therefore been neglected in the Lagrangian (\ref{Spi}). They can be nevertheless straightforwardly included, as done in~\cite{Behbahani:2011it,Achucarro:2012sm}.

In conclusion, with the Lagrangian (\ref{Spi}) one is able to compute
all the observables which are not dominated by the mixing with
gravity, like for example the non-Gaussianities in standard slow-roll
inflation \cite{Maldacena:2002vr,Seery:2006vu}. Notice however that the tilt of the spectrum can be calculated, at leading order, with the Lagrangian (\ref{Spi}).  As we saw earlier, its value can in fact be deduced simply by the power spectrum at horizon crossing computed neglecting the mixing terms.
It is important to stress that our approach does not lose its validity
when the mixing with gravity is important so that the Goldstone action
is not sufficient for predictions. The action (\ref{eq:actiontad}) contains all the
information about the model and can be used to calculate all
predictions even when the mixing with gravity is large.

Let us stress a few points
\begin{itemize}
\item The above Lagrangian is very simple, and it unifies all single-degree-of-freedon inflationary models.
\item It describes the theory of the fluctuations, which is what we are actually testing.
\item It is analogous to the Chiral Lagrangian of particle physics. Indeed, it is telling us that from the experimental point of view, inflation is the theory of a Goldstone boson
\item Since it encodes all possible single-clock models on inflation, it allows to prove theorems on the possible signals.
\item It also allows us to explore all possible signatures.
\item What is forced by symmetries, what are the allowed operators and what is possible to do is made clear. For example, the coefficient of $(\d_i\pi)^2$ is fixed to be $\dot H\mpl^2$. This is not the case for $\dot\pi^2$. This tells us that at leading order in derivatives it is impossible to violate the null energy condition. $\dot H>0$ implies that the spatial kinetic term for $\pi$ has the negative-energy sign, and so it leads to an uncontrollable instability. The EFT also tells you how this problem can be fixed, by adding higher derivative terms. Indeed all currently known ways to violate the null Energy Condition (NEC) that are currently known have been found in this context.
\item This formalism is very prone to do with it what we normally do for the beyond the standard model physics: one can add symmetries to enhance operators with respect to others, or one can try to UV complete some specific models.
\item Being explicitly a theory for the fluctuations, it allows to assess the important of operators very easily. For example, in the standard treatment with scalar fields, an operators $(\d\phi)^8$ contributes to the quadratic action with $\dot\phi_0^6(\d\delta\phi)^2$. This is also very useful for studying loop corrections. At a fixed order in fluctuations and derivatives, in the EFT there is a finite number of counter terms, while this is not so with the scalar field theory. Indeed the EFT formalism was crucial to prove the constancy of $\zeta$ at quantum level~\cite{Pimentel:2012tw,Senatore:2012ya}.
\end{itemize}

\subsection{Rigorous calculation of the power spectrum in $\pi$-gauge}

We are now ready to see the new spectacular signatures of inflation. But I really feel that it is time for us to do a rigorous calculation. Notice that we got so far without having to do one at all. Pretty good I would say. However, there is little more rewarding that seeing your simple estimates being confirmed by a somewhat tricky calculation.

 Let us write the metric in the so-called ADM parametrization
\be
ds^2=-N^2dt^2+h_{ij}\left(d x^i+N^i dt\right)\left(d x^j+N^j dt\right)
\ee
We have to quantize a system with Gauge redundancy. In our case the gauge  freedom (sometimes historically and wrongly called gauge symmetry) is 4-dim diff invariance, out of which time diffs are non-linearly realized. The quantization is tricky, but it is the same as for gauge theories. Just a different symmetry group. The procedure is the following (see Weinberg's QFT I and II books).
\begin{itemize}

\item Expand the action. In ADM parametrization, it reads
\bea\nonumber\label{eq:actiongauge}
&& S=\frac{1}{2}\int\sqrt{h}\left[N R^{(3)}+\frac{1}{N}\left(E_{ij}E^{ij}-E_i^i{}^2\right)\right.\\ \nonumber
&&\left.+2\mpl^2\dot H(t+\pi)\left[-\frac{1}{N}\left(1+\dot\pi\right)^2+\frac{2}{N}(1+\dot\pi)N^i(\d_i\pi)-N(h^{ij} \d_i\pi\d_j\pi)-\frac{1}{N}(N^i\d_i\pi)^2\right]\right.\\ \nonumber
&&\left.-\mpl^2\left(3 H^2(t+\pi)+\dot H(t+\pi)\right)\cdot N+\ldots \right]\\
\eea
where
\be
E_{ij}=\frac{1}{2}\left[\d_th_{ij}+\nabla_i N_j+\nabla_j N_i\right]
\ee
and $\nabla$ is the covariant derivative with respect to $h_{ij}$.

\item  For simplicity, we do the calculation for $M_{2,\ldots}=0$ (this includes slow roll inflation). Let us derive the equations of motion for $N$ and $N_i$.  For this action, the equations of motion for $N$ and $N_i$ take the following form
\bea\label{eq:constraints}
&& \nabla_i \left[N^{-1}\left(E^i_j-\delta^i_j E\right)\right]+\frac{2}{N}\dot H(t+\pi)\left[(1+\dot\pi)\d_i\pi-N^j\d_j\pi \d_i\pi\right]=0\\ \nonumber
&& \mpl^2\left[R^{(3)}-\frac{1}{N^2}\left(E_{ij}E^{ij}-E_i^i{}^2\right)\right]-\mpl^2\left[3H^2(t+\pi)+\dot H(t+\pi)\right]\\ \nonumber
&&+\mpl^2\dot H(t+\pi)\left[\frac{1}{N^2}(1+\dot\pi)^2-\frac{2}{N^2}(1+\dot\pi)N^i\d_i\pi+h^{ij}\d_i\pi\d_j\pi+\frac{1}{N^2}(N^i\d_i\pi)^2\right] =0
\eea
Indexes are lowered and raised with $h_{ij}$.

These two equations are extremely important. Notice that no time derivative acts on $N$ nor on $N_i$. These tells us that $N$ and $N_i$ are {\it constrained} variables: they are known once you specify what the other degrees of freedom do. They are {\it not} independent degrees of freedom. They are very much (and not by chance) like the gravitation potential in Newtonian gravity, or the Electric potential in electrostatic.

\item Let us count the degrees of freedom. We started with the metric, which has 10 components, and with the $\pi$ field. But we have 3-independent gauge generators for the spatial diffs and 1 for time diffs. This means that we can set 4 of these components to any value we want (including 0). This means that they are not degrees of freedom.  For example we can set to zero 4 components of $g_{ij}$. Then from above, we see that $N,N^i$ are 4 constrained variables. So they are also not degrees of freedom. We are left with
\be
{\rm number \ of\ degrees\ of \ freedom}=11-4-4=3
\ee
Does this work? We should have the two elicities of the graviton and the matter degree of freedom (equivalent to $\pi$): 3. Ok, we are on!

\item We now fix a gauge: let us fix the gauge to the so called $\pi$-gauge, where where the space time diffs are fixed by imposing the spatial metric $h_{ij}$ to take the following form
\be
h_{ij}=\delta_{ij} a^2\ .
\ee

\item The constrained variables $N$ and $N^i$ are constrained, and so we can solve for them in terms of the only remaining degree of freedom: $\pi$. The solution reads
\be
N=1-\frac{\dot H}{H}\pi\ , \qquad \d_i N_i= \frac{\dot H}{H^2} \d_t(H\pi)
\ee
Notice how the sourcing of the metric fluctuations are suppressed by at least a slow roll factor.

\item Plug back this values for $N$ and $N^i$ in the action. Notice, we can do this only because they are constrained variables.
The action now reads
\be
S=\int d^4x\, a^3 (-\dot H \mpl^2)\left[\dot\pi^2-\frac{1}{a^2}(\d_i\pi)^2-3 \dot H \pi^2\right]
\ee

\item We need to connect $\pi$, for which we have just derived the action, to $\zeta$, which is the quantity we wish to compute. The rigorous way to find the relationship between $\pi$ and $\zeta$ is to perform a time-diff to go from $\pi$-gauge, where the spatial metric is $h_{ij}=a^2\delta_{ij}$, to $\zeta$-gauge, where the spatial metric is $h_{ij}=a^2 e^{2\zeta}\delta_{ij}$. The time diff has parameter $\delta t =\pi$. For the two-point function, it is enough the relationship at linear level. As it is quite intuitive, this is
\be\label{eq:matching}
\zeta(\vec x,t)=-H(t)\,\pi(\vec x,t)
\ee

\item Let us quantize the system. Follow textbook: find
\be
\Pi_\pi=\frac{\delta{\cal L}}{\delta \dot\pi}=-2a^3\mpl^2 \dot H\dot\pi
\ee
and impose
\be
\left[\pi,P_\pi\right]=i
\ee
This is a quadratic Lagrangian, so we simply expand the fourier components of $\pi$ in annihilation and creation operators
\be
\hat \pi_{\vec k}(t)=\pi_{\vk}^{cl}(t) a^\dag_{\vk}+\pi_{{\vec k}}^{cl}{}^\star(t)a_{-\vk}
\ee
with $\pi^{cl}$ satisfying the equation of motion (Heisemberg equation for $\hat \pi$)
\be\label{eq:pi}
0=\frac{\delta L}{\delta \pi}=\frac{d\left(-a^3\dot H\dot\pi^{cl}_k\right)}{dt}+\dot Ha k^2\pi^{cl}_k
\ee
This is a second order equation, that requires two initial conditions. This condition can be found in the following way. We define the vacuum state as the state annihilated by~$a_{\vk}$:
\be
a_{\vk}|0\rangle=0
\ee
but what this state is actually depends on what we choose as $\pi^{cl}$. How do we choose it? Well, we know that at early times, the mode $k/a\gg 1$, so we would like the solution to be the same as in Minkowski space (this is GR!). In other words, the vacuum state for modes well inside $H^{-1}$ should be the same as in flat space. This give the following condition
\be\label{eq:pi_boundary}
\pi^{cl}_k(-k\eta\gg 1)\sim \frac{-i }{(2\epsilon)^{1/2}\mpl a(\eta)^3}\frac{1}{(2k/a(\eta))^{1/2}}e^{i k \eta}\quad {\rm for}\quad \frac{k}{a H}=-k\eta\gg 1 
\ee
Notice that the exponential reads $k\eta\simeq \frac{k}{a} a\eta\simeq k_{\rm phys} t$. The prefactor come from the canonical normalization. This is the solution that we would get for an harmonic oscillator $1/\sqrt{2\omega}$ after we take into account of the rescaling to make the field canonical.

\item At this point we simply need to solve eq.~(\ref{eq:pi}) with boundary condition (\ref{eq:pi_boundary}). Solving this exactly is unfortunately very hard  because of the time-dependent coefficients. However, these are very slow varying coefficients. We can use this fact if we realize that modes inside the Hubble scale oscillate very fast, and so one can use an adiabatic or WKB approximation for solving the equation in that regime. This approximation becomes not good when the mode becomes much longer than $H$, as its frequency drops to zero, and the time-dependence of the coefficients cannot be neglected anymore. However, at this point one can realize the following: one can simply convert the $\pi$ fluctuation   into a $\zeta$ fluctuation using (\ref{eq:matching}) evaluated at the time of freeze out. This is justified because we know that $\zeta$ is constant on super-Hubble scale. In doing so, we can therefore solve for the $\pi$ equation neglecting slow-roll corrections and then match the solution that we find using (\ref{eq:matching}) evaluate at $t=t_{f.o.}$. Notice that in particular, this means that we can neglect the mass term in the $\pi$ Lagrangian: $m^2\sim \epsilon H^2$. This was the only term in the action that was resulting from having taken care carefully of the metric fluctuations. And now we are seeing that the contribution is slow roll suppressed and can actually be neglected at leading order. This is the formal justification of why metric fluctuations can be neglected when working with $\pi$, when the leading effect does not come from the mixing with gravity.

\item Solving the differential equation for $\pi$, we get
\be
\pi^{cl}_k(\eta)=- \frac{1}{(2\epsilon)^{1/2}\mpl}\frac{1}{(2k)^{3/2}}(1-i k\eta)e^{i k \eta}\ , 
\ee
This solution is valid until times of order freeze out: $k\eta_{f.o.}\sim 1$. By switching to $\zeta$, which is constant in time after freeze out, we can get a solution that is valid at late times:
\be
\zeta^{cl}_k(\eta)= \frac{H_{f.o.}}{(2\epsilon_{f.o.})^{1/2}\mpl}\frac{1}{(2k)^{3/2}}(1-i k\eta)e^{i k \eta}\ , 
\ee

\item We can now compute the power spectrum:
\bea
&&\langle0|\zeta_{\vk}(\eta)\zeta_{\vk'}(\eta')|0\rangle=(2\pi)^3\delta^3(\vk+\vk')\\ \nonumber
&& \qquad\qquad\times\frac{H_{f.o.}}{(2\epsilon_{f.o.})^{1/2}\mpl}\frac{1}{(2k)^{3/2}}(1-i k\eta)e^{i k \eta}\frac{H_{f.o.}}{(2\epsilon_{f.o.})^{1/2}\mpl}\frac{1}{(2k)^{3/2}}(1+i k\eta')e^{-i k \eta'}
\eea
when $k\eta\ll1$ and $k\eta'\ll1$, we obtain
\be
\langle\zeta_{\vk}\zeta_{\vk'}\rangle_{\rm late}=(2\pi)^3\delta^3(\vk+\vk')\; \frac{1}{k^3}\cdot\frac{H_{f.o.}^4}{4(-\dot H_{f.o.})\mpl^2}
\ee

Which nicely reproduces the results we found with our estimates (but now we even got the factor of 4!).

\end{itemize}

\subsection{Rigorous calculation of the power spectrum in Unitary gauge}

We just saw that we could neglect metric perturbations for standard slow roll inflation. And indeed we did the correct calculation neglecting them. Additionally, we saw that using $\pi$ makes it explicit this fact. We even did the rigorous calculation to see that this approach works. In order to see that we did not loose anything, it is instructive to perform the calculation in a different, un-intuitive gauge: the so called $\zeta$-gauge or Maldacena-gauge. This is one of the gauges that are possible in our unitary gauge. Even though it is unintuitive, it is good for something. Indeed, it is the absolutely best gauge to study the tricky infrared properties of $\zeta$, the variable we ultimately need to compute. In this gauge, we will see that in the infrared $\zeta$ becomes constant. Unfortunately, as we discussed, unitary gauges are the worst possible gauges to see the decoupling of matter perturbations from metric perturbations. I am not aware of a gauge which is equally nice both in the UV and the IR at the same time. We just did the calculation with $\pi$, let us do it now with $\zeta$.

We said that we want to compute the correlation function of $\zeta$. Let us write again the metric in the so-called ADM parametrization
\be
ds^2=-N^2dt^2+h_{ij}\left(d x^i+N^i dt\right)\left(d x^j+N^j dt\right)\ .
\ee

\begin{itemize}

\item Expand the action. We get (\ref{eq:actiongauge})

Now take equations of motions with respect to all fluctuating variables.
\be
\frac{\delta S}{\delta\; \delta g^{\mu\nu}}=0\ ,\ ,
\ee

\item For semplicity, we do the calculation for $M_{2,\ldots}=0$ (this includes slow roll inflation). In this case the equations of motion for $N$ and $N_i$ take the form in (\ref{eq:constraints}).

\item Fix a gauge.  
Fix the spatial diffs by fixing the spatial metric to be
\be
h_{ij}=a^2\delta_{ij} e^{2\zeta}
\ee
while time-diffs are fixed by imposing 
\be
\pi=0
\ee
I am neglecting tensor perturbations here, because as said at quadratic level they do not mix.
This gauge is called $\zeta$-gauge or Maldacena-gauge. 
The constrain equations read
\bea
&& \nabla_i\left[N^{-1}\left(E^i_j-\delta^i_j E\right)\right]=0\\ \nonumber
&& \mpl^2\left[R^{(3)}-\frac{1}{N^2}\left(E_{ij}E^{ij}-E_i^i{}^2\right)\right]-(3H^2+\dot H)+2\mpl^2\dot H\cdot\frac{1}{N^2}=0
\eea

\item In this gauge you can clearly see why $\zeta=\delta a/a$. Assuming that $N$ and $N_i$ go to their unperturbed value when $k/(aH )\to 0$~\footnote{Indeed this will be true because $N,N_i$ are constrained variables sourced by the gradients of $\zeta$.}, then we see that, for $\zeta=$const, we are in an perturbed FRW (as $\delta\phi=0$), with just a $\delta a$. 

\item The constrained variables $N$ and $N^i$ are constrained, and so we can solve for them in terms of the only remaining degree of freedom: $\zeta$. The solution reads
\be
N=1+\frac{\dot\zeta}{H}\ , \qquad N_i=\d_i\left(-\frac{1}{a^2}\frac{\dot\zeta}{H}-\frac{\dot H}{H^2}\frac{1}{\d^2}\dot\zeta\right)
\ee

\item Plug back this values for $N$ and $N^i$ in the action. Notice, you can do this only because they are constrained variables.
The action now reads
\be
S=\int d^4x\, a^3 \left(-\frac{\dot H}{H^2}\right)\mpl^2\left[\dot\zeta^2-\frac{1}{a^2}(\d_i\zeta)^2\right]
\ee
Now it looks like the action of a massless scalar field, but notice how much work was necessary to get it.

\item Let us quantize the system. Follow textbook: find
\be
\Pi_\zeta=\frac{\delta{\cal L}}{\delta \dot\zeta}=-2a^3\mpl^2 \left(\frac{\dot H}{H^2}\right)\dot\zeta
\ee
\be
\left[\zeta,P_\zeta\right]=i
\ee

This is a quadratic Lagrangian, so we simply expand the fourier components of $\zeta$ in annihilation and creation operators
\be
\hat \zeta_{\vec k}(t)=\zeta_{\vk}^{cl}(t) a^\dag_{\vk}+\zeta_{cl}^\star(t)a_{-\vk}
\ee
with $\zeta^{cl}$ satisfying the equation of motion (Heisemberg equation for $\hat \zeta$)
\be
0=\frac{\delta L}{\delta \zeta}=\frac{d\left(a^3 \left(-\frac{\dot H}{H^2}\dot\zeta^{cl}_k\right)\right)}{dt}+\frac{\dot H}{H^2}a k^2\zeta^{cl}_k
\ee
As before, and for the same reasons as before, we choose the following initial condition at early time
\be
\zeta^{cl}_k(-k\eta\gg 1)\sim \frac{-i }{(2\epsilon)^{1/2}\mpl a(\eta)^3}\frac{H}{(2k/a(\eta))^{1/2}}e^{i k \eta}\quad {\rm for}\quad \frac{k}{a H}=-k\eta\gg 1 
\ee

\item Now we can solve the linear equation. Since at early times the Hubble expansion is negligible, and at late times $\zeta$ goes to a constant, we can neglect the time dependence of $H,\dot H$, and evaluate those terms at freeze out (it is possible to solve that equation exactly at first order in slow roll parameters. You can do this yourself). Using Mathematica, the solution reads 
\be
\zeta^{cl}_k(\eta)= \frac{H}{(2\epsilon)^{1/2}\mpl}\frac{1}{(2k)^{3/2}}(1-i k\eta)e^{i k \eta}\ , 
\ee

\item We can now compute the power spectrum:
\bea
&&\langle0|\zeta_{\vk}(\eta)\zeta_{\vk'}(\eta')|0\rangle=(2\pi)^3\delta^3(\vk+\vk')\\ \nonumber
&& \qquad\qquad\times\frac{H_{f.o.}}{(2\epsilon_{f.o.})^{1/2}\mpl}\frac{1}{(2k)^{3/2}}(1-i k\eta)e^{i k \eta}\frac{H_{f.o.}}{(2\epsilon_{f.o.})^{1/2}\mpl}\frac{1}{(2k)^{3/2}}(1+i k\eta')e^{-i k \eta'}
\eea
when $k\eta\ll1$ and $k\eta'\ll1$, we obtain
\be
\langle\zeta_{\vk}\zeta_{\vk'}\rangle_{\rm late}=(2\pi)^3\delta^3(\vk+\vk')\; \frac{1}{k^3}\cdot\frac{H_{f.o.}^4}{4(-\dot H_{f.o.})\mpl^2}
\ee

Which nicely reproduces the results we found with our estimates (but now we even got the factor of 4!).

\item One can compute correlation functions not on the vacuum state. Vacuum is somewhat better justified, though generalizations have been considered (see for example~\cite{Agarwal:2012mq}).

\item Notice how simpler was the calculation with $\pi$: no metric perturbations and constraint equations were necessary.

\end{itemize}

\subsubsection{The various limits of single field inflation}

\subsubsection*{Slow-roll inflation and high energy corrections}
The simplest example of the general Lagrangian (\ref{eq:actiontad}) is obtained by keeping only the first three terms, which are fixed once we know the background Hubble parameter $H(t)$, and setting to zero all the other operators of higher order: $M_2 = M_3 = \bar M_1 =\bar M_2 \ldots =0$. In the $\phi$ language, this corresponds to standard slow-roll inflation, with no higher order terms.  We have already done this case, both using $\pi$ or using $\zeta$.

Notice however that not all observables can be calculated from the $\pi$ Lagrangian (\ref{Spi}): this happens when the leading result comes from the mixing with gravity or is of higher order in the slow-roll expansion.  For example, as the first two terms of eq.~(\ref{Spi}) do not contain self-interactions of $\pi$, the 3-point function $\langle \zeta(\vec k_1) \zeta(\vec k_2) \zeta(\vec k_3) \rangle $would be zero. One is therefore forced to  look at subleading corrections, taking into account the mixing with gravity in eq.~(\ref{Smixed}). 

Obviously our choice of setting to zero all the higher order terms
cannot be exactly true. At the very least they will be radiatively
generated even if we put them to zero at tree level. The theory is
non-renormalizable and all interactions will be generated with
divergent coefficients at sufficiently high order in the perturbative
expansion. As additional terms are generated by graviton loops, they
may be very small. For example it is straightforward to check that
starting from the unitary gauge interaction $M_{\rm Pl}^2 \dot H g^{00}$ a
term of the form $(\delta g^{00})^2$ will be generated with a
logarithmically divergent coefficient $M_2^4 \sim \dot H^2 \log
\Lambda$. This implies that one should assume $M^4_2 \gtrsim \dot H^2$
(\footnote{The explicit calculation of logarithmic divergences in a theory
  of a massless scalar coupled to gravity has been carried out a
  long time ago in \cite{'tHooft:1974bx}.}). This lower limit is however very small. For example the dispersion relation of $\pi$ will be changed by the additional contribution to the time kinetic term: this implies, as we will discuss thoroughly below, that the speed of $\pi$ excitations deviates slightly from the speed of light, by a relative amount $1-c_s^2 \sim M_2^4/(|\dot H| M_{\rm Pl}^2) \sim |\dot H|/M_{\rm Pl}^2$. Using the normalization of the scalar spectrum, we see that the deviation from the speed of light is $\gtrsim \epsilon^2 \cdot 10^{-10}$. A not very interesting lower limit.

The size of the additional operators will be much larger if additional
physics enters below the Planck scale. In general this approach gives
the correct parametrization of all possible effects of new physics. As
usual in an effective field theory approach, the details of the UV
completion of the model are encoded in the higher dimension
operators. This is very similar to what happens in physics beyond the
Standard Model. At low energy the possible effects of new physics are
encoded in a series of higher dimensional operators compatible with
the symmetries \cite{Barbieri:2004qk}. The detailed experimental study of the Standard model allows us to put severe limits on the size of these higher dimensional operators. The same can be done in our case, although the set of conceivable observations is unfortunately much more limited.

\subsubsection*{Small speed of sound and large non-Gaussianities}
The Goldstone action (\ref{Spi}) shows that the spatial kinetic term $(\partial_i \pi)^2$ is completely fixed by the background evolution to be $M_{\rm Pl}^2 \dot H (\partial_i\pi)^2$. In particular only for $\dot H <0$, it has the ``healthy" negative sign. This is an example of the well studied relationship between violation of the null energy condition, which in a FRW Universe is equivalent to $\dot H<0$, and the presence of instabilities in the system. Notice however that the wrong sign of the operator  $(\partial_i \pi)^2$ is not enough to conclude that the system is pathological: higher order terms like $\delta K^\mu {}_\mu {}^2$ may become important in particular regimes, as we will discuss thoroughly below. 

The coefficient of the time kinetic term $\dot{\pi}^2$ is, on the other hand, not completely fixed by the background evolution, as it receives a contribution also from the quadratic operator $(\delta g^{00})^2$. In eq.~(\ref{Spi}) we have 
\begin{equation}
\left(-M_{\rm Pl}^2 \dot{H} + 2 M_2^4 \right) \dot\pi^2 \;.
\end{equation}
To avoid instabilities we must have $-M_{\rm Pl}^2 \dot{H} + 2 M_2^4 >0$ .
As time and spatial kinetic terms have different coefficients, $\pi$
waves will have a ``speed of sound'' $c_s \neq 1$. This is expected as
the background spontaneously breaks Lorentz invariance, so that
$c_s=1$ is not protected by any symmetry. As we discussed in the last
section, deviation from $c_s=1$ will be induced at the very least by
graviton loops \footnote{If we neglect the coupling with gravity and
the time dependence of the operators in the unitary gauge Lagrangian
(so that $\pi \to \pi + {\rm const}$ is a symmetry),
$c_s=1$ can be protected by a symmetry $\partial_\mu\pi \to
\partial_\mu\pi + v_\mu$, where $v_\mu$ is a constant vector. Under
this symmetry the Lorentz invariant kinetic term of $\pi$ changes by a
total derivative, while the operator proportional to $M_2^4$ in
eq.~(\ref{Spi}) is clearly not invariant, so that $c_s=1$. Notice that
the theory is not free as we are allowed to write interactions with
more derivatives acting on $\pi$. This symmetry appears in the study of the brane
bending mode of the DGP model.}.
The speed of sound is given by
\be
c_s^{-2} = 1-\frac{2 M_2^4}{M_{\rm Pl}^2 \dot H} \;.
\ee
This implies that in order to avoid superluminal propagation we must
have $M_2^4 >0$ (assuming $\dot H <0$). Superluminal propagation would
imply that the theory has no Lorentz invariant UV completion \cite{Adams:2006sv}. In the following we will concentrate on
the case $c_s \le 1$.

Using the equation above for $c_s^2$ the Goldstone action can be
written at cubic order as

\begin{eqnarray}\label{Spi_cs}
S_{\rm \pi} = \int \!  d^4 x  \:  \sqrt{- g} \left[
-\frac{M^2_{\rm Pl} \dot{H}}{c^2_s} \left(\dot\pi^2-c^2_s
\frac{(\partial_i\pi)^2}{a^2}\right)
+M_{\rm Pl}^2 \dot H \left(1-\frac{1}{c^2_s}\right)
\left(\dot{\pi}^3-\dot\pi\frac{(\partial_i\pi)^2}{a^2}
\right)- \frac43 M_3^4 \dot\pi^3... \right].
\end{eqnarray}

From the discussion in section (\ref{sec:Goldstone}) we know that the
mixing with gravity can be neglected at energies $E \gg E_{\rm mix}
\simeq \epsilon H$.

The calculation of the 2-point function follows closely the case
$c_s=1$ if we use a rescaled momentum $\bar k=c_s k$ and
take into account the additional factor $c_s^{-2}$ in front of the
time kinetic term. We obtain
\be
\label{eq:spectrumzetacs}
\langle \zeta(\vec k_1) \zeta(\vec k_2)\rangle = (2 \pi)^3 \delta(\vec
k_1 + \vec k_2)  \frac1{c_{s*}} \cdot \frac{H^4_*}{4 M_{\rm Pl}^2 |\dot H_*|}\frac{1}{k_1^3} = (2
\pi)^3 \delta(\vec k_1 + \vec k_2) \frac1{c_{s*}} \cdot \frac{H_*^2}{4 \epsilon_* M_{\rm Pl}^2} \frac{1}{k_1^3} \;.
\ee
The variation with time of the speed of sound introduces an additional
contribution to the tilt
\begin{equation}
n_s=\frac{d}{d \log k}\log \frac{H^4_{*}}{|\dot{H}_*|
c_{s*}}=\frac{1}{H_*}\frac{d}{dt_*}\log \frac{H^4_{*}}{|\dot{H}_*|
c_{s*}} =4\frac{\dot{H_*}}{H^2_*}-\frac{\ddot{H}_*}{\dot{H}_*
H_* }-\frac{\dot{c}_{s*}}{c_{s*}H_{*}} \;.
\end{equation}

From the action (\ref{Spi_cs}) we clearly see that the same operator
giving a reduced speed of sound induces cubic couplings of the
Goldstones of the form $\dot{\pi}(\nabla\pi)^2$ and $\dot\pi^3$. The
non-linear realization of time diffeomorphisms forces a relation
between a reduced speed of sound and an enhanced level of
the 3-point function correlator, {\em i.e.} non-Gaussianities. Indeed remember that the $\phi$-wavefunction was a Gaussian in the vacuum state simply because the action was quadratic in the fields. Interactions will lead to deviation from a Gaussian wavefunction: i.e. non-Gaussianities.

To estimate the size of non-Gaussianities, one has to compare
the non-linear corrections with the quadratic terms around freezing, $\omega \sim H$. We have to evaluate qualities at the time of freezing because the interaction operators have derivatives acting on each fluctuation, so that the interaction effectively shut down after freezing. 
In the limit
$c_s\ll 1$, the operator $\dot\pi(\nabla{\pi})^2$ gives the leading
contribution, as the quadratic action shows that a mode freezes with $k/a
\sim H/c_s$, so that spatial derivatives are enhanced with respect to
time derivatives. 
Notice indeed that
\be
H\sim \omega\sim c_s \frac{k}{a} \ , \quad\Rightarrow\quad \frac{k}{a(t_{f.o.})}\sim \frac{H}{c_s}\gg H\ .
\ee

The level of non-Gaussianity will thus be given by the
ratio:
\begin{equation}
\frac{{\cal L}_{\dot\pi(\nabla \pi)^2}}{{\cal L}_{2}}\sim \frac{H
\pi \left(\frac{H}{c_s} \pi\right)^2}{H^2 \pi^2} \sim \frac{H}{c^2_s}
\pi \sim \frac{1}{c^2_s} \zeta \, ,
\end{equation}
where in the last step we have used the linear relationship
between $\pi$ and $\zeta$. Taking $\zeta \sim
10^{-5}$ we have an estimate of the size of the non-linear correction.
 Usually the
magnitude of non-Gaussianities is given in terms of the parameters $f_{\rm
NL}$, which are parametrically of the form: 
\be\label{eq:estimate_one}
\frac{{\cal L}_{\dot\pi(\nabla \pi)^2}}{{\cal L}_2}\sim f_{\rm NL} \zeta
\ee 
The leading contribution
will thus give 
\begin{equation}
\label{eq:nabla2dot}
f^{\rm equil.}_{\rm NL, \;\dot\pi(\nabla\pi)^2}\sim \frac{1}{c^2_s} \;.
\end{equation}
The superscript ``equil.'' refers to the momentum dependence of the
3-point function, which in these models is of the so called equilateral or orthogonal form. This is physically clear in the Goldstone
language as the relevant $\pi$ interactions contain derivatives, so
that they die out quickly out of the horizon; the correlation is only
among modes with comparable wavelength.

In the Goldstone Lagrangian (\ref{Spi_cs}) there is an
additional independent operator, $-\frac43 M_3^4 \dot\pi^3$,
contributing to the 3-point function, coming from the unitary gauge
operator $(\delta g^{00})^3$. We thus have two contributions of the form
$\dot\pi^3$ which give
\begin{equation}\label{fnldotpicube}
f^{\rm equil.}_{\rm NL,\ \dot\pi^3}\sim 1-\frac{4}{3}
\frac{M_3^4}{M_{\rm Pl}^2 |\dot H| c_s^{-2}} \;.
\end{equation}
The size of the operator $-\frac43 M_3^4 \dot\pi^3$ is not constrained
by the non-linear realization of time diffeomorphisms: it is a free
parameter. In DBI inflation \cite{Alishahiha:2004eh} we have 
$M_3^4 \sim M_{\rm Pl}^2 |\dot H| c_s^{-4}$, so that its contribution to
non-Gaussianities is of the same order as the one of 
eq.~(\ref{eq:nabla2dot}). The same approximate size of the $M_3^4$ is
obtained if we assume that both the unitary gauge operators 
$M_2^4 (\delta g^{00})^2$ and $M_3^4 (\delta g^{00})^3$ become strongly coupled
at the same energy scale.

\subsubsection*{Cutoff and Naturalness}
As discussed, for $c_s <1$ the Goldstone action contains
non-renormalizable interactions. Therefore the self-interactions among
the Goldstones will become strongly coupled at a certain energy scale, which
sets the cutoff of our theory. This cutoff can be estimated looking at
tree level partial wave unitarity, {\em i.e.} finding the maximum energy
at which the tree level scattering of $\pi$s is unitary. The
calculation is straightforward, the only complication coming from the
non-relativistic dispersion relation. The cutoff scale $\Lambda$
turns out to be
\be
\label{eq:cutoff}
\Lambda^4 \simeq 16 \pi^2 M_2^4 \frac{c_s^7}{(1-c_s^2)^2} \simeq 16
\pi^2 M_{\rm Pl}^2 |\dot H|
\frac{c_s^5}{1-c_s^2} \;.
\ee
The same result can be obtained looking at the energy scale where loop
corrections to the $\pi \pi$ scattering amplitude become relevant.
As expected the theory becomes more and more strongly coupled for
small $c_s$, so that the cutoff scale decreases. On the other hand, for
$c_s \to 1$ the cutoff becomes higher and higher. This makes sense as
there are no non-renormalizable interactions in this limit and the
cutoff can be extended up to the Planck scale. This cutoff scale is
obtained just looking at the unitary gauge operator $(\delta g^{00})^2$;
depending on their size the other independent operators may give an
even lower energy cutoff. Notice that the scale $\Lambda$ indicates the maximum energy at which
our theory is weakly coupled and make sense; below this scale new
physics must
come into the game. However new physics can appear even much below
$\Lambda$. 

If we are interested in using our Lagrangian for making predictions for
cosmological correlation functions, then we need to use it at a
scale of order the Hubble parameter $H$ during inflation. We therefore
need that this energy scale is below the cutoff, $H \ll
\Lambda$. Using the explicit expression for the cutoff (\ref{eq:cutoff}) in the case $c_s
\ll 1$ one gets
\be
H^4 \ll M_{\rm Pl}^2 |\dot H| c_s^5
\ee 
which can be rewritten using the spectrum normalization
(\ref{eq:spectrumzetacs}) as an inequality for the speed of sound
\be
c_s \gg P_\zeta^{1/4} \simeq 0.003 \;. 
\ee
A theory with a lower speed of sound is strongly coupled at $E \simeq H$. Not surprisingly this value of the speed of sound also corresponds to the value at which non-Gaussianity are of order one: the theory is strongly coupled at the energy scale $H$ relevant for cosmological predictions.

Let us comment on the naturalness of the theory. One may wonder
whether the limit of small $c_s$ is natural or instead loop
corrections will induce a larger value. The Goldstone
self-interactions, $\dot\pi(\nabla\pi)^2$ and $(\nabla\pi)^4$ for
example, will
induce a radiative contribution to $(\nabla\pi)^2$. It is easy to
estimate that these contributions are of order $c_s^{-5}
\Lambda^4/(16 \pi^2 M_2^4)$, where $\Lambda$ is the UV cutoff, {\em i.e.} the
energy scale at which new physics enters in the game. We can see that
it is impossible to have large radiative contribution; even if we take
$\Lambda$ at the unitarity limit (\ref{eq:cutoff}), the effect is of
the same order as the tree level value. This makes sense as
the unitarity cutoff is indeed the energy scale at which loop
corrections become of order one.

 We would like also to notice that the action (\ref{Spi}) is
natural from an effective field theory point of view
\cite{Polchinski:1992ed}. The relevant operators are in fact protected
from large renormalizations if we assume an approximate shift symmetry
of $\pi$. In this case the coefficients of the relevant operators are sufficiently
small and they will never become important for observations as cosmological correlation functions
probe the theory at a fixed energy scale of order $H$: we never go
to lower energy. Clearly here we are only looking at the period of
inflation, where an approximate shift symmetry is enough to make the
theory technically natural; providing a graceful exit from inflation
and an efficient reheating are additional requirements for a working
model which are not discussed in our formalism.

\subsubsection*{De-Sitter Limit and the Ghost Condensate}

In the previous section we saw that the limit $c_s \to 0$ is
pathological as the theory becomes more and more strongly
coupled. However we have neglected in our discussion the higher derivative
operators in the unitary gauge Lagrangian (\ref{eq:actiontad})
\be
\int d^4 x \sqrt{-g} \left(-\frac{\bar M_2(t)^2}{2} \delta K^\mu {}_\mu {}^2
-\frac{\bar M_3(t)^2}{2} \delta K^\mu {}_\nu \delta K^\nu {}_\mu
\right) \;.
\ee
These operators give rise in the Goldstone action to a spatial kinetic term of the
form
\begin{equation}
\label{di_pi2}  \int  \! d^4x \: \sqrt{-g} \left[ - \frac{\bar M^2}{2} \, \frac{1}{a^4}(\di_i^2 \pi)^2
   \right] \; ,
\end{equation}
where $\bar M^2 = \bar M_2^2+  \bar M_3^2$. Notice that we obtain the very non-relativistic dispersion relation
\be
\omega^2\sim \frac{k^4}{M^2}\ .
\ee
This models naturally leads to large non-Gaussianities.

\subsubsection*{De-Sitter Limit without the Ghost Condensate}
In this section we want to study the effect of the operator
\be
\label{eq:leooperator}
\int \! d^4 x \: \sqrt{- g} \:
\left(-\frac{\bar M_1(t)^3}{2} \delta g^{00}\delta K^\mu {}_\mu \right)\;.
\end{equation}
 on
the quadratic $\pi$ action. We will see that, if the coefficient of
this operator is sufficiently large, we obtain a different de Sitter limit,
where the dispersion relation at freezing is of the form $\omega^2
\propto k^2$, instead of the Ghost Condensate behavior $\omega^2
\propto k^4$.

For simplicity we can take $\bar M_1$ to be time
independent. Reintroducing the Goldstone we get a 3-derivative term of
the form $-\bar M_1^3 \dot\pi\nabla^2\pi/a^2$ (\footnote{The operator
  gives also a contribution to $\dot\pi^2$ proportional to $H$. We
  will assume that this is small compared to $M_2^4 \dot\pi^2$.}). This would be a total
time derivative without the time dependence of the scale factor
$a(t)$ and of the metric determinant. Integrating by parts we get a standard 2-derivative spatial
kinetic term
\be
\label{eq:byparts}
-\int d^4 x \sqrt{-g} \,\frac{\bar M_1^3 H}{2}
\left(\frac{\partial_i}{a}\pi\right)^2 \;.
\ee
In the exact de Sitter limit, $\dot H =0$, and taking $M_2 \sim \bar
M_1 \sim M$, this operator gives a dispersion relation of the form
\begin{equation}
c^2_s=\frac{H}{M}\ll 1 \;.
\end{equation}
and naturally to large non-Gaussianities. 

This and the Ghost condensate case are finally the only known ways to violet the Null Energy Condition in a stable way~\cite{Creminelli:2006xe}.

\subsection{Summary of Lecture 3}

\begin{itemize}
\item Acoustic oscillations of the CMB tell us that primordial perturbations were super-Hubble 
\item The CMB temperature 2-point is fitted by using just two numbers from the inflationary theory.
\item Detection of scale invariant $B$-modes would teach us of a primordial de Sitter like epoch. 
\item There are many more models of Inflation beyond slow-roll.
\item A General description is offered by the Effective Field Theory of Inflation, that parametrizes inflation as the theory of space-time diff.s spontaneously broken to time-dependent spatial diffs.
\item The Lagrangian of the associated Goldstone boson is very simply and allows us to learn the relevant physics straightforwardly.
\item We see that there are interaction operators that are allowed to be large, and they will potentially lead to detectable non-Gaussianities in the CMB.
\item We learnt how to accurately compute the power spectrum.
\end{itemize}

\newpage
\section{Lecture 4: Non-Gaussianity: who are you?}

We have seen in the former section that we can have inflationary models with large self-interactions. We said that they produce some non-Gaussianity. Indeed we saw that in the limit of free-theory the vacuum wavefunction was a Gaussian. This was because the Lagrangian was quadratic and each fourier mode was like an harmonic oscillator. But if the action is slightly non-linear, than we can imagine some slight non-Gaussianity. Something like, just symbolically:

\be
|0\rangle_{k_i/a \ll H}\sim\prod_{\{\zeta_{\vk}\}}\; e^{-\frac{\zeta_{\vk_i}^2}{\sigma_{\zeta_{\vk_i}}}- \frac{\zeta_{\vk_i}\zeta_{\vk_j}\zeta_{\vk_i+\vk_j}}{C(\vk_1,\vk_2,\vk_1+\vk_2)}} |\{\zeta_{\vk}\}\rangle
\ee
This would mean that a signal like the three-point function
\be
\langle\zeta_{\vec k_1}\zeta_{\vec k_2}\zeta_{\vec k_3}\rangle=(2\pi^3)\delta^{(3)}(\vec k_1+\vec k_2+\vec k_3) F(\vk_1,\vk_2,\vk_3)
\ee
would not be zero.
Current limits set the skewness of the distribution to satisfy
\be
\frac{\langle\zeta^3\rangle}{\langle\zeta^2\rangle^{3/2}}\lesssim 10^{-2,-3}\sim \frac{1}{N_{\rm pix}^{1/2}}
\ee
which is a very small number! Look at the plot.

\begin{figure}[h!]
\begin{center}
\includegraphics[width=8cm]{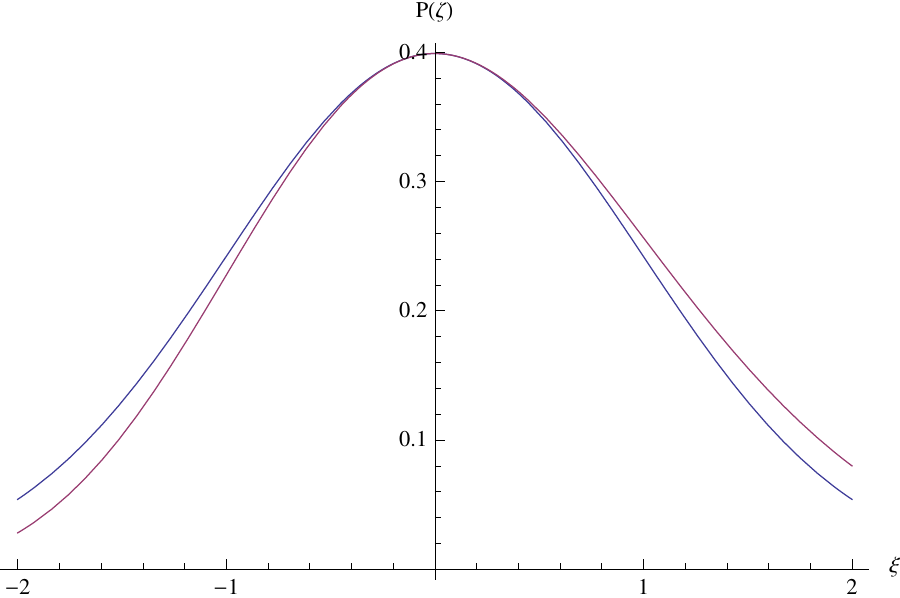}
\caption{\label{fig:two-gauss} \small Plot of a Gaussian Distribution and of a non-Gaussian distribution that has skewness approximately equal to 1. Can you tell which one is which? Cosmological observations are constraining the skewness of the distribution of the CMB radiation to be less than a percent of the one of the figure. Wow! }
\end{center}
\end{figure}

Being a limit on a statistics, the limit scale as $N_{\rm pix}^{-1/2}$. For WMAP, we have indeed about $10^5$ modes.

But what this tells us is that a detection of non-Gaussianities would be associated to the interacting part of the Lagrangian, which is really the interesting part of the Lagrangian! And we are talking of interactions at extremely high energies! Interactions contain so much more information that they would allow us to learn about the real dynamics that drove inflation.

Clearly, since non-Gaussianities are small, it is expectable that the leading signature will appear in the 3-point function. 

Let us look at the function $F$. So far it depends on 9 variables. But let us use the symmetries of the problem. By the cyclic invariance of the correlation function (remember that at late times we are semiclassical), we can set $k_1\geq k_2\geq k_3$. Translation invariance forces the sum of the three momenta to be zero: they must form a closed triangle. We are down to 6.  We can use to rotational invariance to point $\vec k_1$ in the $\hat x$ direction, and $\vk_2$ in the $x-y$ plane. We are down to 3 variables. Additionally, the 3-point function should be scale invariant, because two triplet of modes, one an overall rescaling of the other, see approximately the same history. We can use this to set the modulus of $k_1=1$. The overall $k_1$ dependence has to be $1/k_1^6$, so that the real space 3-point function
\be
\langle\zeta(x)^3\rangle=\int d^3k_1 d^3 k_2 d^3k_3 \; \langle\zeta_{\vec k_1}\zeta_{\vec k_2}\zeta_{\vec k_3}\rangle
\ee
receives the same contribution from each logarithmic interval.

So, in terms of degrees of freedom, we are down to two variables
\bea
&&\langle\zeta_{\vec k_1}\zeta_{\vec k_2}\zeta_{\vec k_3}\rangle=(2\pi^3)\delta^{(3)}(\vec k_1+\vec k_2+\vec k_3)\frac{1}{k_1^6}F(x_2,x_3)\\ \nonumber
&& x_2=k_2/k_1\ , \quad x_3=k_3/k_1\ , \quad x_3\leq x_2\leq 1 \ ,\quad x_3\leq x_2 
\eea

This is a {\it huge} amount of information. Remember that because of the various symmetries, the 2-point function had to go as $1/k^3$, and so it dependent only on one {\it number}. Because of the slight deviation of scale invariance, we had also the tilt, which is just a second {\it number}. Here with non-Gaussianities, we are talking about a {\it function of 2 parameters}. This is $\infty$ numbers! this is a huge amount of information, incomparable with respect to the information contained in the 2-point function. Indeed, it has the same amount of information as a 2-2 scattering as a function of angles. And this is not little thing: we learn about spin of particles and nature of interactions from this. Let us plot $F$. A useful quantity to plot is a quantity the resembles the signal to noise ratio in each triangular configuration~\cite{Babich:2004gb}. It is 
\be
\left.\frac{S}{N}\right|_{\rm triangle}\sim x^{2}_2 x_3^2 F(1,x_2,x_3)
\ee
 which is a function of the triangular shape. A typical shape is the following:

\begin{figure}[h!]
\begin{center}
\includegraphics[width=12cm]{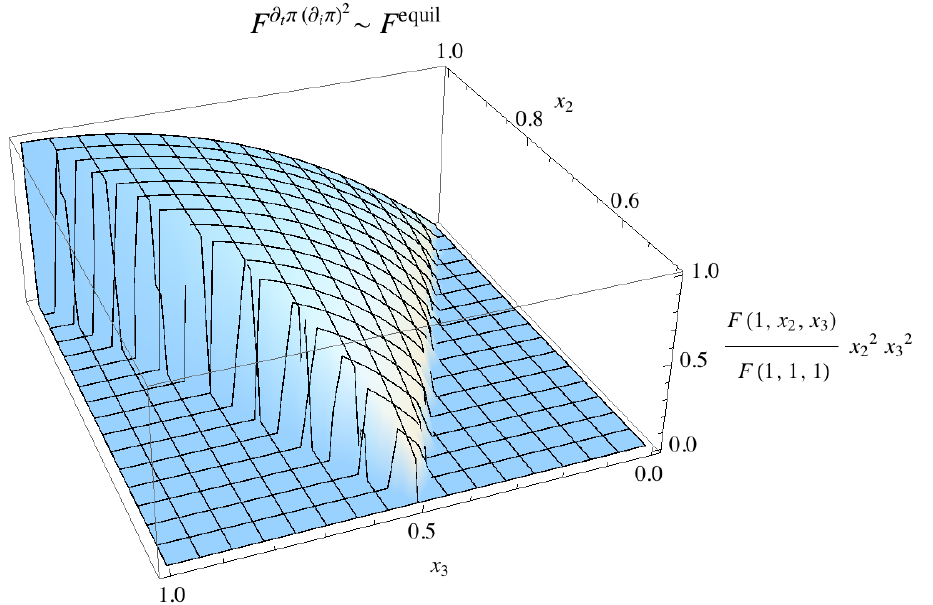}
\caption{\label{fig:shape} \small A shape of non-gaussianities. We will explain the details more later, but you can see that it contains very non-trivial information.}
\end{center}
\end{figure}

Isn't this a beauty? It has a lot of information. Such a detection would really make us confident that something very non-trivial was going on in the sky. It would also teach a lot about the dynamics that drove inflation.

\subsection{Estimating the size of non-Gaussianities}

Let us first learn how to estimate the size of non-Gaussianities. In the EFT of inflation, we have seen that at leading order in derivatives we have two interaction operators: $\dot\pi^3$ and $\dot\pi(\d_i\pi)^2$. Let us consider the case of the operator $\dot\pi(\d_i\pi)^2$. We already learned how to estimate $f_{\rm NL}$ for this operator~(\ref{eq:estimate_one}):

\be\label{eq:estimate_onebis}
 f_{\rm NL} \zeta\sim \left.\frac{{\cal L}_{\dot\pi(\nabla \pi)^2}}{{\cal L}_2}\right|_{E\sim H}
\ee 

It is useful to derive this in a different way. We can follow the manipulations of the single field Lagrangian of~\cite{Senatore:2010jy}. Let us consider, in the limit $c_s\ll 1$ for simplicity,
\begin{eqnarray}\label{eq:Spics}
S_{\rm \pi} =\int d^4 x\,   \sqrt{- g} \left[ -\frac{M^2_{\rm Pl}\dot{H}}{c_s^2} \left(\dot\pi^2-c_s^2\frac{ (\partial_i \pi)^2}{a^2}\right)
-\frac{\mpl^2\dot H}{c_s^2}
\dot\pi\frac{(\partial_i\pi)^2}{a^2} -\frac{2}{3}  \frac{\tilde c_3}{c_s^4}\mpl^2\dot H \dot{\pi}^3 \right]\ ,
\eea
where we have used
\be
c_s^{-2} \sim \frac{2 M_2^4}{-M_{\rm Pl}^2 \dot H} \;.
\ee
 and we have redefined $M_3^4=\tilde c_3M_2^4/c_s^2$. We can perform a transformation of the spatial coordinates:
\be
\vec x\quad \rightarrow \quad \vec{ \tilde x}=\vec x/c_s\ ,
\ee
and canonically normalize the field $\pi$
\be
\pi_c=(-2\mpl^2 \dot H c_s)^{1/2}\;\pi\ ,
\ee
to obtain
\begin{eqnarray}\label{eq:Spicscan}
S_{\rm \pi}=\int dt\, d^3\tilde x\,   \sqrt{- g} &&\left[ \frac{1}{2} \left(\dot\pi_c^2-\frac{ (\tilde \partial_i \pi_c)^2}{a^2}\right)\right.\\ \nonumber
&&\left.-\frac{1}{\left(8|\dot H |\mpl^2c_s^5\right)^{1/2}}
\dot\pi_c\frac{(\tilde \partial_i\pi_c)^2}{a^2} -\frac{2}{3}  \frac{\tilde c_3}{\left(8|\dot H|\mpl^2c_s^5\right)^{1/2}} \dot\pi_c^3 \right]\ ,
\eea
where $\tilde \d_i=\d/\d\tilde x^i$. Notice that since we did not rescale time so the unitarity bound in energy can directly read off from (\ref{eq:Spicscan}). Since the kinetic part of the Lagrangian (\ref{eq:Spicscan}) is now Lorentz invariant, it is  easy to read off the unitarity bound
\be
 \Lambda^4_U\sim 16\pi^2 c_s^5|\dot H|\mpl^2 \sim16\pi^2 c_s^7 M_2^4 \ .
\ee
The factors of $c_s$ came out automatically.
The importance of the interaction operators becomes smaller and smaller as we move to lower energies, as typical for dimension 6, irrelevant, operators.
The size of non-Gaussianity is determined by the size of this operator at freezing $\omega\sim H$. Forgetting factors of $\pi$'s and numerical factors, we have
\be\label{eq:estimate_two}
 f_{\rm NL} \zeta\sim \left.\frac{{\cal L}_{\dot\pi(\nabla \pi)^2}}{{\cal L}_2}\right|_{E\sim H}\sim \frac{H^2}{\Lambda_U^2} .
\ee 
We therefore relate the detection or limits on non-Gaussianities, to limits or measurements of $\Lambda_U$, the scale suppressing the higher dimensional operators in the Effective Theory.

\subsection{Computation of the 3-point function}

Let us see how to compute this $\langle\zeta^3\rangle\sim F$ precisely. In the EFT of inflation, we have seen that at leading order in derivatives we have two interaction operators: $\dot\pi^3$ and $\dot\pi(\d_i\pi)^2$. Let us compute the shape due to the first, as an example.

This is nothing by a QFT exercise, just follow the rules. 
\begin{itemize}

\item We have an interacting theory. Very much as we do when computing scattering amplitudes or correlation functions in Minkowski, we go to the interaction picture. We split the Hamiltonian in
\be
H=H_0+ H_{\rm int}
\ee
and evolve the operators with $H_0$ and the state with $H_{\rm int}$. Since the evolution under $H_0$ is completely understood, we need simply to evolve the state with the interaction picture evolutor 
\be
U_{\rm int}(t,\tin)=U_0(\tin,t)U(t,\tin)U_0(t,\tin)=T e^{-i \int_{\tin}^t dt' H_{\rm int}(t')} 
\ee
where $U(t,t')$ is the full evolutor, and $U_0(t,t')$ is the free theory evolutor, while $T$ denotes time ordering.

\item  What we would like to compute is the expectation value of $\zeta_{\vk_1}\zeta_{\vk_2}\zeta_{\vk_3}$ evaluated on the initial state of the theory, which is the vacuum $|\Omega(\tin)\rangle$, evolved to time $t$.
\be
|\Omega(t)\rangle=U_{\rm int}(t,\tin)|\Omega(\tin)\rangle\ .
\ee
We then have:
\be
\langle\Omega(t)|\zeta_{\vk_1}\zeta_{\vk_2}\zeta_{\vk_3}|\Omega(t)\rangle=\langle\Omega(\tin)|\left(\bar T e^{i \int_{\tin}^t dt' H_{\rm int}(t')} \right)\zeta^{\rm int}_{\vk_1}\zeta^{\rm int}_{\vk_2}\zeta^{\rm int}_{\vk_3}\left(Te^{-i \int_{\tin}^t dt' H_{\rm int}(t')}\right) |\Omega(\tin)\rangle
\ee
with $\bar T$ representing anti-time ordering and $\zeta^{\rm int}$ the interaction picture operator.

Notice that this expectation value is taken between two $in$ states. This is why it is called in-in formalism. Notice that this is different than what one usually does in scattering amplitudes, where one computes in-out correlation functions. This is the source of a series of differences with scattering amplitude. For example, the results are not independent of field redefinitions. We wish to compute correlation functions of $\zeta$.

\item How do we compute the vacuum state? We know how to express well states in the Fock base, so, it would be good to express $|\Omega(t)\rangle$ in this base. It is possible to express $|\Omega(t)\rangle$ in terms of the free theory Bunch Davies vacuum with a simple rotation in the complex plane of the contour of integration of the evaluator operator~\footnote{I thank Luca Delacretaz and Matt Lewandowski for collaborating in formalizing the careful construction of the interacting vacuum that I present here.}. We are interested in defining carefully the vacuum at some early time. For the problem of defining the initial vacuum, we can therefore assume that the Hamiltonian is time-independent. The interaction picture evolutor becomes
\be
\uint(t,t')=e^{i H_0 (t-t')}\, e^{-i H (t-t')}\, e^{-i H_0(t'-t_0)}
\ee
As mentioned, we have, for an operator ${\cal O}$ of which we want to compute:
\be
\langle\Omega|{\cal O}_H(t)|\Omega\rangle=\langle\Omega|\uint(\tin,t){\cal O}_{\rm int}(t)\uint(t,\tin)|\Omega\rangle
\ee
where ${}_H$ stays for Heisemberg picture and $_{\rm int}$ for interaction picture.

On the ket side, let us write $\uint ( t , \tin ) = \uint( t , \tin) \uint ( \tin , \tin ( 1 - i \epsilon)  ) \uint (\tin( 1 - i \epsilon) , \tin ) $, with $\epsilon$ being a small positive number, and then look at 
\begin{align}
\uint (  \tin( 1 - i \epsilon) , \tin )  \vacright & = e^{ i H_0 (  \tin(1 - i \epsilon) - \tin) } e^{- i H (  \tin(1 - i \epsilon) - \tin) } \vacright \\ \nn
& = e^{ \epsilon H_0 \tin }  \vacright  e^{- \epsilon E \tin  } \\ \nn
& =\sum_n  e^{ \epsilon\, \tin\, ( E_n - E ) } \vacleft n\rangle |n\rangle
\rightarrow \freevacright  \freevacleft \Omega \rangle e^{ \epsilon\, \tin\, ( E_0 - E )  }
\end{align}
Here $|n\rangle$ is the fock basis in the free theory, $H \vacright = E \vacright$ and $H_0 \freevacright = E_0 \freevacright$.  In the last passage, we have expanded the interacting vacuum in a superposition of free states, and noticed that as $\tin\to-\infty$, the exponentially larger term is the one that overlaps with the free vacuum $\freevacright$.
Similarly on the bra side, we can write $U_{\rm int} ( \tin , t ) =    \uint ( \tin , \tin ( 1 + i \epsilon) )          \uint( \tin ( 1 + i \epsilon), \tin )  \uint( \tin , t) $ and then look at 
\begin{align}
\vacleft \uint ( \tin , \tin( 1 + i \epsilon ) ) & = \vacleft e^{- i H ( \tin - \tin ( 1 + i \epsilon) ) } e^{ - i H_0 ( \tin (1 + i \epsilon ) - \tin ) } \\ \nn
& = e^{-  E \tin  \epsilon  } \vacleft e^{   H_0  \tin \epsilon  } \\ \nn
& \rightarrow  e^{ \epsilon \tin ( E_0 - E) }   \vacleft 0 \rangle  \freevacleft
\end{align}
as $\tin \rightarrow - \infty$. 

\item We are therefore led to compute
\bea
&&\langle\Omega|{\cal O}_H(t)|\Omega\rangle=\langle\Omega|\uint(\tin,t){\cal O}_{\rm int}(t)\uint(t,\tin)|\Omega\rangle=\\ \nonumber
&&=e^{ 2\epsilon\, \tin\, ( E_0 - E) }  | \vacleft 0 \rangle|^2  \freevacleft   \uint( \tin ( 1 + i \epsilon), \tin )  \uint( \tin , t) {\cal O}_{\rm int}(t) \uint( t , \tin) \uint ( \tin , \tin ( 1 - i \epsilon)  )\freevacright
\eea
The prefactor of the expectation value is actually equal to
\bea \freevacleft \uint (\tin ( 1 + i \epsilon ) , \tin( 1 - i \epsilon) )  \freevacright \rightarrow e^{  2 \tin \epsilon (E_0 - E ) } | \freevacleft \Omega \rangle |^2 
\eea

We can therefore write
\bea
&&\langle\Omega|{\cal O}_H(t)|\Omega\rangle=\\ \nn
&&=\lim_{\epsilon\to 0}\lim_{\tin\to -\infty}\frac{  \freevacleft   \uint( \tin ( 1 + i \epsilon), \tin )  \uint( \tin , t) {\cal O}_{\rm int}(t) \uint( t , \tin) \uint ( \tin , \tin ( 1 - i \epsilon)  )\freevacright}{\freevacleft \uint (\tin ( 1 + i \epsilon ) , \tin( 1 - i \epsilon) )  \freevacright }
\eea

\item The denominator is nothing but the sum over the bubble diagrams. As usual, their contribution resums as an exponential prefactor of the whole expression. We can therefore write
\bea
&&{}_{\rm in}\langle\Omega|{\cal O}_H(t)|\Omega\rangle_{\rm in}=\\ \nn
&&=\lim_{\epsilon\to 0}\lim_{\tin\to -\infty} {}_{\rm in}\freevacleft   \uint( \tin ( 1 + i \epsilon), \tin )  \uint( \tin , t) {\cal O}_{\rm int}(t) \uint( t , \tin) \uint ( \tin , \tin ( 1 - i \epsilon)  )\freevacright_{\rm in,\ no\ bubbbles}
\eea
where we have introduced the subscript $_{\rm in}$ to remind that these vacua are defined on the initial time.

We finally notice that, by the composition rule of the $U$'s, 
\bea
&&\uint( t , \tin) \uint ( \tin , \tin ( 1 - i \epsilon)  )=\uint( t , \tin ( 1 - i \epsilon)  )\;,\\ \nn
&&  \uint( \tin ( 1 + i \epsilon), \tin )  \uint( \tin , t) = \uint( \tin ( 1 + i \epsilon),t)
\eea
 Notice that  $\uint( t , \tin ( 1 - i \epsilon)  )$ can be thought just as a rotation of the countour of integration of the time-evolution $\int^t_{\tin\to -\infty (1-i\epsilon)} dt' H_{\rm int}(t')$, and $ \uint( \tin ( 1 + i \epsilon),t)=(U(t,\tin ( 1 - i \epsilon)))^\dag$. We therefore realize that the shift in time $\tin (1-i\epsilon)$ can be thought as an analytic rotation of the $t'$ contour  of integration. 

\item We therefore write our final expression
 \bea\label{eq:final}
 {}_{\rm in}\langle\Omega|{\cal O}_H(t)|\Omega\rangle_{\rm in}= {}_{\rm in}\freevacleft   (U(t,-\infty^-))^\dag  {\cal O}_{\rm int}(t) U(t,-\infty^-)\freevacright_{\rm in,\ no\ bubbbles}
\eea
where the integration contour has been rotated to approach $-\infty$ with a positive complex imaginary part on the right, and with a negative imaginary part on the left~\footnote{An equivalent way to write this expression is by an observation that, as far as I know, is originally due to Kendrick Smith and to~\cite{Behbahani:2012be}. One notices that we can perform the full 90-degrees rotation in the $t$-countour, so that the time integration is done with an Euclidean time that moves parallel the imaginary axis from $+\infty$ to $-\infty$. In this Euclidean time, the operators appear therefore as anti-time-ordered. So we can write the expression~(\ref{eq:final}) as
 \bea\label{eq:final2}
 {}_{\rm in}\langle\Omega|{\cal O}_H(t)|\Omega\rangle_{\rm in}= {}_{\rm in}\freevacleft \bar T \left[  {\cal O}_{\rm int}(t)\; {\rm Exp}\left(-\int_{-\infty}^{+\infty} dt_E'\; H_{\rm int}(t'+i\, t_E')\right)\right]\freevacright_{\rm in,\ no\ bubbbles}
\eea
where $\bar T$ is anti-time ordered. In this way, one performs Wick contraction of anti-time-ordered products of fields evaluated at times $t$ or $t'+i\, t_E'$. 
}.

\item At leading order in $H_{\rm int}$, we can Taylor expand the exponential to obtain
\be
\langle\Omega(t)|\zeta_{\vk_1}\zeta_{\vk_2}\zeta_{\vk_3}|\Omega(t)\rangle\simeq-2 {\rm Re}\left[\int_{-\infty(1-i\epsilon)}^\tau d\tau' \langle0|\zeta^{\rm int}_{\vk_1}(\tau)\zeta^{\rm int}_{\vk_2}(\tau)\zeta^{\rm int}_{\vk_3}(\tau) H_{\rm int}(\tau') |0\rangle\right]
\ee

\item At this order in perturbation theory, $H_{\rm int}=-\int d^3x {\cal L}_{\rm int}$. Pay attention, this is partially non trivial!
Our ${\cal L}_{\rm int}$ is given by
\bea
&&{\cal L}_{\rm int}=-\frac{4}{3} M_{3}^4 \int d^3x\; a^4 \left(\frac{1}{a(\tau)}\frac{\d\pi(\vec x,\tau)}{\d\tau}\right)^3=\\ \nonumber
&&\qquad=-\frac{4}{3} M_{3}^4\int d^3k_1\, d^3 k_2\, d^3 k_3\; a\; \delta^3(\vk_1+\vk_2+\vk_3)\; \pi^{\rm int}_{\vk_1}{}'(\tau)\pi^{\rm int}_{\vk_2}{}'(\tau)\pi^{\rm int}_{\vk_3}{}'(\tau)
\eea
The factor $a^4$ is due to the fact that we are integrating in conformal time. 

\item Use that $\zeta=-H\pi$ and that
\be
 \pi^{\rm int}_{\vec k}(\tau)=\pi_{\vk}^{cl}(\tau) a^\dag_{\vk}+\pi^{cl}{}^\star_{\vk}(\tau)a_{-\vk}
\ee
with
\be
\pi^{cl}_k(\tau)=-\frac{1}{H} \frac{c_s}{(2\epsilon)^{1/2} \mpl}\frac{1}{(2 c_s k)^{3/2}}(1-i c_s k\tau)e^{i c_s k \tau}\ , 
\ee

\item Perform the Wick contraction, and then perform the integral. The integral reads:
\bea
&&\langle\Omega(t)|\zeta_{\vk_1}\zeta_{\vk_2}\zeta_{\vk_3}|\Omega(t)\rangle=(- H^3)(-6) \times 2\times \frac{4}{3} M^4_3\\ \nonumber
&&\qquad\times {\rm Re}\left[\pi^{cl}_{\vk_1}(\tau)^\star\pi^{cl}_{\vk_2}(\tau)^\star\pi^{cl}_{\vk_3}(\tau)^\star \int_{-\infty(1-i\epsilon)}^\tau \pi^{cl}_{\vk_1}{}'(\tau') \pi^{cl}_{\vk_2}{}'(\tau') \pi^{cl}_{\vk_3}{}'(\tau') a(\tau')\,d\tau'  \right]\\
 \eea
 The results gives
 \be
\langle \Phi_{\vec k_1} \Phi_{\vec k_2} \Phi_{\vec k_3} \rangle=(2\pi)^3 \delta^{(3)}( \sum_i \vec{k}_i ) F(k_1,k_2,k_3)\ .
\ee
\be
F_{\dot\pi^3}(k_1,k_2,k_3)=\frac{20}{3}\left(1-\frac{1}{c_s^2}\right)\,\tilde c_3\,\cdot\Delta_\Phi^2\cdot\frac{1}{k_1 k_2 k_3(k_1+k_2+k_3)^3}\ .
\ee
where
\be
\Phi=\frac{3}{5}\zeta\ ,
\ee
 \be
 \Delta_\Phi=\frac{9}{25}\frac{H^2}{4 \epsilon\, c_s \mpl^2 }\ , \qquad M_3^4=\frac{\dot H \mpl^2}{c_s^4}\tilde c_3\ .
 \ee
 For $\tilde c_3\sim 1$, we have that the unitarity bound associated to the operator in $M_3$ is the same as the one from the operator in $M_2$.
 
 \item  The standard definition of $f_{\rm NL}$ is
\be
F(k,k,k)=f_{\rm NL}\cdot\frac{6\Delta_\Phi^2}{k^6} \ ,
\ee
 This allows us to define
\bea\label{eq:newfnl}
&&f_{\rm NL}^{\dot\pi(\d_i\pi)^2}=\frac{85}{324}\left(1-\frac{1}{c_s^2}\right) \ , \\ \nonumber
&&f_{\rm NL}^{\dot\pi^3}=\frac{10}{243}\left(1-\frac{1}{c_s^2}\right)\left(\tilde c_3+\frac{3}{2}c_s^2\right)\ ,
\eea

\end{itemize}

\subsubsection{Shape of Non-Gaussianities}

\begin{itemize}

\item {\bf Huge information} 

We see that at leading order in derivatives we have two operators $\dot\pi^3$ and $\dot\pi(\d_i\pi)^2$.. Let us see the plots. We clearly see that there is a huge amount of information contained in the 3-point function. These are functions, not just numbers: they have maxima, minima, asymptotic behaviours, etc. For example, since there are two operators at leading order in derivatives, we get any linear combination of two different shapes.
%\vspace{7cm}

\begin{figure}[h!]
\begin{center}
\includegraphics[width=12cm]{equilateral_shape.pdf}
\includegraphics[width=12cm]{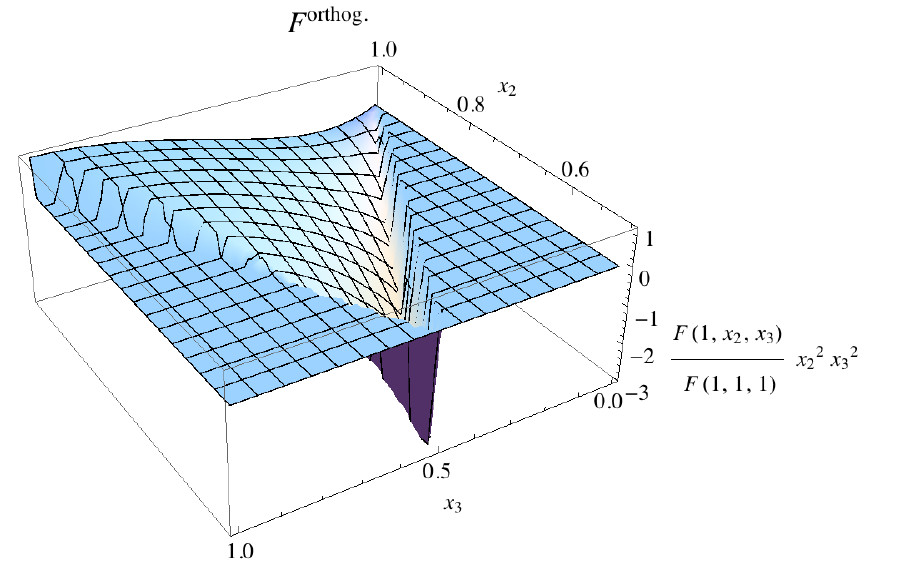}
\caption{\label{fig:shapes} \small Different shapes of the three-point function are obtained as we change the relative size of the operators $\dot\pi^3$ and $\dot\pi(\d_i\pi)^2$. The shape can peak on equilateral triangles, on flattened triangles~\cite{Creminelli:2005hu}, or on both, as in the case of the orthogonal shape~\cite{Senatore:2009gt}.}
\end{center}
\end{figure}

\item{\bf Local Shape:}

As we can see, the non-Gaussian signal from these models is always very small in the squeezed limit $k_3\ll k_1,k_2$. This is indeed a theorem due to Maldacena~\cite{Maldacena:2002vr,Creminelli:2004yq,Cheung:2007sv}. In reality, in some humble sense we are now beyond that theorem, because we have the Lagrangian for any single-degree-of-freedom inflationary model. We have therefore access to all the shapes that single-clock inflation can do: if we see something different, we exclude single-degree-of-freedom inflation. But still it is a remarkable feature of single degree of freedom inflation that in the squeezed limit the signal is so small. Can there be inflationary models that give large 3-point function in that limit? Yes, multi filed inflation can do that.

%\vspace{7cm}

\begin{figure}[h!]
\begin{center}
\includegraphics[width=12cm]{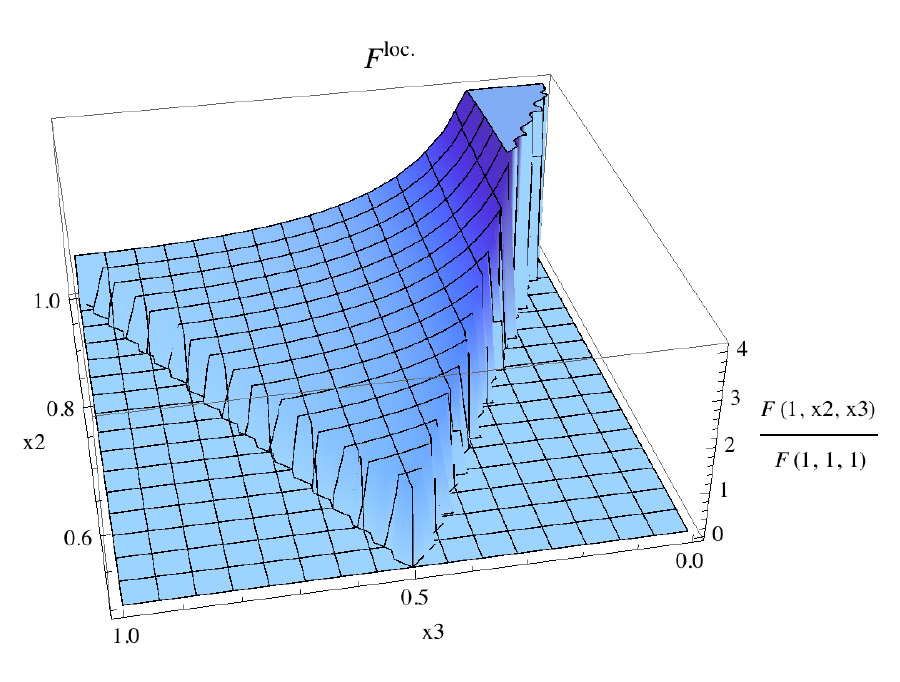}
\caption{\label{fig:local_shape} \small The local shape has a signal peaked on the squeezed triangles. It can be produced only in multi field inflationary models. See for example~\cite{Senatore:2010wk}.}
\end{center}
\end{figure}

A shape with a lot of signal there is a shape where the fluctuation $\zeta$ is defined in real space with the help of an auxiliary gaussian field:
\be
\zeta(\vec x)=\zeta_{gaussian}(\vec x)+\frac{6}{5}f_{\rm NL}^{local}\left(\zeta_{gaussian}(\vec x)^2-\langle\zeta_{gaussian}(\vec x)^2\rangle\right)
\ee
Its $F$ reads something like
\be\label{eq:local}
F_{local}(k_1,k_2,k_3)=\frac{1}{k_1^3 k_2^3}+\frac{1}{k_2^3 k_3^3}+\frac{1}{k_1^3 k_3^3}
\ee

Such a non-Guassianity is generated for example when the duration of inflation depends on a second field which fluctuates during inflation. For example, this could happen if the decay rate $\gamma$ of the inflation is determined by a coupling that depends in turns from a light field $\sigma$.

\begin{figure}[h!]
\begin{center}
\includegraphics[width=8cm]{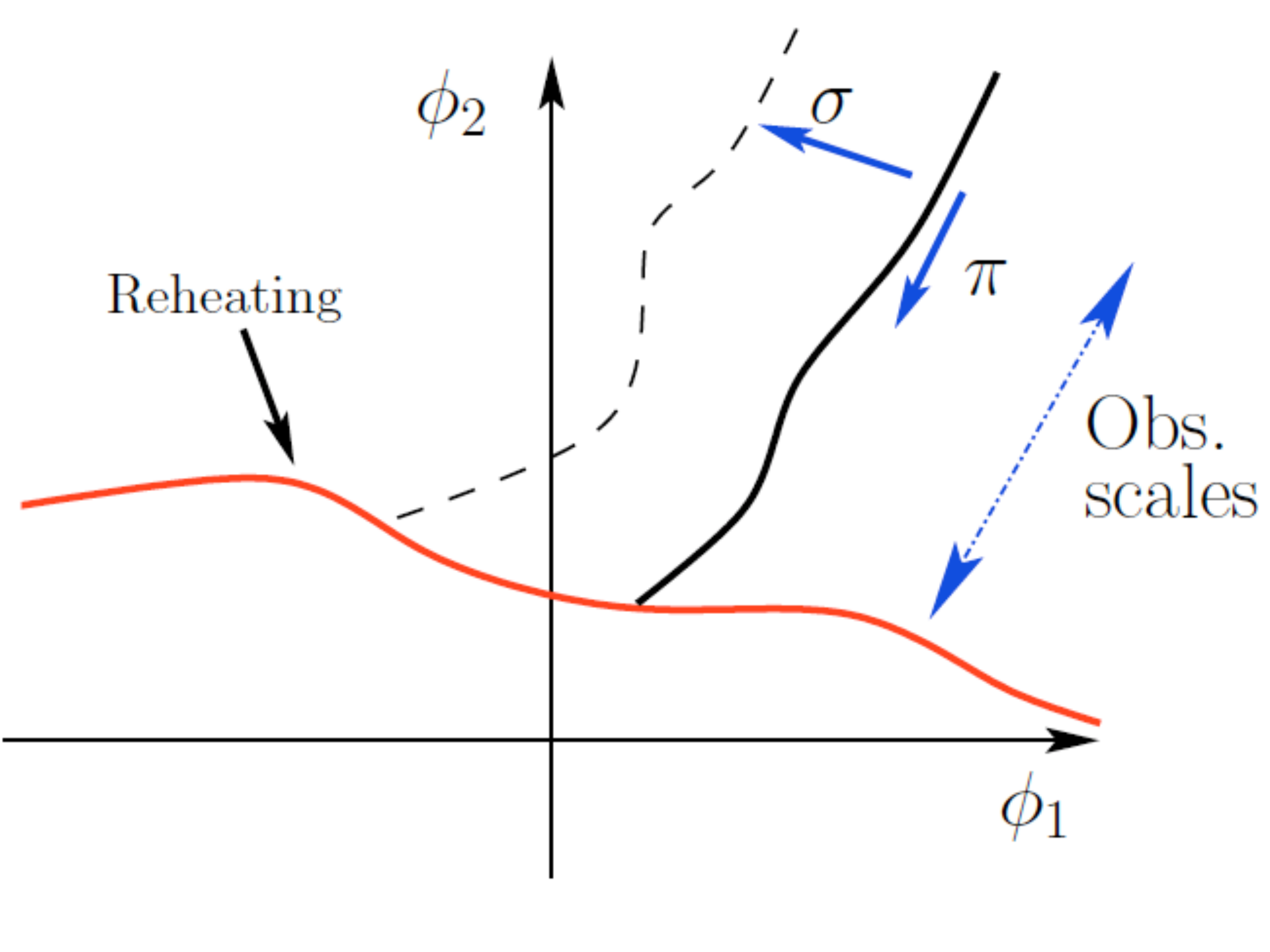}
\caption{\label{fig:two-gauss} \small Plot of typical multi field inflationary potential. Fluctuations of the second $\sigma$ field affect the duration of inflation and therefore the curvature perturbation of the universe at reheating. If the relationship between $\sigma$ and $\zeta$ is non-linear, then non-Gaussianities of the local kind are produced.}
\end{center}
\end{figure}

In this way:
\be
\frac{\delta a}{a}= \zeta(\vec x)= f(\Gamma(\{\sigma\}))
\ee
Since the conversion of the $\sigma$ fluctuations into $\delta a/a$ happens when all the interesting modes are outside of $H^{-1}$, the relation above must be local in real space:
\be
\zeta(\vec x)= f(\Gamma(\sigma(\vec x)))
\ee
Since the non-gaussianities are quite small, the linear term must dominate. We can taylor expand $f$:
\be
\zeta(\vec x)\simeq a_0+ a_1 \sigma(\vec x)+ a_2\sigma(\vec x)^2\equiv \zeta_{gaussian}(\vec x)+\frac{3}{5}f_{\rm NL}^{local}\left(\zeta_{gaussian}(\vec x)^2-\langle\zeta_{gaussian}(\vec x)^2\rangle\right)
\ee

\item Another interesting option to generate detectable non-Gaussian signal is if during the epoch of inflation there is a sector of particles that are not heavier than the Hubble scale. If these particles do not affect the duration of the inflationary epoch directly, but rather interact with the inflaton, and if they have a non-negligible mass or spin, they induce a peculiar non-Guassian signal. In particular, if the exchanged particle has scalar mass $m$, the squeezed limit is given by~\cite{Chen:2009zp}
\be
\lim_{\{k_1/k_2,k_1/k_3\}\to 0,\ k_2\simeq k_3}\langle\zeta_{\vec k_1}\zeta_{\vec k_2}\zeta_{\vec k_3}\rangle\propto \frac{1}{k_1^3 k_2^3}\left(\frac{k_1}{k_2}\right)^{\frac{3}{2}-\sqrt{\frac{9}{4}-\frac{m^2}{H^2}}}  (2\pi)^3\delta^{(3)}\left(\vec k_1+\vec k_2+\vec k_3\right) \ .
\ee
Notice that for $m\simeq 0$, we obtain the same squeezed limit as in~(\ref{eq:local}), but now there is a whole range of possible power laws, which somewhat covers the intermediate range in squeezed limits between multifield inflation and single field inflation, which as the following squeezed limit~\cite{Creminelli:2011rh,Creminelli:2013cga} 
\be
\lim_{\{k_1/k_2,k_1/k_3\}\to 0,\ k_2\simeq k_3}\langle\zeta_{\vec k_1}\zeta_{\vec k_2}\zeta_{\vec k_3}\rangle\propto \frac{1}{k_1^3 k_2^3}\left(\frac{k_1}{k_2}\right)^{2}  (2\pi)^3\delta^{(3)}\left(\vec k_1+\vec k_2+\vec k_3\right) \ .
\ee
Similar squeezed limits, with a somewhat different range, are present when the exchanged particles are strongly coupled~\cite{Green:2013rd}.
Even more interesting squeezed limits are obtained when considering the exchange of particles with spin, as recently described in~\cite{Arkani-Hamed:2015bza}. This is particularly interesting in the sense that these particles do not have a scale invariant spectrum of perturbations, so that they can lead to a visible signal only through the effect that comes from exchanging them.

\item {\bf Particle Physics Knowledge} 

Limits on non-Gaussian signatures get translated into limits onto limits of the parameters of the inflationary Lagrangian. See Fig.~\ref{fig:tcountour}. Cosmological observations are mapped directly into parameters of a fundamental physics Lagrangian$\ldots$ the sky is like a particle accelerator!  This approach was developed in~\cite{Senatore:2009gt}.

%\vspace{7cm}

\begin{figure}[h!]
\begin{center}
\includegraphics[width=15cm]{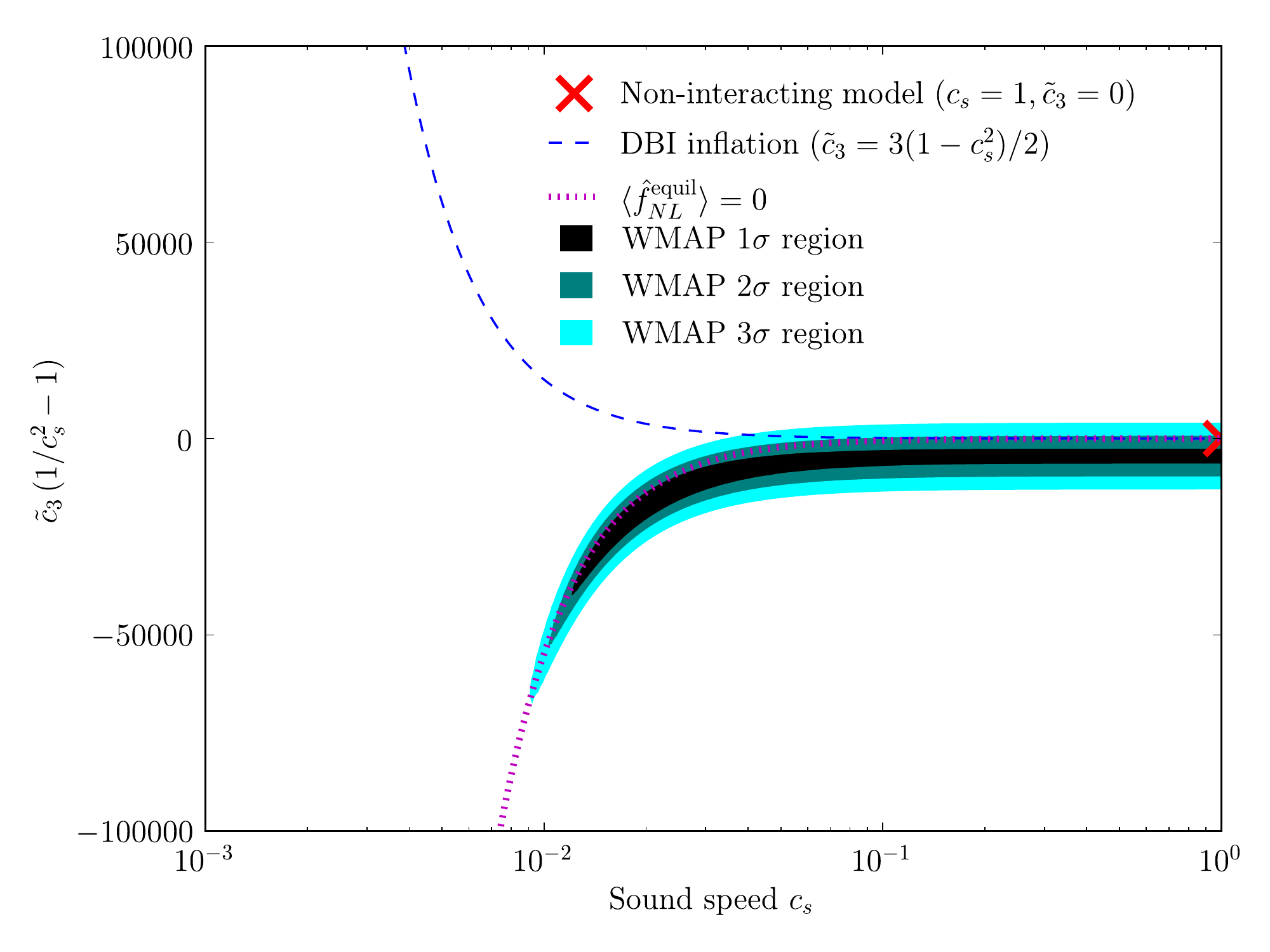}
\caption{\label{fig:tcountour} \small Contour plot for the parameters of the EFT Lagrangian $c_s$ and $\tilde c_3$ from WMAP7yr. Figure takes from~\cite{Senatore:2009gt}.}
\end{center}
\end{figure}

\end{itemize}

There is really a lot more to say about non-Guassianities and the EFT of Inflation. Non-Gaussianities have really become a large field in inflationary cosmology, and maybe this is happening also for the EFT of inflation, as this is the ideal set up to study interactions. Indeed, many additional developments have been made in this field, that I have no time to mention: EFT of multi field inflation, impose additional symmetries on $\pi$. such as  Supersymmetry, discrete shift symmetry, parity, etc. É.. roughly, all what we have been doing in Beyond the Standard Model physics has now motivation to be applied to inflation and the EFT of inflation offers the simple connection.

%I leave you with the current Planck constraints at $2\sigma$~\cite{Bennett:2012fp}:
%\bea
%&&-221<f_{\rm NL}^{\rm equil.}<323\ , \\ \nonumber
%&&-445<f_{\rm NL}^{\rm orthog.}<-45\ , \\ \nonumber
%&&-3<f_{\rm NL}^{\rm loc.}<77\ . \nonumber
%\eea
%We have a $2.5\sigma$ evidence that the orthogonal shape is non-zero. Further data, and in particular Planck, will tell us.
%Here is the plot countour plot of parameters of the EFT of Inflation from the WMAP team himself in Fig.~\ref{fig:wmap9ng}~\cite{Bennett:2012fp}. They use the EFT of Inflation to interpret their results on non-Gaussianities.

I leave you with the current Planck constraints at $2\sigma$~\cite{Ade:2013ydc}
\bea
&&-156<f_{\rm NL}^{\rm equil.}<124\ , \\ \nonumber
&&-100<f_{\rm NL}^{\rm orthog.}<32\ , \\ \nonumber
&&-9<f_{\rm NL}^{\rm loc.}<14\ . \nonumber
\eea
We see that there is no evidence of non-zero $f_{\rm NL}$.
Even the Planck team uses the EFT of Inflation to interpret their non-Gaussianity constraints. Here is the contour plot of the parameters of the EFT Lagrangian from the Planck team, Fig.~\ref{fig:planckng}~\cite{Ade:2013ydc}. Already the WMAP team had used the EFT to put their (weaker but earlier) limits~\cite{Bennett:2012fp}.

\begin{figure}[h!]
\begin{center}
\includegraphics[width=8cm]{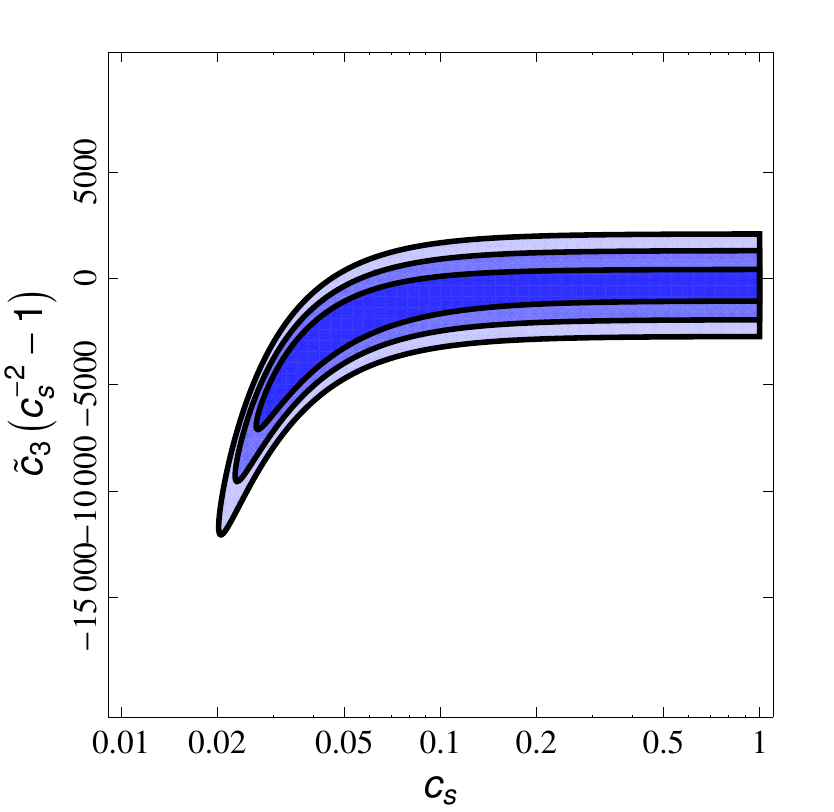}
\caption{\label{fig:planckng} \small Contour plots of parameters of the EFT of Inflation Lagrangian from the Planck team~\cite{Ade:2013ydc,Ade:2015ava}. Figure from~\cite{Ade:2015ava}.}
\end{center}
\end{figure}

\begin{figure}[h!]
\begin{center}
\includegraphics[width=5cm]{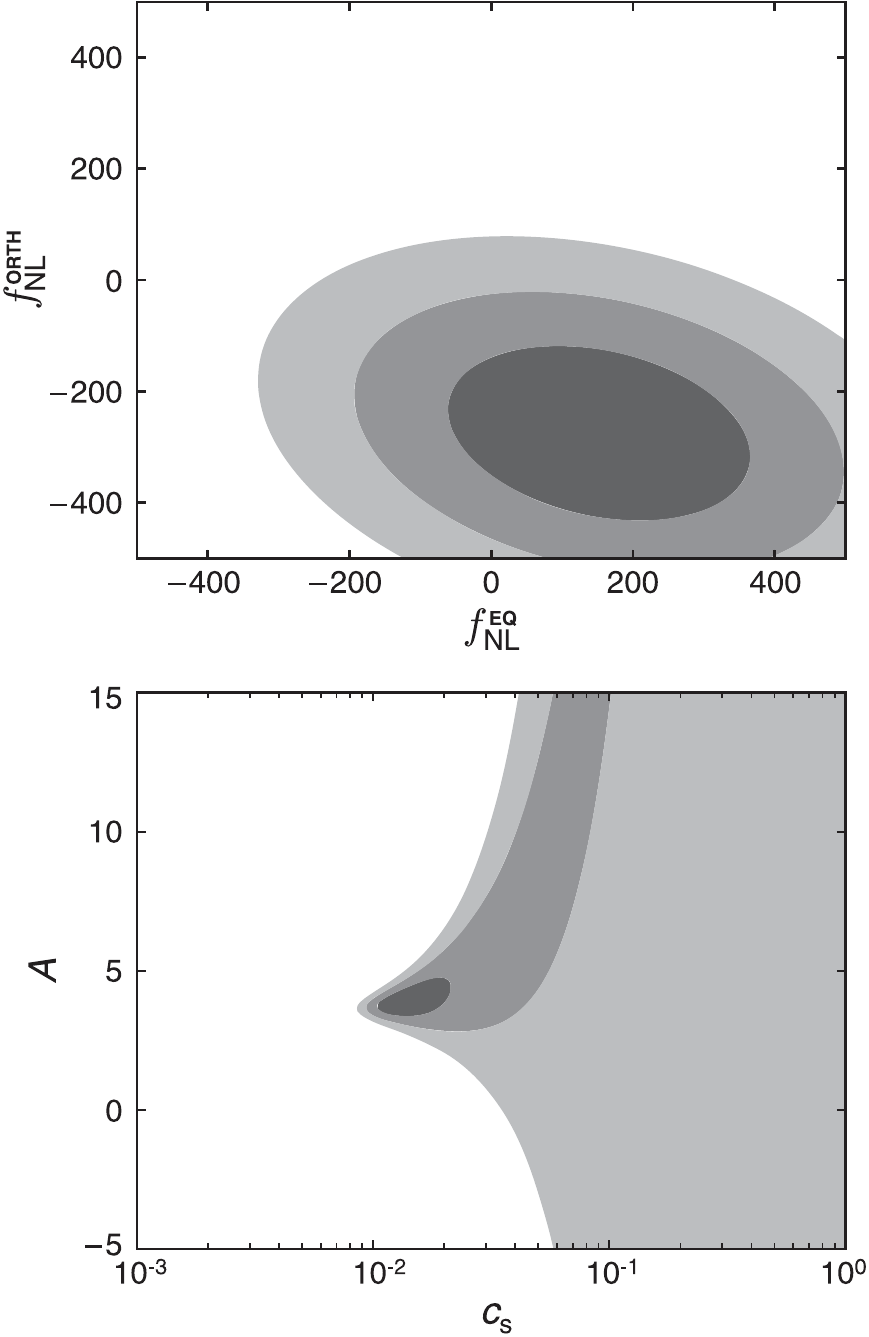}
\caption{\label{fig:wmap9ng} \small Contour plots of parameters of the EFT of Inflation Lagrangian from the earlier WMAP team~\cite{Bennett:2012fp}. The coefficient $A$ is related to $\tilde c_3$. Figure from~\cite{Bennett:2012fp}.}
\end{center}
\end{figure}

Given the absence of detection non-Gaussianities from Planck, one might wonder if non-Gaussianities are by now very much constrained. This is not quite so. The amount of non-Gaussianities is dictated by the size of the mass scale $\Lambda_U$ suppressing the interaction, dimension 6, operators. It should be made clear that we do not have any strong theoretical prior of what this mass scale should be. It is arguably one of the greatest results of the EFT to show that it is possible to have large non-Gaussianities. However, this does not mean that we have a strong theoretical prior in favor of having a small $\Lambda_U$. For this dimension 6 operators, $\Lambda_U^4\sim c_s^5 \dot H\mpl^2$. A natural scale to wonder about is if $\Lambda_U$ can be made greater than $\dot H \mpl^2$, something that would require to constrain $c_s\sim 1$, or $f_{\rm NL}\sim 1$. This threshold is interesting because if we take standard slow roll inflation, we have $\dot\phi^2\sim \dot H\mpl^2$. So, if we were able to show, by bounding non-Gaussianities, that $\Lambda_U^4\gg \dot H\mpl^2$, we would know that standard slow roll inflation would be an allowed UV complition of the EFT of inflation. However, we are currently very far from this. Very roughly, we have that $\Lambda_U^2\sim 10^{3} H^2$, or equivalently $f_{\rm NL}\sim 10^2$. Notice that Planck improvements did not change much this estimate: in going from WMAP to Planck, error bars on $f_{\rm NL}$ shrinked by a factor of 3, so $\Lambda_U$ went up by a mere factor of $\sqrt{3}$. Given that we had no strong theoretical prior, no nearby threshold for $\Lambda_U$ to cross, this is not a such an improvement that can change the theory. Planck results could have been a great opportunity to learn a lot of new physics from a detection of non-Gaussianity; absence of detection is not changing the theory~\footnote{This is to be contrasted with LHC, where absence of a detection of an Higgs  or something like that that would unitarize $WW$ scattering at high energies, would have forced us to change quantum mechanics. The point is that a few hundreds GeV energy was a very strong threshold for the theory of the Standard Model of Particle Physics.}. In order to change the theory, observational progress must be greater, and, unfortunately, this is not easy at all! Still, in the  next decade, there will be large scale structure surveys such as LSST or Euclid, that can potentially promise to decrease our limits on $f_{\rm NL}$ by, in the some optimistic estimates, even by a factor of 10 or so.

\subsection{Summary of Lecture 4}

\begin{itemize}
\item We learnt what non-Gaussianities mean.
\item We learnt how to estimate their size,
\item And how to compute them accurately.
\item Non-Gaussianities contain a huge amount pod information. They represent a very non-trivial signal.
\item They teach us about the interacting part of the theory, and, thanks to the EFT of inflation, their measurement can be mapped into measurements of parameters of a fundamental Lagrangian.

\end{itemize}

\newpage
\section{Lecture 5: Eternal Inflation}

\subsection{Slow-roll Eternal Inflation}

I would like to give you a brief introduction to eternal inflation. This is one of the most fascinating solutions of general relativity, in which quantum effects make a otherwise classically ending inflationary solution, actually never ending and eternal. Let us start with slow roll eternal inflation, and work in the context of a slow rolling scalar field for simplicity. Let us consider a classical slow rolling inflationary solution, as the one represented in Fig.~\ref{fig:inflation}. In the typical time scale of the problem, which is Hubble, and in the typical patch of volume of order $H^{-3}$, the field performs a classical advancement of order 
\be
\Delta\phi_{\rm cl}\sim \dot\phi_{0} H^{-1}\sim \frac{(-\dot H\mpl^2)^{1/2}}{H}\ .
\ee 
In the same time, the field undergoes a quantum fluctuation of order $H$
\be
\Delta\phi_{\rm quantum}\sim H\ .
\ee
It is pretty clear that as $\Delta\phi_{\rm cl}\ll \Delta\phi_{\rm quantum}$, there is an equal probability of going backwards in the potential as in going forward. Given that if the inflaton goes backwards it takes some time to get to the starting point, and in the meantime the volume expands exponentially, creating many new patches that undergo the same jumps, it becomes pretty clear that in the regime  $\Delta\phi_{\rm cl}\ll \Delta\phi_{\rm quantum}$ there is some chances for not all the spacetime points reaching the end of inflation, and therefore for inflation to become never ending. This is called slow roll eternal inflation.

We therefore expect that slow-roll eternal inflation to happen when the potential is very flat. We expect a phase transition as soon as 
\be
\Delta\phi_{\rm cl}\lesssim \Delta\phi_{\rm quantum}\quad\Rightarrow\quad \frac{-\dot H \mpl^2}{H^4}\lesssim 1\ ,
\ee
or, in terms of slow roll parameters,
\be
\epsilon\lesssim \epsilon_c\sim \frac{H^2}{\mpl^2}\ . 
\ee
Notice some peculiarities of this regime. When the potential is very flat and $H\ll \mpl$, the slow roll parameters are very small, and therefore the self interactions of the inflation become very small and the metric fluctuations are small. This means that the inflationary regime is very well described by a free scalar field living in unperturbed quasi de Sitter space. Notice that since there is a constant drift towards the bottom of the potential, every point will sooner or later exit the inflationary region. Inflation will be eternal simply because each patch produces many other patches before exiting inflation. The situation for the space time is quite different in those region that have exited inflation at a given time. Since $\zeta\sim\delta\rho/\rho|_{\rm after\ inflation}$ goes as ${H^4}/(\dot H \mpl^2)$, we will have in those regions the overdensities are of order one, and therefore the description of the spacetime, locally after the de Sitter epoch, will be very complicated. The situation can be represented in this Fig.~\ref{fig:eternal}. One sees that in each each region of space inflation ends, but the amount of time inflation lasts at each point is very different. In particular, one can concentrate on the well defined surface of constant $\phi=\phi_{\rm reheating}$ and compute its volume. As we approach the eternal regime, the effect of the quantum fluctuations on the duration of inflation becomes larger and larger, and so the volume of the reheating surface is not always the same, and it is therefore better described by a probability distribution. This probability distribution has exactly zero support at infinite volume for $\epsilon\gtrsim \epsilon_c$. As we approach the critical $\epsilon_c$, the typical volume begins to grow, and at the critical point given by, up to subleading slow roll corrections,
\be
\Omega=\frac{4\pi^2}{3}\frac{\dot H\mpl^2}{H^4}=1\ ,
\ee
the probability distribution of having infinite volume becomes non-zero: $P(V=\infty)\neq 0$. This is the onset of slow roll eternal inflation. I would like you to realise how non-trivial and actually beautiful this fact is. Quantum effects, that are usually relegated to the world of small distances, are here having a huge effect on the largest possible distances, actually distances of order of the whole universe. The only other solution of general relativity I am aware of where quantum physics have effect on Astrophysical scales is the Black Hole evaporation by Hawking~\cite{Hawking:1974sw}, according to which otherwise eternal Black Holes slowly evaporate through a quasi thermal radiation. This has been of course a fantastic theoretical discovery. Now, the slow roll eternal inflationary solution is in my opinion even more spectacular. Since the Black Hole evaporation happens, for Astrophysical Black Holes, through the emission of a huge number of photons, the geometry of the space is always described by a well determined metric. In the case of slow roll eternal inflation, the whole spacetime is no more described by a determined manifold, but rather by a stochastic, semiclassical, one. This is, to me, beautiful. This approach to the study of slow roll eternal inflation and the first studies of this probability distribution were developed first in~\cite{Creminelli:2008es}, where the first quantitative understanding of slow roll eternal inflation was made since the first seminal papers of~\cite{Guth:1980zm,Linde:1982ur,Steinhardt:1982kg,Vilenkin:1983xq,Linde:1986fd,Goncharov:1987ir}. 

\begin{figure}[h!]
\begin{center}
\includegraphics[width=12cm]{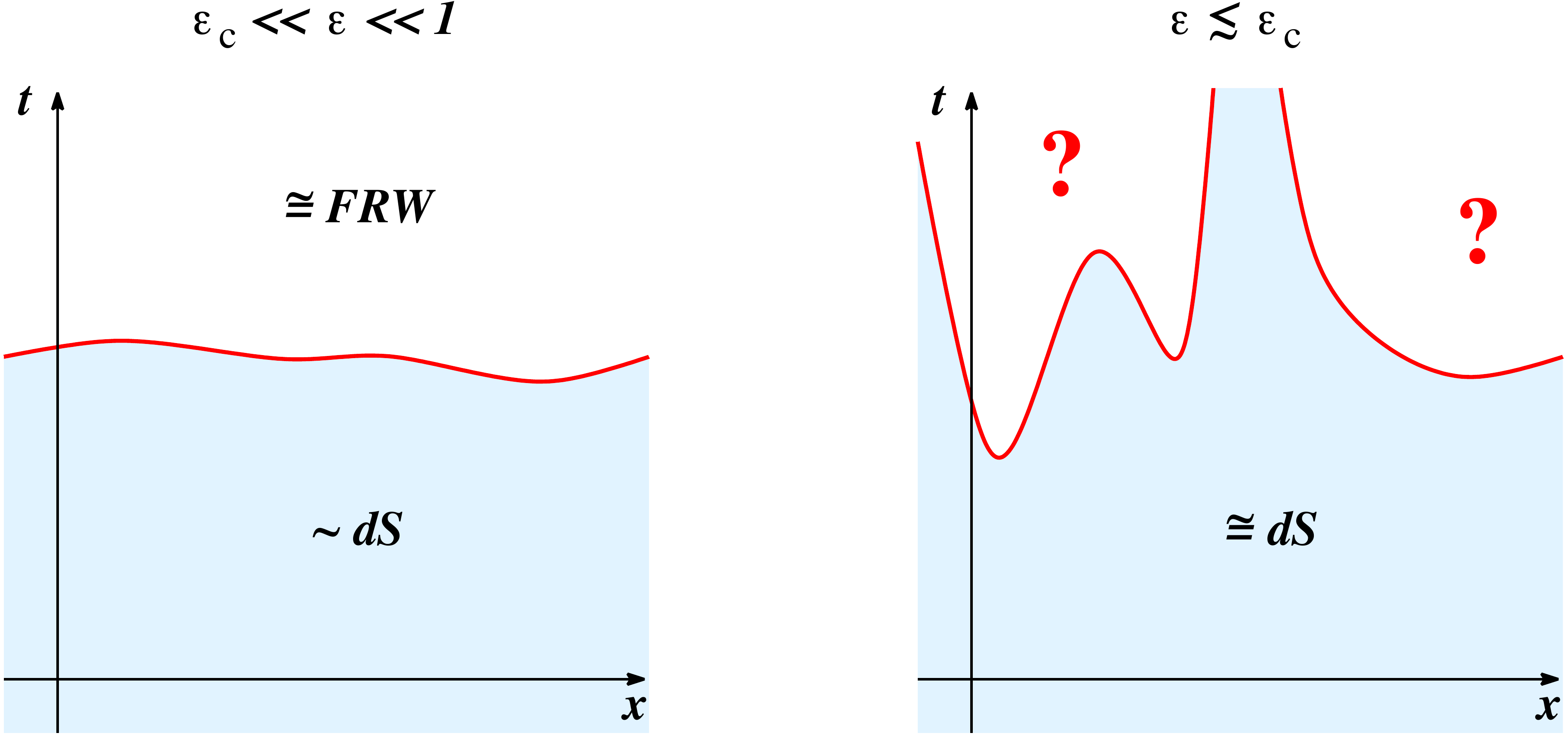}
\caption{\label{fig:eternal} \small As the slow roll parameter becomes smaller and smaller, the spacetime during inflation becomes more and more unperturbed, and the one after inflation more and more perturbed. The surface of constant $\phi=\phi_{\rm reheating}$, in red, becomes more and more perturbed, curved, reaching infinite volume  with a non-zero probability in the eternal regime. Figure from~\cite{Creminelli:2008es}.}
\end{center}
\end{figure}

Notice that the situation with slow roll eternal inflation is very similar to what happens with the evolution of a population with infinite resources: everybody die, but on average everybody make some children. If the average number of children is higher than a critical number, the population will become eternal and will have an infinite number of elements; otherwise it will extinct. Therefore, by using techniques shared by this field, it is possible to actually study the probability distribution of the volume of the reheating surface~\cite{Dubovsky:2008rf}. If we call $N_{\rm cl}$ the classical number of $e$-foldings so that, neglecting quantum fluctuations, the volume of the reheating surface $V$ is of order  Exp$[3N_{\rm cl}]$, we find the following quite remarkable result. As the slow roll parameter gets smaller and smaller and approaches $\epsilon_c$, the width of the distribution passes from being very small to be of order one, so that it is still quite peaked around the average; while the average moves from  Exp$[3N_{\rm cl}]$ to Exp$[6N_{\rm cl}]$ at the phase transition. This is a huge boost in the overall volume: a factor of 2 in the exponent. Then, as we pass beyond the phase transition at $\epsilon\lesssim \epsilon_c$, we develop some finite probability of infinite volume, while the average of the finite volume does not increase, and it actually starts receding towards smaller values. This happens as, with a very flat potential, it becomes very improbable to make a large volume that is not infinite. See Fig.~\ref{fig:rhoV-proto}.

\begin{figure}[h!]
\begin{center}
\includegraphics[width=11cm]{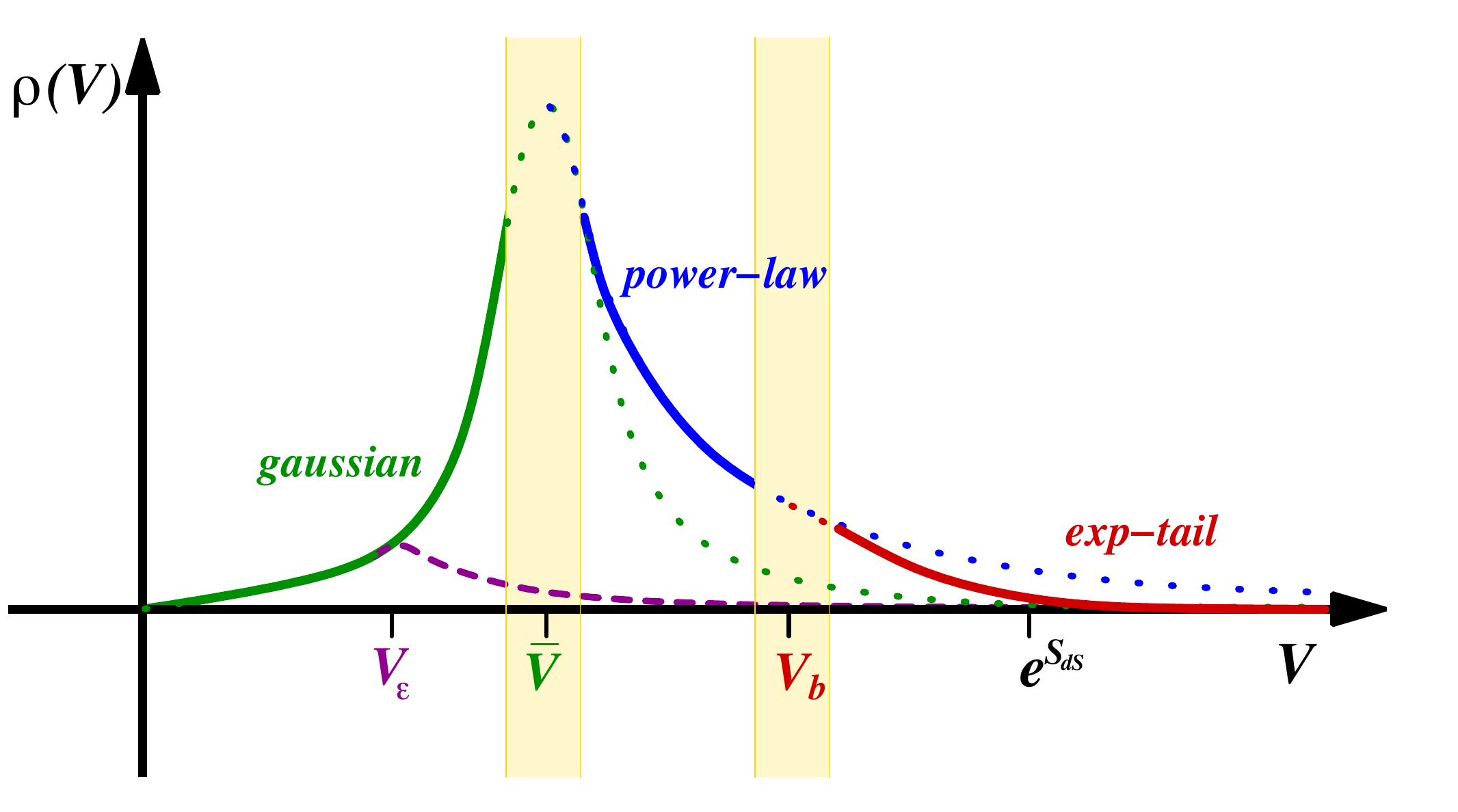} 
\caption{\label{fig:rhoV-proto} \small Typical shape for the probability distribution of the volume $\rho(V)$.
For small volumes the behavior is gaussian with the number of $e$-foldings ($\rho\sim e^{-c(N-\overline N)^2}$); for volumes larger
than the average value $\overline V$, $\rho(V)$ follows a power law in the volume ($\rho\sim 1/V^\alpha$) that eventually
turns into an exponential law ($\rho \sim e^{-{\rm const}\cdot V}$) at large enough volumes ($V\gtrsim V_b$, with $V_b$ representing the classical volume obtained if the inflation started from a barrier at the top of the potential). When $\epsilon<\epsilon_c$ the exponential tail starts earlier at $V\simeq V_\epsilon=e^{\pi/(2\sqrt{1-\Omega})}$ and the integral of the probability distribution for finite volumes becomes smaller than one. The average volume is $\bar V={\rm Exp}\left[\tfrac{2 N_{\rm cl}}{1+\sqrt{1-\Omega^{-1}}}\right]$. We see that as $\Omega$ starts very large in standard inflation and approaches the phase transition, the average volume interpolates between Exp$[3N_{\rm cl}]$ to Exp$[6N_{\rm cl}]$. Figure from~\cite{Dubovsky:2008rf}.}
\end{center}
\end{figure}

It is quite remarkable that there is a maximum finite volume that can be produced by inflation, the probability of obtaining a volumes larger than the average being exponentially small. Interesting, this maximum value can be expressed in terms of the de Sitter entropy, to give 
\be
V_{\rm finite,\; max}\lesssim e^{S_{\rm dS}/2}
\ee
with $S_{\rm dS}=\pi \mpl^2/H^2$ representing the de Sitter entropy of the inflationary space at reheating~\cite{Dubovsky:2008rf}. This bound generalizes at quantum level the classical one which gives $V_{\rm finite,\; max}\lesssim e^{S_{\rm ds}/4}$~\cite{ArkaniHamed:2007ky}. Even more remarkably, this bound remains unchanged as we change the number of space dimensions, the number of fields involved in inflation, and also the higher derivative correction to the theory of inflation and gravity~\cite{Dubovsky:2011uy}. Somehow, the volume produced by inflation, when finite, it is always smaller than Exp$[S_{\rm dS}/2]$. The  sharp physical interpretation of this bound, which is sharp and universal, is still unknown, though its universality seems to suggest a possibly deep meaning. It is fair to say that it looks like we just scratched the surface of this very interesting quantum mechanical solution of general relativity.

\subsection{False-vacuum Eternal Inflation}

There is another kind of eternal inflation, called false vacuum eternal inflation~\cite{Guth:1982pn}. This happens in the following context. Suppose we have a potential of the following form, Fig.~\ref{fig:false_vaccum}.  There is a false vacuum and a true vacuum.  The false vacuum is classically stable, but metastable thought quantum tunneling. This is a non-perturbative process, usually dominated by the Coleman de-Luccia instanton~\cite{Coleman:1980aw}. Because it is non-perturbative, this is usually a very slow process compared to the typical mass scales in the potential. If we call the decay rate per unit space-time volume, $\Gamma$, this will go more or less as the inverse of the exponential of the action associated to the instanton that interpolates between the two vacua:  Exp$[-S_{\rm instanton}]$. If the inflaton happens to find himself in the false vacuum, it will stay there for a relatively long amount of time. Notice now that it can be, as it is in Fig.~\ref{fig:false_vaccum}, that the energy of the false vacuum configuration,  $\Lambda_{\rm false}$, is positive. Because of gravity, if the inflaton will happen to be in the false vacuum, the universe will start expanding in a de Sitter like manner, with an Hubble rate of oder $\Lambda_{\rm false}/\mpl^2$. Since the false vacuum configuration is metastable, each point of space will sooner or later decay to the true vacuum. It will do so by producing a bubble of true vacuum, than that expand at approximately the speed of light into the false vacuum, similarly  to bubbles of water when boiling. However, since the false vacuum is expanding itself, if two bubbles will be nucleated too far part, farther than about $H^{-1}$, they will never collide. There is therefore the possibility that, if the decay rate is sufficiently slow, bubbles get continuously produced, but the space between bubbles expands more rapidly, so that the asymptotic situation is that an infinite volume with infinite bubbles is produced. This is represented in Fig.~\ref{fig:eternal_false}. Bubbles that nucleated earlier have had more time to expand, but as time goes on, the volume of the false vacuum grows as well.

\begin{figure}[h!]
\begin{center}
\includegraphics[width=7cm]{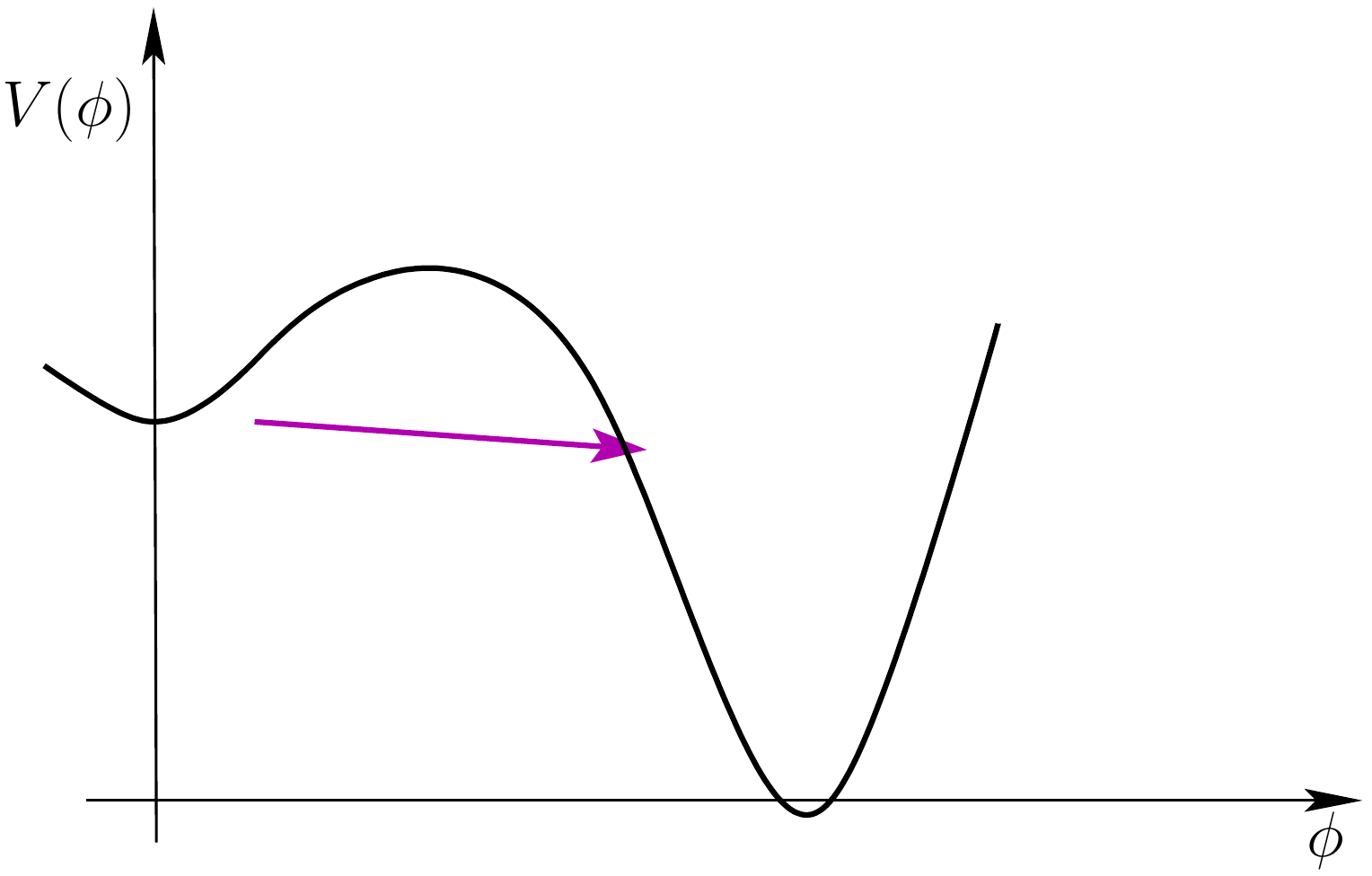} 
\caption{\label{fig:false_vaccum} \small A potential with an absolute minimum and a de Sitter local minimum. If the inflaton happens to be in the false vacuum and the tunnelling rate is sufficiently small, this situation leads to false vacuum eternal inflation.}
\end{center}
\end{figure}

\begin{figure}[h!]
\begin{center}
\includegraphics[width=7cm]{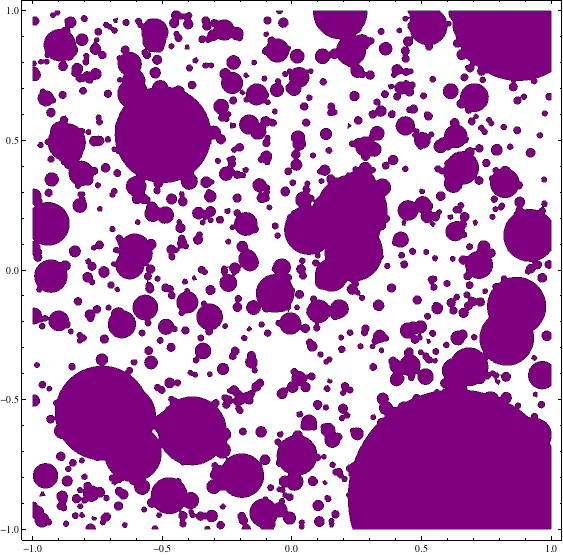} 
\caption{\label{fig:eternal_false} \small A representation of a region of universe that is undergoing false vacuum eternal inflation. Bubbles that nucleated earlier have had more time to expand, but as time goes on, the volume of the false vacuum grows as well.
 Figure from~\cite{Kleban:2011pg}.}
\end{center}
\end{figure}

The transition between eternal and non-eternal inflation is governed by the ratio $\Gamma/H^{4}$. We actually encounter two phase transitions as we start from a very small $\Gamma$ and we increase it to cross the region $H^{4}$~\cite{Guth:1982pn}. Both of these two phase transitions happen for $\Gamma\sim H^4$. First, $\Gamma$ will become fast enough so that bubbles will percolate, that is they become able to form chains that connect the two sides of the box, even though not the whole space will be filled by true vacuum. Then, as we increase $\Gamma$ even more, the true vacuum bubbles will fill up the whole space, and no false vacuum region will remain. The actual numerical point at which this transition happen is still unknown, to my knowledge, though some numerical studies have been done.

\subsection{Summary of Lecture 5}

\begin{itemize}
\item Eternal Inflation is a solution of GR and Quantum Mechanics where quantum effects change the asymptotic space-time.
\item Classically finite universes become infinite.
\item There are two kinds of Eternal Inflation: False Vacuum and Slow Roll.
\item Their current understanding in only partial.
\end{itemize}

\newpage
\section{Summary}

This is all Guys. 

In these lectures we have started from the shortcomings of Big Bang Cosmology that motivated inflation. We have seen how a period of accelerated expansion fixes all these problems. With simple estimates that are helpful to develop intuition, we have seen how inflation produces a quasi scale-invariant, quasi-Gaussian, stochastic but classical, spectrum of density perturbations, and how some qualitative predictions of inflation have been confirmed in the data. We have also seen that it would be great to have something more to look for. For this reason, we have introduced the Effective Field Theory of Inflation, which shows that Inflation is essentially a theory of a Goldstone boson. We have seen that there are new spectacular signatures in inflation: the non-Gaussianity of the density perturbation. They contain a huge amount of information, and they represent the interactions, and therefore the non-trivial dynamics, of the inflationary Lagrangian.

Inflationary physics is very ample, and there are many aspects that we could not touch. For example we did not discuss how some inflationary models are embedded in string theory, or, in any detail, that beautiful phase called eternal inflation, according to which quantum effects change the asymptotic of the space-time, arise.

In any event, for all what concerns the phenomenology of Inflation and its connection to the data, you should be good to go.

Thank you very much for your attention and your interactions. Teaching here has been a wonderful experience for me, and it has been a pleasure to have you around and discuss with you. I hope you'll find these lectures useful for your future research in Physics and Cosmology. It is a great moment for our field.

My best wishes.

\section*{Acknowledgments}
This work is supported by DOE Early Career Award DE-FG02-12ER41854 and by NSF grant PHY-1068380.  I thank Luca Delacretaz and Matt Lewandowski for collaborating in formalizing the careful construction of the interacting vacuum that is presented in lecture 4, and in general all the students who gave me feedback.

  \newpage
 \begingroup\raggedright\endgroup

%\bibliography{trapbib}

\end{document}